\documentclass[Journal]{IEEEtran} %draftclsnofoot

\usepackage{cite} 
\usepackage{graphicx}
\usepackage[english]{babel}
\usepackage{amsmath}
\usepackage{amsthm}
\usepackage[table,xcdraw]{xcolor}
\usepackage{subfig}
\usepackage[linesnumbered, ruled]{algorithm2e}
\usepackage[noend]{algpseudocode}
\usepackage[font=footnotesize]{caption}
\usepackage{epsfig}
\usepackage{xfrac}
\usepackage{nicefrac}
\usepackage{psfrag}
\usepackage{times}
\usepackage{balance} 
\usepackage{amssymb}
\usepackage{amsfonts}
\usepackage{bm}
\usepackage[bottom]{footmisc}
\usepackage{mwe}
\usepackage{color}
\usepackage{booktabs,multirow}
\usepackage{tabularx}
\usepackage{tikz}
\usetikzlibrary{patterns,arrows}

\usepackage{algorithm}

\usepackage{verbatim}

\SetKwRepeat{Do}{do}{while}
\SetKwInOut{Input}{Input}
\SetKwInOut{Output}{Output}

\definecolor{color1}{RGB}{228,26,28}
\definecolor{color2}{RGB}{55,126,184}
\definecolor{color3}{RGB}{77,175,74}
\definecolor{color4}{RGB}{152,78,163}
\definecolor{color5}{RGB}{255,127,0}

\begin{document}

%%%%%%%%%%%%%%%%%%%%%%%%%%%%%%%%%%%%%%%%%%%%%%%%%%%%%%%%%%%%%%%%%%%%%%%%%%%%%%%%%%%%%%%%%%%%%%%%%%%%%%
%%%%%%%%%%%%%%%%%%%%%%%%%%%%%%%%%%%%%%%%%%%%%%%%%%%%%%%%%%%%%%%%%%%%%%%%%%%%%%%%%%%%%%%%%%%%%%%%%%%%%%
% \title{\fontsize{21}{24}\selectfont Joint Satellite Association and UAV Mobility Management for Non-Terrestrial Network via Multi-Agent Reinforcement Learning }
% \title{\fontsize{21}{24}\selectfont Dynamic Planning of LEO Satellite and UAV for Non-Terrestrial Networks via Multi-Agent Reinforcement Learning }
% \title{\fontsize{21}{24}\selectfont Relaying while Flying: Dynamic Planning of LEO Satellite and UAV via Multi-Agent Reinforcement Learning}
% \title{\fontsize{24}{24}\selectfont Integrating LEO Satellite and UAV for Energy-Efficient Non-Terrestrial Network Design via Multi-Agent Reinforcement Learning }

% \title{\fontsize{23}{28}\selectfont Energy-Efficient Multi-UAV Reinforcement Learning for LEO Satellite Integrated Non-Terrestrial Networks}

% Integrating LEO Satellite and UAV Relaying for Hybrid FSO/RF Non-Terrestrial Networks

\title{\fontsize{21}{28}\selectfont Integrating LEO Satellites and Multi-UAV Reinforcement Learning for Hybrid FSO/RF Non-Terrestrial Networks}

% \title{\fontsize{21}{24}\selectfont Relaying while Flying: Energy-Efficient Non-Terrestrial Network Design via Multi-Agent Reinforcement Learning}
% \title{\fontsize{21}{24}\selectfont Energy-Efficient Design of LEO Satellites and UAVs for Non-Terrestrial Networks via Multi-Agent Reinforcement Learning}
% \title{Energy-Efficient Non-Terrestrial Network Design via Multi-Agent Reinforcement Learning}

% \author{ 
% Ju-Hyung Lee$^\dag$, Jihong Park$^*$, Mehdi Bennis$^\ddag$, and Young-Chai Ko$^\dag$ \\

% 	\small $^\dag$Electrical and Computer Engineering, Korea University,  
% 	\small Seoul, Korea \\
% 	\small $^*$ School of Information Technology, Deakin University, Geelong, VIC 3220, Australia\\
% 	\small $^\ddag$Centre for Wireless Communications, University of Oulu, 90014 Oulu, Finland \\
% 	\small leejuhyung@korea.ac.kr, jihong.park@deakin.edu.au, mehdi.bennis@oulu.fi, koyc@korea.ac.kr 
% 	}

\author{Ju-Hyung Lee,~\IEEEmembership{Student Member,~IEEE,}
		Jihong Park,~\IEEEmembership{Member,~IEEE,}\\
		Mehdi Bennis,~\IEEEmembership{Senior Member,~IEEE,}
        and~Young-Chai Ko,~\IEEEmembership{Senior Member,~IEEE}% <-this % stops a space
\thanks{J.-H. Lee, and Y.-C. Ko are with the School of Electrical and Computer Engineering, Korea University, Seoul, Korea (Email: leejuhyung@korea.ac.kr; koyc@korea.ac.kr).

$^\dagger$J. Park was with the University of Oulu, Finland, and is now with the School of Information Technology, Deakin University, Geelong, VIC 3220, Australia (Email: jihong.park@deakin.edu.au).

$^\ddagger$M. Bennis is with the Centre for Wireless Communications, University of Oulu, Oulu 90014, Finland (Email: mehdi.bennis@oulu.fi).

Part of this work is to be presented in 2020 IEEE Global Communications Conference (GLOBECOM).
}% <-this % stops a space
% \thanks{Manuscript received April XX, 20XX; revised January XX, 20XX.}
}
\maketitle
%%%%%%%%%%%%%%%%%%

\begin{abstract}

A mega-constellation of low-altitude earth orbit (LEO) satellites (SATs) and burgeoning unmanned aerial vehicles (UAVs) are promising enablers for high-speed and long-distance communications in beyond fifth-generation (5G) systems. Integrating SATs and UAVs within a non-terrestrial network (NTN), in this article we investigate the problem of forwarding packets between two faraway ground terminals through SAT and UAV relays using either millimeter-wave (mmWave) radio-frequency (RF) or free-space optical (FSO) link. Towards maximizing the communication efficiency, the real-time associations with orbiting SATs and the moving trajectories of UAVs should be optimized with suitable FSO/RF links, which is challenging due to the time-varying network topology and a huge number of possible control actions. To overcome the difficulty, we lift this problem to multi-agent deep reinforcement learning (MARL) with a novel action dimensionality reduction technique. Simulation results corroborate that our proposed SAT-UAV integrated scheme achieves $1.99$x higher end-to-end sum throughput compared to a benchmark scheme with fixed ground relays. While improving the throughput, our proposed scheme also aims to reduce the UAV control energy, yielding $2.25$x higher energy efficiency than a baseline method only maximizing the throughput. Lastly, thanks to utilizing hybrid FSO/RF links, the proposed scheme achieves up to $62.56$x higher peak throughput and $21.09$x higher worst-case throughput than the cases utilizing either RF or FSO links, highlighting the importance of co-designing SAT-UAV associations, UAV trajectories, and hybrid FSO/RF links in beyond-5G NTNs.

% Furthermore, we answer the three questions; 
% 1) how much throughput is achieved for the LEO SAT constellation, 2) what is the efficient way of improving throughput for the LEO SAT-based NTN, and 3) which connectivity links (e.g., optical communication, RF communication) are more favorable for NTN based scenarios.

\end{abstract}

\begin{IEEEkeywords}	
% \tred{Satellite internet constellation, high-altitude platform, non-terrestrial network, hybrid FSO/RF, beyond 5G, association, path-planning.}
LEO satellite, UAV, non-terrestrial network, hybrid FSO/RF, multi-agent deep reinforcement learning.
\end{IEEEkeywords}

%%%%%%%%%%%%%%%%%%%%%%%%%%%%%%%%%%%%%%%%%%%%%%%%%%%%%%%%%%%%%%%%%%%%%%%%%%%%%%%%%%%%%%%%%%%%%%%%%%%%%%%%%%%%%%%%%%%%%%%%%%%%%%%%%%%%%%%%%%%%%%%%%%%%%%%%%%%%%%%%%%%%%%%%%%%%%%%%%%%

%%%%%%%%%%%%%%%%%%%%%%%%%%%%%%%%%%%%%%%%%%%%%%%%%%%%%%%%%%%%%%%%%%%%%%%%%%%%%%%%%%%%%%%%%%%%%%%%%%%%%%%%%%%%%%%%%%%%%%%%%%%%%%%%%%%%%%%%%%%%%%%%%%%%%%%%%%%%%%%%%%%%%%%%%%%%%%%%%%%
% INTRODUCTION %
%%%%%%%%%%%%%%%%%%%%%%%%%%%%%%%%%%%%%%%%%%%%%%%%%%%%%%%%%%%%%%%%%%%%%%%%%%%%%%%%%%%%%%%%%%%%%%%%%%%%%%%%%%%%%%%%%%%%%%%%%%%%%%%%%%%%%%%%%%%%%%%%%%%%%%%%%%%%%%%%%%%%%%%%%%%%%%%%%%%
\section{Introduction} 

% \tred{[JH: Fill in the missing REFs]}

Witnessing the recent developments in non-terrestrial network (NTN) \cite{White_Loon, Deployment_OneWeb}, we are at the cusp of a communication revolution where space is envisaged to meet the ground through the sky \cite{Intro_1_3, Intro_1, DRL_JH}. On the one hand, space is emerging as the new frontier in beyond fifth-generation (5G) communication, through which the high-speed wireless coverage is extended to remote areas even in the ocean \cite{SAT_Coverage}. This trend has been encouraged by the recent launches of low-altitude earth orbit (LEO) satellite (SAT) mega-constellations \cite{Oneweb, Kuiper, Telesat}. In particular, SpaceX's 713 {Starlink} SATs have already been provisioning $>100$ Mbps data rate with $<20$ ms latency for ground terminals \cite{FCC_Starlink} while orbiting at an altitude of $550$~km~\cite{Starlink}. The data rate is comparably fast, and the latency is even lower than wired connections thanks to the direct links between ground terminals and overflying SATs. Such SAT-assisted systems have great potential particularly in enabling low-latency and long-range communications \cite{UCL_2, UCL_1}. The key challenge is to overcome the time-varying SAT network topology 
% due to the orbiting SATs that are hardly controlled 
for improving communication efficiency.

On the other hand, the sky is a burgeoning frontier wherein unmanned aerial vehicles (UAVs), ranging from small drones to high altitude platforms (HAPs), can assist terrestrial communication systems in 5G and beyond \cite{UAV_Survey1,UAV_Survey2,UAV_Survey3}. There are a variety of UAV-assisted communication applications such as offloading ground terminal traffic in urban areas \cite{UAV_Offloading}, providing emergency networks for disaster sites \cite{UAV_Disaster}, and extending the wireless coverage to rural areas (e.g., Google's \textit{Loon} \cite{Loon}, Facebook's \textit{Aquila} \cite{Aquila}, and HAPSMobile's \textit{HAWK30} \cite{HAPSMobile}). The benefits of these applications are rooted in the flexible maneuvering capability of UAVs. However, the coverage per UAV is smaller than SATs due to the low altitude, limiting the scalability and applicability within relatively short-range communications.

% https://mashable.com/article/spacex-starlink-internet-speed-latency/
% https://ecfsapi.fcc.gov/file/109041365616217/SpaceX%20Degani%20Ex%20Parte%20(9-4-20).pdf

\begin{figure*}
  % \vspace{-1em}
  \centering
  \includegraphics[width=\textwidth]{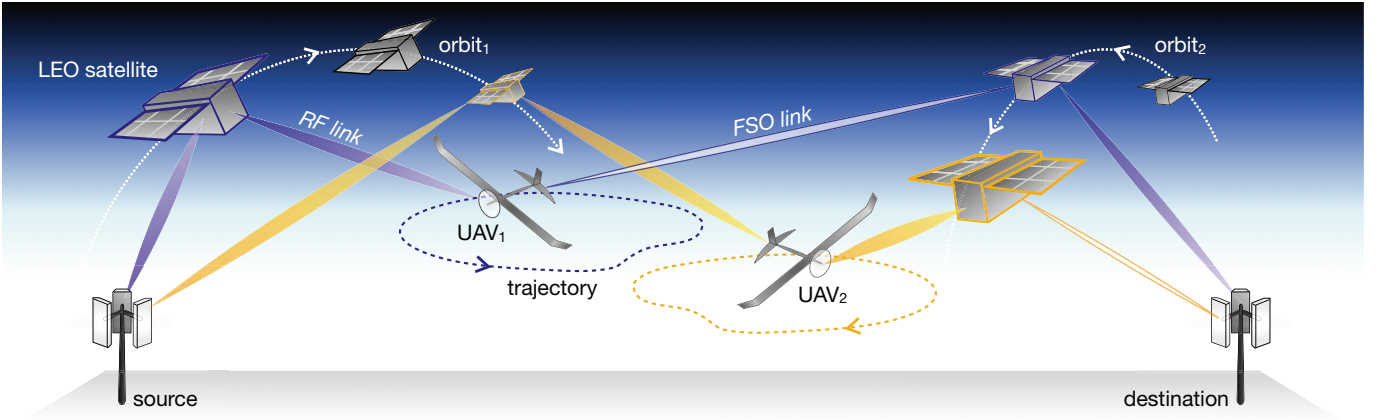}
  \caption{An illustration of an SAT-UAV integrated NTN wherein packets are forwarded between ground source and destination terminals through the two (purple and yellow) paths formed by SAT and UAV relayins utilizing hybrid FSO/RF (filled/void) links.}
  \label{Illustration}
  % \vspace{-.5em}
\end{figure*}

Motivated by the aforementioned complementary advantages of SATs and UAVs, in this article we study the problem of forwarding packets between two faraway ground terminals through multiple SATs and UAVs, as depicted in Fig.~\ref{Illustration}. The long-distance communications therein are enabled by the multiple orbital lanes of SATs while the communication efficiency is improved by controlling UAVs relaying the inter-orbit traffic. To this end, for the given SAT orbiting dynamics, UAVs optimize their locations and determine the associations with SATs and ground terminals in real time, so as to maximize the long-term average end-to-end data rate while minimizing the UAV control energy. To cope with the time-varying SAT topology, a novel multi-agent deep reinforcement learning (MARL) framework is applied for UAV control and associations. Furthermore, to take into account different SAT and UAV altitudes and channel characteristics, free-space optical (FSO) and millimeter wave (mmWave) radio frequency (RF) communication links are considered for maximizing the data rates. More details on the backgrounds and contributions of this work are elucidated in the following subsections.

% source terminal (Src) and destination terminal (Dst), relayed through an LEO satellite (SAT) and a high altitude platform (HAP) such as a fixed-wing UAV or an airship drone, as illustrated in Fig. 1. To maximize the end-to-end (E2E) data rate of the Src-SAT-HAP-Dst link, we aim to optimize the Src- SAT-HAP association while adjusting the HAP location in real time. 

% Solving this problem is non-trivial due to the orbiting SATs and the resultant time-varying network topology.

% , 74 \textit{OneWeb} SATs at $1,200$ km \cite{Oneweb}) 

% wherein LEO SATs, UAVs, and ground terminals are seamlessly integrated.

% such as the low-altitude earth orbit (LEO) satellite (SAT) mega-constellations  

% While SAT has until now been an assistant for wireless connectivity, the recent launches of the SpaceX Starlink constellation \cite{Starlink}, and the growing publicity of the plans of other companies for major constellations of thousands of LEO SAT \cite{Oneweb, Kuiper, Telesat}, have shown an new potential of SAT In beyond fifth-generation (B5G) communication.

% With the widely observed industrial projects, leveraging network flying platforms (NFPs) for the purpose of internet connectivity, we envision that the NTN, including LEO SATs and UAVs, extends base station connectivity from one station to another.

%%%%%%%%%%%%%%%%
\subsection{Related Works}
% \tred{[JH: Based on the revised introduction above, revise the subsections A and B accordingly]}

\subsubsection{Non-Terrestrial Network}

%%%%%% UAV통신은 성능과 통신시나리오마다의 최적화는 제안되어있다.
Ground-based wireless connectivity has already been extending towards the sky by integrating UAVs.
As opposed to fixed ground base stations, these UAV terminals can be mobile and flexibly deployed, provisioning large-scale three-dimensional (3D) wireless connectivity.
Many studies, for instance \cite{UAV_Multiple3, UAV_Multiple1, UAV_3D_Online}, have carried out the optimization for multi-UAV communication.
% These works have improved various target performances. 
% with certain optimization variables.
For instance in \cite{UAV_Multiple3}, network throughput has been maximized by optimizing the association and power for UAVs allocation.
% Particularly, the authors of \cite{UAV_Multiple3} have focused on the interference channels incurred from the multi-UAV which may represent the downlink (DL) scenario, and multi-users which may represent the uplink (UL) scenario, respectively.
Besides in \cite{UAV_Multiple1}, latency has been minimized by optimizing the cell association and multi-UAV location.

%%%%%% LEO 통신의 가능성에 대한 연구와 성능분석이 밝혀져 있다.
% Towards delivering high speed internet access across the globe, SAT constellation has been dealt with in the industry (such as SpaceX's \textit{Starlink} \cite{Starlink}, \textit{OneWeb} \cite{Oneweb}, Amazon's \textit{Kuiper} \cite{Kuiper}, and \textit{Telesat} \cite{Telesat}) as well as in the academia. 
Likewise, towards provisioning high-speed global Internet access, SAT constellation is currently under heavy investigation by the industry (such as SpaceX's \textit{Starlink} \cite{Starlink}, \textit{OneWeb} \cite{Oneweb}, Amazon's \textit{Kuiper} \cite{Kuiper}, and \textit{Telesat} \cite{Telesat}) and academia.
Recent research works \cite{UCL_2, UCL_1} have advocated that exploiting SAT relays can achieve faster communication for long distances $>3000$ km compared to terrestrial optical fiber links. 
% It is noted that in \cite{UCL_2}, the effectiveness of a fixed ground relay in-between SAT links has been studied, as opposed to this work considering a moving UAV relay between SAT links.
It is noted that in \cite{UCL_2}, lower latency performance can be achieved through additional fixed ground relays in-between the SAT relays.

%%%%%% 하지만, 이 다양한 타입의 NFP를 통합하는 경우에 대한 연구가 부족하며, 특히 제일 대두되는 HAP 와 LEO SAT를 통합하는 연구가 필요하다.
NTN, including UAVs and SATs, has been studied in academia and formally organized by 3GPP \cite{Intro_4}.
However, existing works have focused only on NTN with UAVs or NTN with SATs, even though NTN deployment scenarios include both space and airborne platforms. 
These two aerial platforms have been separately investigated, in contrast to our work in which SAT and UAV relays are jointly optimized. 
% In other words, it is not yet established how to efficiently integrate the space-borne vehicles and airborne vehicles for more practical NTN scenarios.

\subsubsection{Deep Reinforcement Learning}

%%%%%% NTN에서 다양한 목적 성능을 최적화하기 위해, 이러이러한 방식들이 사용되었다.
Towards improving the communication efficiency of NTNs, a variety of optimization techniques have been applied, such as successive convex approximation (SCA) method \cite{SCA_YZ, SCA_JH}, block coordinate descent (BCD) method \cite{BCD, BCD_JH}, and graph theoretical \cite{Graph} method.
% for efficient network design.
%%%%%% 하지만, 이런 방식들은 ... 의 한계점을 가지고 있다.
These approaches, however, require the global knowledge of the system parameters, which is often not feasible in practice. Furthermore, these methods commonly incur huge computing overhead and/or rely on multiple approximations to tackle the non-convex nature of the problems.

% Even with perfect information of all the relevant system parameters, optimization problems of current communication scenarios are highly non-convex and challenging to solve.

%%%%%% 그렇기에 다양한 작업들이 RL로 접근한다.
To overcome the aforementioned limitations, several recent works have proposed deep reinforcement learning (DRL) based solutions \cite{DRL_UAV, MADRL_UAV}, a machine learning based decision-making framework. In \cite{MADRL_UAV}, DRL for multi-agent, so-called multi-agent deep reinforcement learning (MARL), is proposed for the distributed trajectory design of cooperative multi-UAV. 
% While DRL has attracted growing attention in NTN design particularly, to our best knowledge, its application of jointly designing SATs association as well as UAVs path-planning in NTN  has not been reported.
While DRL has attracted growing attention in designing NTNs, to our best knowledge, the problem of jointly designing SAT associations and UAV path-planning (See Fig. \ref{Illustration}) has not been studied in this literature.

\subsubsection{FSO/RF Link}

Unlike terrestrial networks, wireless communication systems operating at very high frequencies, such as mmWave and FSO communications, are widely considered in NTN. 
% Unlike terrestrial networks where it is difficult to secure line-of-sight (LoS) link conditions due to the multi-path fading and blockage, wireless communication system on very high frequency, such as millimeter-wave communication (mmWave) and free-space optical communication (FSO), is widely considered in NTN.
% Accordingly, as the channel characteristics of NTN is distinguished from these of terrestrial networks, various link candidates is considered in the NTN. 
For instance, \textit{Aquila} \cite{Aquila} considered FSO for ground-to-air and air-to-air links, and \textit{Starlink} \cite{Starlink} considered both RF and FSO for the inter-satellite links (ISLs). 
In addition to industrial projects, academia have expressed a significant interest in FSO-based NTN, by leveraging the inherent characteristics of FSO communication, e.g., unlicensed broad-spectrum, immunity to electromagnetic interference, and security \cite{BackhaulFronthaul_UAV_FSO, Backhaul_Hybrid_RFFSO_1, Backhaul_Hybrid_RFFSO_3}.
% NTN, which serve as relay terminals (e.g., eNodeB) in the broad internet access, is connected by the ground-to-air and air-to-air links.
% Furthermore, the communication through the aerial relay terminals may propagate over very long link distance, from tens to thousands of [km].
% % Thus, according to the different channel condition in NTN compared with terrestrial networks, various link options should be considered.
% Accordingly, as the channel characteristics of NTN is distinguished from these of terrestrial networks, various link candidates, such as FSO and hybrid FSO/RF, can be considered in the NTN. 
Nevertheless, few studies \cite{Fronthaul_FSO_UAV, Hybrid_RFFSO_UAV} have presented an analysis of the possible link types for NTN, and those few studies have not considered time-varying scenarios of SATs.
% In other words, it is not yet established enough which link type is effective in NTN.

Motivated by the potential of LEO SAT-based internet constellation, our previous works of \cite{DRL_JH} investigated SAT supported by high-altitude platform (HAP) relaying with RF links. 
The previous work, however, only considered a single UAV for only one selected SAT. 
Also, there was no consideration of other link options of SAT-UAV communication. % 
Accordingly, several questions still remain regarding whether integrating multi space-borne vehicles and multi airborne vehicle is feasible and efficient, and whether RF or other link candidates are effective in this scenario. 
In this work, we answer these challenging questions by jointly optimizing the SAT associations and UAV locations in time-varying multi-topologies, using a novel MARL approach of distributed-Actor and centralized-Critic.
Besides, we present key insights concerning which relaying scheme is efficient and which link candidate is a better option for the NTN scenario.

% \tred{some questions [JH: Let's avoid `some' particularly for our key remarks. Clarify what the key questions are]}
% Accordingly, the following questions for a practical NTN scenario are remaining; whether integrating multi space-borne vehicles and multi airborne vehicles is feasible and efficient, and which link type is effective. 
% In this work, we answer the challenging questions by jointly optimizing the SAT associations and UAV locations, with a novel MARL approach of distributed-Actor and centralized-Critic.

% Besides, we present interesting points concerning which relaying scheme is efficient and which link candidate is a better option for the NTN scenario.

%%%%%%%%%%%%%%%%
\subsection{Contributions and Paper Organization}
The main contributions of this work are summarized as follows.
\begin{itemize}
\item  A novel problem has been formulated ($\mathrm{P1}$ in Sec. \ref{Body}), which jointly optimizes multi-SAT association and multi-UAV trajectory under a time-varying network topology, so as to maximize the system throughput of long-range non-terrestrial communication while minimizing the energy consumption of the system. 
% Specifically, we take the communication between London and New York into account as our network scenario.
% Looking especially at trans-oceanic scale (such as between London and New York) communication with few relays,
Considering RF and FSO links, we examine the performance of both link candidates for NTN.
To the best of our knowledge, this is the first work taking into account the joint association of SATs in multi-orbit, and the mobility management for multi-UAV, in the context of non-terrestrial communication.

\item A MARL based solution has been proposed (see Sec.~\ref{Body_ML}). 
Dealing with the scenario of distributed multi-agents (e.g., multi-SAT and multi-UAV), we propose a network structure of distributed-Actor and centralized-Critic.
To cope with a large action space due to time-varying network topology, a novel action dimension reduction technique has been proposed, which focuses only on a couple of SATs proximal to Src and Dst.

\item 
In the optimization problem, we examine the feasibility and the performance of the LEO SAT mega-constellation based NTN, with the following perspectives; 
Firstly, considering the end-to-end (E2E) throughput, the SAT-based NTN is investigated and propose possible improvements.
Besides, the possible link types for NTN (e.g., RF, FSO links) are analyzed, and the E2E throughput of each candidate link is compared in our NTN scenario.

\item 
Numerical results corroborate that the ground-relaying achieves an average system throughput $2.883$x higher than a baseline of SAT-Only (which uses only SAT-based relaying). 
Furthermore, the mobile UAV relaying obtains $1.988$x higher than the case of fixed ground relaying, while the case in which the association is optimized acquires $1.988$x higher throughput than otherwise.
This suggests using additional relays in-between SATs to provide high-throughput wide area networking.
Furthermore, this highlights the importance of SAT association and UAV mobility management in enabling high-throughput non-terrestrial communication.
\end{itemize}

The remainder of this article is organized as follows. 
In Sec.~\ref{System Model}, the SATs-UAVs assisted NTN architecture and hybrid FSO/RF channel model are presented. 
In Sec.~\ref{Body}, the energy efficiency maximization problem, in which the E2E system throughput is maximized while the energy consumption of UAV relay is minimized, is formulated. In Sec.~\ref{Body_ML}, a MARL based solution is proposed.
In Sec.~\ref{Numerical Result}, simulation results are provided, followed by concluding remarks in Sec.~\ref{conclusion}.

%%%%%%%%%%%%%%%%%%%%%%%%%%%%%%%%%%%%%%%%%%%%%%%%%%%%%%%%%%%%%%%%%%%%%%%%%%%%%%%%%%%%%%%%%%%%%%%%%%%%%%%%%%%%%%%%%%%%%%%%%%%%%%%%%%%%%%%%%%%%%%%%%%%%%%%%%%%%%%%%%%%%%%%%%%%%%%%%%%%

%%%%%%%%%%%%%%%%%%%%%%%%%%%%%%%%%%%%%%%%%%%%%%%%%%%%%%%%%%%%%%%%%%%%%%%%%%%%%%%%%%%%%%%%%%%%%%%%%%%%%%
% SYSTEM MODEL %
%%%%%%%%%%%%%%%%%%%%%%%%%%%%%%%%%%%%%%%%%%%%%%%%%%%%%%%%%%%%%%%%%%%%%%%%%%%%%%%%%%%%%%%%%%%%%%%%%%%%%%
\section{System Model} \label{System Model}

%%% (시나리오)
The network under study consists of two SAT constellations rotating around a given orbital lane and additional UAV relay terminal in-between SATs, as illustrated in Fig. \ref{Illustration}. 
In particular, we consider a multi-hop communication link forwarding packets from source terminal (Src) to destination terminal (Dst), via SAT and UAV relays.
For the multi-hop communication, hybrid FSO/RF communication links are considered for the ground-to-air channel as well as the air-to-air channel.

%%% Index
We denote a set of SAT constellation network $\mathcal{I}_{k}=\{i_{k} = 1,2,\ldots,I\}, k=1,2, \ldots K$ in an orbital lane $k$. 
A set of UAVs, which relay between the SAT constellation network, is denoted as $\mathcal{J}=\{j=1,2,\ldots,J\}$.
Two terrestrial terminals, such that one transmits while another receives, are denoted by $\mathcal{S}$ and $\mathcal{D}$, respectively.
Particularly, $\mathcal{S}$ refers to a Src and $\mathcal{D}$ refers to a Dst.

%%%%%%%%%%%%%%%%%%%%%%%%%%%%%%%%%%%%%%%%%%%%
% \tred{[JH: Let's remove subsubsections. We can have the subsections A, B, and C, by combining A and 1), and making 2) as B and B as C.]}

% \subsection{System Model for SAT and UAV}
% \subsubsection{System Model for Joint SAT-UAV Relaying}

% \tred{[JH: This part may sound confusing. We assume fixed altitudes separately for SATs and UAVs, and consider their 2D coordinates that are associated with the fixed altitudes (i.e., 2D coordinates marked with fixed altitudes).]}
Based on three dimensional Cartesian coordinates for the location of the terminals, we assume that the backhaul terminal and the terrestrial base station are located at position $\mathbf{q}_{\mathcal{S}}$ and $\mathbf{q}_{\mathcal{D}}$, respectively, while SAT $i$ orbits at the fixed altitude of $H_{\mathrm{L}}$
% \footnote{
% Satellite constellation Starlink by \textit{Space X} has launched the vehicle to orbit altitude of $550$ [km] for operation \cite{Starlink}.
% } 
with constant speed $\mathbf{v}^{k}_{\mathrm{L}}$ following a predetermined orbital lane $k$.
On the orbital lane, SATs are spaced at equal intervals and circulated.
Particularly, we consider $22$ SATs in the orbital lane and the orbital lane circumference.
In addition, we consider that UAV $j$ flies horizontally in the $xy$-plane with a fixed altitude $H_{\mathrm{U}}$.
The time-varying coordinate of the UAV-relay node can be denoted in [km] as  $\mathbf{q}_{\mathrm{U}}^{j}(t)=[x(t),y(t),H_{\mathrm{U}}]^T \in\mathbb{R}^{3 }$ for $0\leq t \leq T$.
% Two orbital lanes of SATs and two UAV are considered.
Here, the UAVs serve in-between the two orbital lanes of SATs as additional relay terminals.
% \tred{[]JH: Don't we need to elaborate the control range of UAVs (i.e., Does P1 still hold when two UAVs are located at different inter-orbit regions?)]}

% in this network scenario to show a practical situation.

% \subsubsection{Discrete State-Space Model for SAT and UAV}
% \tred{[JH: Discrete state-space is confusing. It should be at least discrete-time (and continuous state values); Right above, we have defined the coordinates based on continuous time although there seems no reason to define them in the continuous space; To avoid any possible confusion, it may be better not to mention the discreateness in the title]}

% For ease of analysis, we consider a discrete-time model as in \cite{BCD, BCD_JH}.  
% The time horizon $T$ is divided into $N$ time intervals each with duration $\delta_{t}$, i.e. $T=N \cdot \delta_{t}$. 
% The duration $\delta_{t}$ is chosen to be sufficiently small so that the positions of UAVs are adequately approximated within each slot.
% Thus, the position of UAV $j$, $\mathbf{q}^{j}_{\mathrm{U}}(t)$, can be approximated in a discrete-time model, i.e. $\mathbf{q}^{j}_{\mathrm{H}}[n] \triangleq \mathbf{q}^{j}_{\mathrm{U}}(n\delta_{t}) = [x_{j}(n\delta_{t}),y_{j}(n\delta_{t}),H_{\mathrm{U}}]^T = [x_{j}[n],y_{j}[n],H_{\mathrm{U}}]^T \in\mathbb{R}^{3}$,  $0 \leq n \leq N+1$.

To obtain a more tractable optimization problem, we apply the discrete linear state-space approximation similarly to \cite{SCA_YZ}. 
Based on the time step size $\delta_{t}$, time $T$ (or $t$) and time slot $N$ (or $n$) can be determined according to $T=\delta_{t} \cdot N$ (or $t=\delta_{t} \cdot n$).
Accordingly, the position of UAV $j$, $\mathbf{q}^{j}_{\mathrm{U}}(t)$ and velocity $\mathbf{v}^{j}_{\mathrm{U}}(t)$ can be well characterized by the discrete-time UAV position vector $\mathbf{q}^{j}_{\mathrm{U}}[n] =\mathbf{q}^{j}_{\mathrm{U}}(n \delta_{t})$ as well as the velocity vector $\mathbf{v}^{j}_{\mathrm{U}}[n] =\mathbf{v}^{j}_{\mathrm{U}}(n \delta_{t})$ and the acceleration vector $\mathbf{a}^{j}_{\mathrm{U}}[n] =\mathbf{a}^{j}_{\mathrm{U}}(n \delta_{t})$ for $n = 0, 1, \cdots, N + 1$.
Therefore, the discrete state of UAV~$j$ can be expressed as
\begin{eqnarray}
\!\!\!\!\! \mathbf{v}^{j}_{\mathrm{U}}[n+1] \!\!\!\!&\!=\!&\!\!\!\! \mathbf{v}^{j}_{\mathrm{U}}[n]\!+\mathbf{a}^{j}_{\mathrm{U}}[n]\delta_t, \ n=0,\cdots, N, \label{C_UAV_v&a} \\
\!\!\!\!\! \mathbf{q}^{j}_{\mathrm{U}}[n+1] \!\!\!\!&\!=\!&\!\!\!\! \mathbf{q}^{j}_{\mathrm{U}}[n]\!+\mathbf{v}^{j}_{\mathrm{U}}[n]\delta_t \!+ \frac{1}{2}\mathbf{a}^{j}_{\mathrm{U}}[n]\delta_t^2,  n=0,\cdots, N,  \label{C_UAV_q&v&a}
\end{eqnarray}
where $\mathbf{q}^{j}_{\mathrm{U}}[0] = \mathbf{q}^{j}_{\mathrm{U}, \mathrm{I}}$ is the initial positions of UAV $j$, and $\mathbf{v}^{j}_{\mathrm{U}}[0] = \mathbf{v}^{j}_{\mathrm{U}, \mathrm{I}}$ is the initial velocity of UAV $j$.

On the other hand, the discrete state of SAT~$i$ in a certain constellation~$k$ can be expressed as
\begin{eqnarray}
\mathbf{q}^{i_{k}}_{\mathrm{L}}[n+1] \!\!&=&\!\! \mathbf{q}^{i_{k}}_{\mathrm{L}}[n]+\mathbf{v}^{k}_{\mathrm{L}}\delta_t, \ n=0,\cdots, N,  \label{C_LEO_q&v_i}
% \mathbf{q}^{k}_{\mathrm{L}}[n+1] \!\!&=&\!\! \mathbf{q}^{k}_{\mathrm{L}}[n]+\mathbf{v}^{k}_{\mathrm{L}}\delta_t, \ n=1,\cdots, N, \ \forall k,  \label{C_LEO_q&v_k}
\end{eqnarray}
where $\mathbf{q}^{i_{k}}_{\mathrm{L}}[0] = \mathbf{q}^{i_{k}}_{\mathrm{L}, \mathrm{I}}$ is the initial position of SAT~$i$ in an orbit plane~$k$, and $\mathbf{v}^{k}_{\mathrm{L}, \mathrm{I}}$ is the orbital speed in an orbit plane~$k$ which is calculated by orbital period.
Note that SAT~$i$ in orbit~$k$ returns to the original position after the orbital period, i.e., $\mathbf{q}^{i_{k}}_{\mathrm{L}}[N+1] = \mathbf{q}^{i_{k}}_{\mathrm{L}}[0]$, since SAT circulates following the predetermined orbit plane~$k$.

%%%%%%%%%%%%%%%%%%%%%%%%%%%%%%%%%%%%%%%%%%
\begin{figure*}[!ht]
\begin{equation}
P_{j}(\mathbf{v}_{j}[n], \mathbf{a}_{j}[n]) = c_1\|\mathbf{v}_{j}[n]\|^3 + \frac{c_2}{\|\mathbf{v}_{j}[n]\|}\left(1+\frac{\|\mathbf{a}_{j}[n]\|^2-\frac{(\mathbf{a}^T[n]\mathbf{v}_{j}[n])^2}{\|\mathbf{v}_{j}[n]\|^2}}{g^2}\right) + \frac{m}{2 \delta_t} \left( \|\mathbf{v}[N+1]\|^2 - \|\mathbf{v}[0]\|^2 \right), \ [\mathrm{Watt}]. \label{UAV_Power} 
\end{equation}
\hrule
\end{figure*}

\subsection{UAV Energy Consumption}
% \tred{[JH: Let's avoid such an ambiguoous expression]}
In the UAV system, the energy consumption model is key for optimizing the service time, as it is not able to recharge propulsion fuel or electric power during a flight.
We consider the propulsion energy for a flight as $E_f$ and the communication energy for a signal processing as $E_c$.
% Note that in the energy consumption model, we assume $E_c$ to be a constant, as it is known to be much smaller than $E_f$ in practical scenarios (i.e., $E_f \gg E_c$) \cite{SCA_YZ}. 
Note that in the energy consumption model, $E_c$ is known to be much smaller than $E_f$ in practical scenarios (i.e., $E_f \gg E_c$) \cite{SCA_YZ}. 
Thus, we simply use total energy as $E_T \simeq E_f$.
Following \cite{Energy_HAP,SCA_YZ}, the power consumption for a fixed-wing type UAV $j$ in [Watt], $P_{j}[n]$, is expressed as \eqref{UAV_Power}. 
% \tred{[JH: Put the floating equation at the same page]}
Note that $c_1$ and $c_2$ are two parameters related to the effect of aircraft weight, its wing area, and the air density.
Here, $g$ is the acceleration due to gravity (9.8 m/s$^{2}$) and $m$ is the mass of the UAV. 

Based on the power model, the energy consumption of UAV~$j$ for time slot $n$ with given time step size $\delta_{t}$ is expressed as 
\begin{equation}
E_{j}[n]= P_{j}[n] \delta_{t} \ [\mathrm{Joule}]. \label{UAV_Energy}
\end{equation}

% For a tractable optimization, the kinetic energy of UAV, $K_{E}=\frac{m}{2} \left( \|\mathbf{v}[N+1]\|^2 - \|\mathbf{v}[0]\|^2 \right)$ will be set to zero given the same initial velocity and final velocity.
% Accordingly, the energy consumption of UAV for the time interval $N$ can be rewritten as \eqref{Energy_NoKinetic}.
% \begin{figure*}[!h]
% \begin{equation}
% % E = \sum_{n=1}^{N} c_1 \delta_t\|\mathbf{v}[n]\|^3 + \frac{c_2 \delta_t}{\|\mathbf{v}[n]\|} \left( 1 +\frac{\|\mathbf{a}[n]\|^2 -\frac{(\mathbf{a}^T[n]\mathbf{v}[n])^2}{\|\mathbf{v}[n]\|^2}}{g^2} \!\right) + \underbrace{\frac{m}{2} \left( \|\mathbf{v}[N+1]\|^2 - \|\mathbf{v}[0]\|^2 \right)}_{K_{E}} \ [\mathrm{Joule}].
% E = \sum_{n=1}^{N} c_1 \delta_t\|\mathbf{v}[n]\|^3 + \frac{c_2 \delta_t}{\|\mathbf{v}[n]\|} \left( 1 +\frac{\|\mathbf{a}[n]\|^2 -\frac{(\mathbf{a}^T[n]\mathbf{v}[n])^2}{\|\mathbf{v}[n]\|^2}}{g^2} \!\right) \ [\mathrm{Joule}]. \label{Energy_NoKinetic}
% \end{equation}
% \hrule
% \end{figure*}

%%%%%%%%%%%%%%%%%%%%%%%%%%%%%%%%%%%%%%%%%%%%

% \tred{[JH: Throughout the paper, be more careful of uncountable and countable expressions, e.g., channel models for RF links (o), for an RF link (o), for RF link (x)]}
\subsection{FSO/RF Channel}
% \tred{[JH: We can avoid too many redundant expressions; e.g., 1) channel model for RF links in B. FSO/RF channel Models in Sec. 2 Syste Model; We can remove all the subordinate `models' if the meaning is already clear]}
\subsubsection{RF link}
Since we consider channel characteristics of ground-to-air link and air-to-air link, we assume LoS links without Doppler effect as in \cite{SCA_YZ}. 
Therefore, the deterministic propagation models are adopted under the position of UAV and attenuation conditions.
The channel gain of RF link $h_{\mathrm{RF}}$ between each terminal at a link distance $d$ can be expressed as $h_{\mathrm{RF}}=\sqrt{{\beta_{0}}/{d^2}},$ where $\beta_{0}$ represents the received power at the reference distance $d_0=1$ [m]. 
Accordingly, the transmission rate in [bps] for the slot $n$ can be expressed as
\begin{equation}
C_{\mathrm{RF}} = B_{\mathrm{RF}} \log_2\left( 1+ \dfrac{\gamma_0}{\| d \|^2}  \right) \ \mathrm{[bps]}, 
\label{Rate_RF}    
\end{equation}
where $B_{\mathrm{RF}}$ represents the RF bandwidth, and $\gamma_0 = \frac{\beta_{0} \cdot P}{\sigma_{\mathrm{RF}}^2}$ indicates the reference SNR with constant transmission power $P$ and noise variance $\sigma_{RF}^2$.

%%%%%%%%%%%%%%%%%%%%%%%%%%%%%%%%%%%%%%%%%%%%%%
\subsubsection{FSO link}
Based on the Beer-Lambert law, which represents the signal attenuation of optics, the channel gain for a FSO link  at a link distance $d$ can be expressed as \cite{Capacity_FSO}
\begin{equation}
h_{\mathrm{FSO}}=e^{-\beta \cdot \|d\|}, \label{h_FSO}    
\end{equation}
where $\beta_{\mathrm{dB}} = \frac{3.91}{V}\left(\frac{\lambda}{550 \ \mathrm{nm}}\right)^{-p} \mathrm{[dB/km]}$ value depends on the wavelength $\lambda$ assumed to be 1550 [nm] in this paper, $V$ is the visibility in [km], and the size distribution coefficient $p$ determines by Kim model \cite{System_2}. 
Note that $\beta = \frac{\beta_{\mathrm{dB}}}{10^4\log_{10}e}  \mathrm{[m^{-1}]}$. 

While the capacity of FSO is unknown in closed-form, the capacity bounds of FSO have been proposed in several papers.
In this paper, the lower bound of FSO capacity introduced in \cite{Capacity_FSO} is used for describing the data rate of FSO link. 
The average optical SNR (ASNR) is denoted as $\gamma_{\mathrm{FSO}}^2=\frac{\varepsilon^2}{\sigma_{\mathrm{FSO}}^2}$ in which $\varepsilon$ and $\sigma_{\mathrm{FSO}}^2$ are the average optical power and noise variance for FSO, respectively. 
The parameters related to ASNR, $k_1$ and attenuation condition, $k_2$ are formulated as \cite{Capacity_FSO}
\begin{eqnarray}
&& k_1 = \begin{cases}
\frac{e^{2\alpha\mu^{*}}}{2\pi e}\left( \frac{1-e^{-\mu^{*}} }{\mu^{*}}\right)^2 \frac{\gamma_{\mathrm{FSO}}^2}{\alpha^2} \ \ \text{if } 0\!<\!\alpha\!<\!\frac{1}{2}, \\
\frac{\gamma_{\mathrm{FSO}}^2}{2\pi e \alpha^2} \ \ \text{if } \frac{1}{2}\!<\!\alpha\!<\!1,      \end{cases}\\
&& k_{2} = 2\beta.
\end{eqnarray}
Here, $\mu$ is the free parameter, which refers to the solution of $\alpha=\frac{1}{\mu^{*}}-\frac{e^{- \mu^{*}}}{(1-e^{-\mu^{*}})}$ in which the average-to-peak ratio (APR) is set to $\alpha=\frac{\varepsilon}{\Lambda}$ where $\Lambda$ is peak optical power. 

With the channel gain for a FSO link in \eqref{h_FSO}, the FSO achievable rate in bits/second [bps] for the time slot $n$ can be expressed as
\begin{equation}
C_{\mathrm{FSO}} = \frac{B_{\mathrm{FSO}}}{2} \log_{2}\left( 1+k_1 e^{-k_{2} \cdot \|d\|} \right) \ \mathrm{[bps]}. 
\label{Rate_FSO}
\end{equation}
Note that $B_{\mathrm{FSO}}$ is the bandwidth of FSO, and $\overline{\gamma}_{\mathrm{FSO}}^2=h_{\mathrm{FSO}}^2\cdot\gamma_{\mathrm{FSO}}^2$ represents the received ASNR.

%%%%%%%%%%%%%%%%%%%%%%%%%%%%%%%%%%%%%%%%%%%%%%
\subsubsection{Hybrid FSO/RF link}
In the multi-hop communication scenario, we employ hybrid FSO/RF communication for each links. 
The hybrid FSO/RF method switches to a link with a better channel condition among the RF link and the FSO link.
Accordingly, the achievable rate of the hybrid FSO/RF method can be expressed as \cite{Hybrid_RF_FSO_2}
\begin{equation}
C = \max \{ C_{\mathrm{FSO}}, C_{\mathrm{RF}} \} \ \mathrm{[bps]}.
\end{equation}

We point out here that the reasons for considering the hybrid FSO/RF link are as follows; 
i) The existing NFP projects, such as Starlink \cite{Starlink} and Aquila \cite{Aquila}, consider both RF and FSO communication for the air-to-ground and air-to-air links, such as inter SAT-UAV link.
However, only few studies focus on the achievable rate of RF and FSO link in the NTN scenario.
We consider the hybrid FSO/RF link to answer the question of which link (RF link or FSO link) is more advantageous.
Adopting the hybrid FSO/RF link enables the comparison study of these two link options which will be given in our numerical results.
ii) Unlike FSO, mm-waves are not attenuated by fog, while mm-waves are highly attenuated by water molecules, such as rainfall \cite{Hybrid_RF_FSO_2}. 
To combat the advantages of both mm-waves and FSO, the hybrid FSO/RF is considered. 
The system switches to FSO during rainy conditions when mm-wave transmission is not favorable due to high rain attenuation and switches to mm-waves during foggy conditions. 
A hybrid FSO/RF link itself can be considered as a potential alternative solution to overcome the link degradation under bad weather conditions in the air-to-ground and air-to-air links \cite{Hybrid_RF_FSO_0}.
We examine the hybrid FSO/RF link as a promising vertical backhauling/fronthauling link configuration.

%%%%%%%%%%%%%%%%%%%%%%%%%%%%%%%%%%%%%%%%%%%%%%%%%%%%%%%%%%%%%%%%%%%%%%

%%%%%%%%%%%%%%%%%%%%%%%%%%%%%%%%%%%%%%%%%%%%%%%%%%%%%%%%%%%%%%%%%%%%%%%%%%%%%%%%%%%%%%%%%%%%%%%%%%%%%%%%%%%%%%%%%%%%%%%%%%%%%%%%%%%%%%%%%%%%%%%%%%%%%%%%%%%%%%%%%%%%%%%%%%%%%%%%%%
%%%%%% Body1 %%%%%%
%%%%%%%%%%%%%%%%%%%%%%%%%%%%%%%%%%%%%%%%%%%%%%%%%%%%%%%%%%%%%%%%%%%%%%%%%%%%%%%%%%%%%%%%%%%%%%%%%%%%%%%%%%%%%%%%%%%%%%%%%%%%%%%%%%%%%%%%%%%%%%%%%%%%%%%%%%%%%%%%%%%%%%%%%%%%%%%%%%
\section{Energy-Efficiency Maximization of a Satellite-UAV Integrated NTN} \label{Body}

This section addresses the energy-efficiency (EE) maximization of the UAV-assisted LEO SAT constellation network.
The connected LEO SATs forward information between the source and destination as a relay node, during SATs circulate rapidly in the predetermined orbital lane.
For efficient relaying, appropriate node-to-node association design are necessary.
Besides, as we consider the UAVs as additional mobile relay nodes in-between the LEO SATs, proper movement control for UAV is also necessary.
Therefore, we optimize the association for SATs and the trajectory of UAVs.

%%%%%%%%%%%%%%%%%%%%%%%%%%%%%%%%%%%%%%%%%%%%%%
\subsection{Multi-Hop Communication in NTN}

As shown in Fig. \ref{Illustration}, multiple SAT and UAV configure the multi-hop communication.
In our NTN scenario, information is transmitted from $\mathcal{S}$ to $\mathcal{D}$ via SAT relay nodes and via UAV relay nodes.
We consider that two distant terminals $\mathcal{S}$ and $\mathcal{D}$ are connected through as many links as the number of UAV relays, $J$.
That is, multiple links support the E2E communication between Src and Dst, if $J \neq 1$.
In addition, we consider two orbital lane, i.e., $K=2$.

According to the network scenario, the instantaneous achievable rates of each link for UAV $j$ in time slot $n$ in bps is given by
\begin{eqnarray}
\!\!&\!\!& \mathit{R}_{\mathcal{S}, i^{j}_{1}}[n] \leq C_{\mathcal{S}, i^{j}_{1}}[n], \label{R_Si1}  \\
\!\!&\!\!& \mathit{R}_{i^{j}_{1}, j}[n] \leq \mathrm{min}\lbrace  C_{i^{j}_{1}, j}[n], \ \mathit{R}_{\mathcal{S}, i^{j}_{1}}[n] \rbrace, \label{R_i1j} \\
\!\!&\!\!& \mathit{R}_{j, i^{j}_{2}}[n] \leq \mathrm{min}\lbrace  C_{j, i^{j}_{2}}[n], \ \mathit{R}_{i, j}[n] \rbrace, \label{R_ji2} \\
\!\!&\!\!& \mathit{R}_{i^{j}_{2}, \mathcal{D}}[n] \leq \mathrm{min}\lbrace C_{i^{j}_{2}, \mathcal{D}}[n], \ \mathit{R}_{j, i^{j}_{2}}[n] \rbrace, \label{R_i2D}
\end{eqnarray}
where $\mathit{R}_{\mathcal{S}, i^{j}_{1}}[n]$, $\mathit{R}_{i^{j}_{1}, j}[n]$, $\mathit{R}_{j, i^{j}_{2}}[n]$, and $\mathit{R}_{i^{j}_{2}, \mathcal{D}}[n]$ denote the achievable rate of $\mathcal{S}-i^{j}_{1}$, $i^{j}_{1}-j$, $j-i^{j}_{2}$, and $i^{j}_{2}-\mathcal{D}$, respectively, for time slot $n$.
% \tred{[JH: Introduce this definition beforehand, and at the beginning of Sec. 2 if useful] Note that the superscript of $j$ is used in $i^{j}_{k} \in \mathcal{I}_{k}$, to clarify which UAV $j$ is associated with SAT $i$ on orbit $k$.}
Note that to clarify which UAV $j$ is associated with SAT $i$ on orbit $k$, the superscript $j$ is used in $i^{j}_{k} \in \mathcal{I}_{k}$, from now on.
In the above equations, the transmission rate of each link for UAV $j$ in bps are represented, respectively, as 
\begin{align}
& C_{\mathcal{S}, i^{j}_{1}}[n] = w_{\mathcal{S}, i^{j}_{1}}[n] C(d_{\mathcal{S}, i^{j}_{1}}[n]),   \label{C_Si1}  \\
& C_{i^{j}_{1}, j}[n] = w_{i^{j}_{1}, j}[n] C(d_{i^{j}_{1}, j}[n]),   \label{C_i1j} \\
& C_{j, i^{j}_{2}}[n] = w_{j, i^{j}_{2}}[n] C(d_{j, i^{j}_{2}}[n]),  \label{C_ji2} \\
& C_{i^{j}_{2}, \mathcal{D}, j}[n] = w_{i^{j}_{2}, \mathcal{D}}[n] C(d_{i^{j}_{2}, \mathcal{D}}[n]),  \label{C_i2D} \end{align}
where $w_{\mathcal{S}, i^{j}_{1}}[n]$, $w_{i^{j}_{1}, j}[n]$, $w_{j, i^{j}_{2}}[n]$, and $w_{i^{j}_{2}, \mathcal{D}}[n]$ represents the association of $\mathcal{S}-i^{j}_{1}$, $i^{j}_{1}-j$, $j-i^{j}_{2}$, and $i^{j}_{2}-\mathcal{D}$, respectively. 
Note that we consider the only one SAT is selected between $i^{j}_{1}-j$ and between $j-i^{j}_{2}$ in each orbit $k$ for time slot $n$, so that it holds $w_{\mathcal{S}, i^{j}_{1}} = w_{i^{j}_{1}, j}$ and $w_{j, i^{j}_{2}} = w_{i^{j}_{2}, \mathcal{D}}$.
In the above operations, the link distance for each link for UAV $j$ is expressed as 
\begin{align}
& d_{\mathcal{S},i^{j}_{1}}[n] = \|\mathbf{q}^{i^{j}_{1}}_{\mathrm{L}}[n] - \mathbf{q}_{\mathcal{S}} \|,  \label{d_Si1}  \\
& d_{i^{j}_{1},j}[n] = \|\mathbf{q}^{j}_{\mathrm{U}}[n] - \mathbf{q}^{i^{j}_{1}}_{\mathrm{L}}[n] \|,   \label{d_i1j} \\
& d_{j,i^{j}_{2}}[n] = \|\mathbf{q}^{i^{j}_{2}}_{\mathrm{L}}[n] - \mathbf{q}^{j}_{\mathrm{U}}[n]\|,  \label{d_ji2} \\
& d_{i^{j}_{2},\mathcal{D}}[n] = \|\mathbf{q}_{\mathcal{D}} - \mathbf{q}^{i^{j}_{2}}_{\mathrm{L}}[n] \|. \label{d_i2D} 
\end{align}
Here, the position of SATs ($\mathbf{q}^{i^{j}_{1}}_{\mathrm{L}}[n]$ and $\mathbf{q}^{i^{j}_{2}}_{\mathrm{L}}[n]$) and the position of UAV ($\mathbf{q}^{j}_{\mathrm{U}}[n]$) are time-varying.

%%%%%%%%%%%%%%%%%%%%%%%%%%%%%%%%%%%%%%%%%%%%%%
\subsubsection{Practical Considerations for Multi-Hop Communication}
In the multi-hop communication, there are two practical considerations. 
Firstly, it is necessary to consider association overlap, since multiple links by UAV relays are considered in the E2E multi-hop communication.
The association overlap happens when multiple UAV associate to the same SAT.
When occurring association overlap, the achievable rate of link should be decreasing, as the resource (e.g., bandwidth, transmission power) is limited for each link and the interference issue may arise.
Thus, we assume that each node can use limited bandwidth and the limited bandwidth is equally divided for association overlapped links.
For instance, if $w_{i^{1}_{1}, 1}[n] = w_{i^{2}_{1}, 2}[n]$, $C_{i^{1}_{1}, 1}[n] = \frac{1}{2}w_{i^{j}_{1}, 1}[n] C(d_{i^{1}_{1}, 1}[n])$ and $C_{i^{1}_{1}, 2}[n] = \frac{1}{2}w_{i^{2}_{1}, 2}[n] C(d_{i^{2}_{1}, 2}[n])$.

%%% Communication between UAVs
% \textbf{Communication between UAVs}.\quad
Secondly, it is necessary to consider communication overhead between UAVs.
The communication between UAVs is needed for avoiding a collision between UAVs and for designing the actions of multi UAV in a cooperative manner which guarantee better network performance.
To deal with the communication overhead issue, we consider the latency between UAVs, $\tau$, which is derived by propagation delay.

\subsubsection{End-to-End summarized Throughput for Multi-Hop Communication}
%%% E2E system throughput + Energy
Without losing generality, assuming decode-and-forward (DF) relaying protocol without buffer, the constraints of \eqref{R_Si1}-\eqref{R_i2D} evolve to $\min \{ \mathit{R}_{\mathcal{S}, i^{j}_{1}}[n], \mathit{R}_{i^{j}_{1}, j}[n], \mathit{R}_{j, i^{j}_{2}}[n], \mathit{R}_{i^{j}_{2}, \mathcal{D}}[n] \}$.
Accordingly, the instantaneous E2E sum throughput for the multi-hop communication in time slot $n$ can be approximated as 
\begin{align}
    \mathit{R}^{j}_{\mathcal{S}, \mathcal{D}}[n] \approx \min \{ \mathit{R}_{\mathcal{S}, i^{j}_{1}}[n], \mathit{R}_{i^{j}_{1}, j}[n], \mathit{R}_{j, i^{j}_{2}}[n], \mathit{R}_{i^{j}_{2}, \mathcal{D}}[n] \}. \label{Rate_E2E}
\end{align}
% \tred{[JH: $\forall n, j$ look also weird]}

%%%%%%%%%%%%%%%%%%%%%%%%%%%%%%%%%%%%%%%%%%%%%%
\subsection{Problem Formulation} \label{Body_1}

For our problem's objective, we consider EE, which includes both the E2E network throughput and the UAVs' energy consumption.
% With the E2E system throughput model and the energy consumption model of UAV, EE can be expressed as 
% \begin{equation}
%     \mathrm{EE} = \dfrac{\sum_{n=1}^{N} \sum_{j\in \mathcal{J}} \mathit{R}^{j}_{\mathcal{S}, \mathcal{D}}[n]}{\sum_{n=1}^{N} \sum_{j\in \mathcal{J}} E_{j}[n]}. \label{EE}
% \end{equation}
The following problem, P1, corresponds to the EE maximization under the constraints related to the practical condition of SAT and UAV.
For mathematical convenience, we define the set of association as $\mathcal{W}=\{ w_{i^{j}_{1}, j}[n], w_{j, i^{j}_{2}}[n], \forall n, i_{k}\in \mathcal{I}_{k}, k=1,2,\ldots,K, j \in \mathcal{J} \}$, and the set of UAV acceleration as $\mathcal{A}=\{\mathbf{a}^{j}_{\mathrm{U}}[n], \forall n, j \}$.
\begin{eqnarray}
(\mathrm{P1}) \!&\!\displaystyle \max_{ \scriptsize \begin{array}{c} \scriptsize \mathcal{W}, \mathcal{A}  \end{array} } \!&\!
	\mathrm{EE} \triangleq \dfrac{\sum_{n=1}^{N} \sum_{j\in \mathcal{J}} \mathit{R}^{j}_{\mathcal{S}, \mathcal{D}}[n]}{\sum_{n=1}^{N} \sum_{j\in \mathcal{J}} E_{j}[n]} \nonumber \label{P1} \\ 
\!&\!\textrm{s.t.}\!&\! \eqref{C_UAV_v&a}-\eqref{C_LEO_q&v_i}, \eqref{R_Si1}-\eqref{R_i2D}, \nonumber \\
\!&\!\!&\! w_{i^{j}_{1}, j}[n] \in \lbrace 0,1 \rbrace, \ \forall n, i^{j}_{1}, j, \label{P1_C_i1j}\\
\!&\!\!&\! w_{j, i^{j}_{2}}[n] \in \lbrace 0,1 \rbrace, \ \forall n, i^{j}_{2}, j, \label{P1_C_ji2}\\
\!&\!\!&\! {\textstyle\sum}_{i^{j}_{1} \in \mathcal{I}_{1}} w_{i^{j}_{1}, j}[n] = 1, \ \forall n, j, \label{P1_C_sum_i1j}\\
\!&\!\!&\! {\textstyle\sum}_{i^{j}_{2} \in \mathcal{I}_{2}} w_{j, i^{j}_{2}}[n] = 1, \ \forall n, j, \label{P1_C_sum_ji2}\\
\!&\!\!&\! \|\mathbf{a}^{j}_{\mathrm{U}}[n]\| \leq A_{\max}, \ \forall n, j, \label{P1_C_Amax} \\
% \!\!\!\!&\!\!&\!\!\!\! E_{j} \leq E_{j}^{\mathrm{given}}, \forall j, \label{P1_C_Energy} \\
\!&\!\!&\! \tau[n] \leq \tau_{\mathrm{max}}, \ \forall n, \label{P1_C_Delay}
\end{eqnarray}
where $N$ is the one orbital cycle of SAT, $A_{\max}$ denotes the maximum acceleration of UAV, and $\tau_{\mathrm{max}}$ indicates the maximum allowable delay between UAVs.

For the maneuverability of UAV, the equality constraints in \eqref{C_UAV_v&a} and \eqref{C_UAV_q&v&a} characterize the discrete-time model of UAV, i.e., the position of UAV $\mathbf{q}^{j}_{\mathrm{U}}[n]$, the velocity of UAV $\mathbf{v}^{j}_{\mathrm{U}}[n]$, as well as the acceleration of UAV $\mathbf{a}^{j}_{\mathrm{U}}[n]$. 
For the orbit of SATs, the equality constraint in \eqref{C_LEO_q&v_i} characterizes the discrete-time model of SAT, i.e., the position of SAT.
In addition, the constraints in \eqref{R_Si1}-\eqref{R_i2D} represent the information-causality constraint in the multi-hop communication.
The information-causality, that relay node can only forward the information which has been previously received from the source, is a inherent constraint of relay node.
For the association of SAT-UAV, the constraints in \eqref{P1_C_sum_i1j}-\eqref{P1_C_sum_ji2} represent that UAV $j$ can be linked by at most one SAT in each orbit, while the constraints of \eqref{P1_C_i1j}-\eqref{P1_C_ji2} compose the association variables as integer variables.
The constraint in \eqref{P1_C_Amax} ensures that the acceleration of UAV is within acceptable acceleration.
% Besides, the constraint in \eqref{P1_C_Delay} limits the latency for link between UAVs, to consider the communication overhead between UAVs.
Besides, the constraint in \eqref{P1_C_Delay} limits the latency for link between UAVs, as UAV should share the information of itself with other UAVs within a finite time.

For solving the optimization problem (P1), however, traditional convex optimization methods (such as SCA, BCD method) face several challenges. 
Firstly, the formulated problem of (P1), which designs the SAT-UAV association and finds the optimal trajectory of UAVs, is obviously non-convex and thus difficult to be directly solved. 
Furthermore, the variables of association and trajectory, which depends on the time-varying network topology (e.g., \eqref{R_Si1}-\eqref{R_i2D}), are challenging to be efficiently relaxed to convex.
Secondly, traditional off-line optimization-based methods require the global knowledge of the system parameters, which is non-trivial to acquire in practice. 
Even given global knowledge of the system parameters, it is difficult to deal with the association overlap issue, such as $w_{i^{1}_{1}, 1}[n] = w_{i^{2}_{1}, 2}[n]$.

%%%%%%%%%%%%%%%%%%%%%%%%%%%%%%%%%%%%%%%%%%%%%%
\section{Multi-Agent Reinforcement Learning for an Energy-Efficient Satellite-UAV Integrated NTN}
\label{Body_ML}

To address the challenges for association and trajectory design in the time-varying NTN topology, we propose a novel MARL method, which only requires observable information of UAVs along with its flight as the input. 
We leverage the multi-agent Actor-Critic method, and further introduce the action dimension reduction technique to improve the learning convergence and performance.
For using the MARL method, the objective, variables, and constraints of (P1) is needed to be mapped into a state, action, and reward function in the environment.
However, the mapping of (P1) is non-trivial due to problematic learning convergence of the MARL method.
% However, the mapping of (P1), which should take into account the learning convergence of the MARL method, is non-trivial.
% due to learning convergence of the MARL method, however, the mapping of (P1) is non-trivial.
% We now describe how the formulated problem (P1) is matched to the MARL approach in detail.
% \tred{[JH: Briefly explain why this problem is non-trivial, requiring major modifications to be elaborated in the following subsections]}

\subsection{Problem Reformulation}
%%%% 왜 P1이 P1*로 바뀌어야 할까?
Before applying a MARL method directly into (P1), we slightly change the original problem for the following fact;
The value of the fractional objective in (P1) fluctuates easily according to the denominator that changes according to the dynamic state of UAV (e.g., $\mathbf{v}_{j}[n], \mathbf{a}_{j}[n]$), such that it may cause the learning convergence problem.
% In addition, it is more efficient to consider the latency constraint in \eqref{P1_C_Delay} as the UAV distance constraint.  

%%%% Re-written Problem Formulation
To this end, motivated by the Dinkelbach's algorithm, a widely-known method to solve large-scale fractional programming for which its optimality and convergence properties are established \cite{Dinkelbach}, we rewrite (P1) as
\begin{eqnarray}
(\mathrm{P1^{\star})} \!&\!\displaystyle \max_{ \scriptsize \begin{array}{c} \scriptsize \mathcal{W}, \mathcal{A}  \end{array} } \!&\!
	\sigma_{\mathrm{R}}\sum_{n=1}^{N} \sum_{j\in \mathcal{J}} \mathit{R}^{j}_{\mathcal{S}, \mathcal{D}}[n] -  \sigma_{\mathrm{E}}\sum_{n=1}^{N} \sum_{j\in \mathcal{J}} E_{j}[n] \nonumber \label{P2} \\ 
\!&\!\textrm{s.t.}\!&\! \eqref{C_UAV_v&a}-\eqref{C_LEO_q&v_i}, \eqref{R_Si1}-\eqref{R_i2D}, \eqref{P1_C_i1j}-\eqref{P1_C_Amax}, \nonumber \\
\!&\!\!&\! d_{j, j^{'}}[n] \leq d_{\mathrm{max}}, \ \forall n, \label{P2_C_Delay}
\end{eqnarray}
where $\sigma_{\mathrm{R}}$ and $\sigma_{\mathrm{E}}$ are the coefficient for the sum of E2E throughput and the sum of energy consumption of UAVs, respectively, and the distance between UAVs is denoted as 
\begin{equation}
    d_{j, j^{'}}[n] = \|\mathbf{q}^{j}_{\mathrm{U}}[n] - \mathbf{q}^{j^{'}}_{\mathrm{U}}[n] \|, \ j\in\mathcal{J}, j^{'}\in\mathcal{J}/j, \label{C_jj'}    
\end{equation}
which is limited by the maximum allowable distance between UAVs, $d_{\mathrm{max}}$.

Now, we can handle the weights for the E2E system throughput and the energy consumption of UAVs by adjusting $\sigma_{\mathrm{R}}$ and $\sigma_{\mathrm{E}}$ in (P1$^{\star}$). 
For example, the throughput performance, $\sum_{n=1}^{N} \sum_{j\in \mathcal{J}} \mathit{R}^{j}_{\mathcal{S}, \mathcal{D}}[n]$, can be mainly optimized by relatively increasing $\sigma_{\mathrm{R}}$, or the energy (related to the service-time), $\sum_{n=1}^{N} \sum_{j\in \mathcal{J}} E_{j}[n]$, can be mainly optimized by relatively increasing $\sigma_{\mathrm{E}}$.
The latency constraint in \eqref{P1_C_Delay} is rewritten as \eqref{P2_C_Delay}.
% which is derived by the propagation delay.
Note that the latency is derived by the propagation delay, as the distance between UAVs is quite far and the amount of information exchanged between UAVs is not considerable.
We now describe how the formulated problem (P1$^{\star}$) is matched to the MARL approach in detail.

\subsection{Markov Decision Process Modeling} 
% To apply DRL approach, we need to frame a problem in a Markov decision process (MDP).
In this subsection, we design our optimization problem of (P1$^{\star}$) into a Markov decision process (MDP) model.
We note that MDP is a discrete-time stochastic control process, which is provides a mathematical framework for modeling decision making in situations where outcomes are partly random and partly under the control of a decision maker.
% \tred{which is prerequisite for DRL approach [JH: can we elaborate the reasoning in more detail?]}.
In the MDP model, the policy corresponds to the probability of choosing an action according to the current state. 
The optimal policy $\pi^{*}$ is the policy that contributes to the maximal long-term system reward. 
Our goal is to find $\pi^{*}$ to maximize the average long-term system reward.

By the dynamic states of UAV $j$ in \eqref{C_UAV_v&a}-\eqref{C_UAV_q&v&a} (e.g., $\mathbf{q}^{j}_{\mathrm{U}}[n]$ and $\mathbf{v}^{j}_{\mathrm{U}}[n]$) and the dynamic states of UAV $j$ in \eqref{C_LEO_q&v_i} (e.g., $\mathbf{q}^{i_{k}}_{\mathrm{L}}[n]$), it holds a Markov characteristics. 
As such, we can map the optimization problem (P1$^{\star}$) as MDP.
Briefly, mapping of (P1$^{\star}$) into MDP model, each system utility in the objective function (such as E2E system throughput, consumption energy of UAVs) corresponds to reward function, while the optimization variables (e.g., $\mathcal{W}, \mathcal{A}$) corresponds to action space. 
A mapping of (P1$^{\star}$) into the environment, state, action, and reward functions in MDP model is presented in detail in the following subsections.

%%% Environment 와 observation space 관련
\subsubsection{Environment}

\begin{figure}
    % \vspace{-1em}
    \centering
    \includegraphics[width=\linewidth]{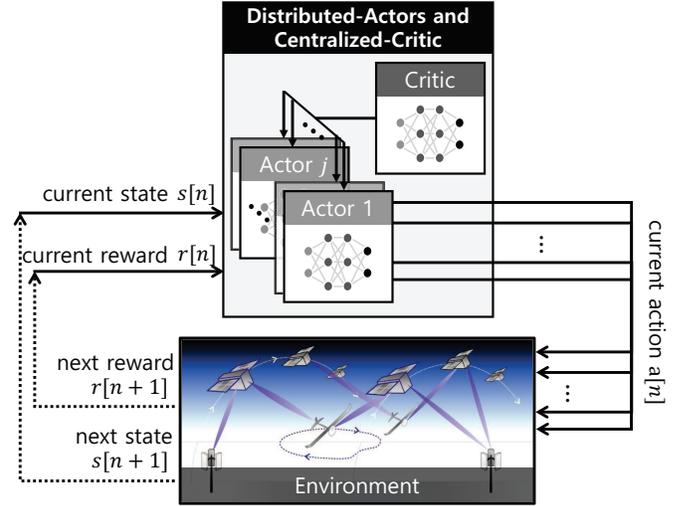}
    \caption{Proposed multi-agent Actor-Critic structure for vertical multi-hop communication in LEO SAT constellation network.}
    \label{Structure}
    % \vspace{-1.em}
\end{figure}

% In the MDP model of (P1$^{\star}$), each UAV $j$ is considered as an agent which interacts with an environment, and accordingly, this problem is composed of multi-agents. 
As shown in Fig. \ref{Structure}, the MARL framework for joint SAT-UAV multi-hop communication is based on the MDP model of (P1$^{\star}$), wherein multi-agents, corresponding to multiple UAVs, interact with an environment. In this scenario, the environment includes everything related to the multi-hop communication.
At each time $n$, each UAV $j$, as a agent, observes a state $s_{j}[n]$ from the state space $S$, and accordingly takes an action $a_{j}[n]$ from the action space $A$ selecting the association action set and the acceleration action set based on the policy $\pi$. 
% The decision policy $\pi$ is determined by value functions $V^{\pi_{\theta}}(s[n], a[n])$. 
Following the action, the state of the environment transitions to a new state $s_{j}[n+1]$ and the agent $j$ receives a reward $r_{j}[n]$ which is determined by the E2E system throughput in the multi-hop communication and the energy consumption of UAVs.

%%% State space 관련
\subsubsection{State}

In our system, the state observed by each UAV node for characterizing the environment consists of several parts: the position of SAT in the two orbital lane $\mathbf{q}^{i_{k}}_{\mathrm{L}}[n]\in\mathbb{R}^{I \times 3}, i\in \mathcal{I}, k=1,2$, the position of UAV $j$ $\mathbf{q}^{j}_{\mathrm{U}}[n]\in\mathbb{R}^{J \times 3}$, the link distance for each link $d_{j}[n] = \lbrace d_{\mathcal{S}, i^{'}_{1}}[n], d_{i^{'}_{1}, j}[n], d_{j, i^{'}_{2}}[n], d_{i^{'}_{2}, \mathcal{D}}[n] \rbrace\in\mathbb{R}^{4}$, the energy consumption of UAV $j$ for the predetermined time $P_{j}[n]\delta_{t}$, and the time slot $n$. Thus, the state of agent $j$ is given by
\begin{align}
s_{j}[n] = \lbrace \mathbf{q}^{i_{k}}_{\mathrm{L}}[n], \mathbf{q}^{j}_{\mathrm{U}}[n], \mathbf{v}^{j}_{\mathrm{U}}[n], d_{j}[n], P_{j}[n]\delta_{t}, n \rbrace.
\end{align}
In the state space, $n$ is used as a fingerprint, considering two methods proposed in \cite{Stabilising} for stabilizing experience replay in the MARL; i) using a multi-agent variant of importance sampling to naturally decay obsolete data, and ii) conditioning each agent’s value function on a fingerprint that disambiguates the age of the data sampled from the replay memory.
% \tred{[JH: `Note that' is too often used throughout the paper. You can simply remove or paraphrase them, e.g., here, in the above equation, in the aforementioned operations, ...]}

% \tred{[JH: I removed $\forall n, i, k, j$ in (33), as it sounds weird to say `for all' the parameters when it comes to agent $j$'s state]}

%%% Action space 관련
\subsubsection{Action}
The action space in our system includes two actions, i.e., associations $\mathcal{W}$ and accelerations $\mathcal{A}$.
% Firstly, for the association of $i^{j}_{1}-j$ and $j-i^{j}_{2}$, the agent $j$ chooses a SAT among SATs on the orbital lane $k=1$ and $k=2$ with $w_{i^{j}_{1}, j}[n]$ and $w_{j, i^{j}_{2}}[n]$, respectively. 
Firstly, with $w_{i^{j}_{1}, j}[n]$ and $w_{j, i^{j}_{2}}[n]$, the agent $j$ choose one SAT to associate with among SATs on the orbital lane $k=1$ and $k=2$ , respectively, for the association of $i^{j}_{1}-j$ and $j-i^{j}_{2}$. 
Note that we consider that the decision on $w_{\mathcal{S}, i^{j}_{1}}[n]$ follows $w_{i^{j}_{1}, j}[n]$, and the decision on $w_{i^{j}_{2}, \mathcal{D}}[n]$ follows $w_{j, i^{j}_{2}}[n]$.
% Note that our fully observable scenario operates in an offline-manner, so that it is able to optimize $w_{\mathcal{S}, i}[n]$ (and $w_{i, k}[n]$) for each satellite $k$.
To design the action set of association, $a_{j}^{\mathcal{W}}[n] \in\mathbb{R}^{I \times 2}, \forall j$, we use the one-hot encoding.

Secondly, for trajectory design of UAV, UAV chooses the acceleration action set in each time slot.
While there are various ways to efficiently handle a continuous-state action, the most straightforward approach is discretizing it to form a finite-state. 
Hence, we consider the action set of acceleration, $a_{j}^{\mathcal{A}}[n] \in\mathbb{R}^{2 \times (2D+1)}$, by uniformly discretizing between $-A_{\max}$ and $A_{\max}$ as follows.
% \tred{[JH: isn't it between $-A$ and $A$?]}
\begin{equation}
\begin{split}
    a_{j}^{\mathcal{A}}[n] \in & \lbrace -A_{\max}, \cdots, \frac{-A_{\max}}{D}, 0, \frac{A_{\max}}{D}, \cdots, A_{\max};\\
    & -A_{\max}, \cdots, \frac{-A_{\max}}{D}, 0, \frac{A_{\max}}{D}, \cdots, A_{\max} \rbrace. \label{action_acceleration_discrete}    
\end{split}
\end{equation}
Here, the acceleration action space can be managed by a discretization level $2D+1$ with a positive integer $D$. Accordingly, the action space is given by 
\begin{equation}
a_{j}[n] = \{ a_{j}^{\mathcal{W}}[n], a_{j}^{\mathcal{A}}[n] \}.   
\end{equation}

%%% Reward 관련
\subsubsection{Reward}
We design the reward function that will drive UAV to find optimal actions, which maximize the E2E system throughput while minimizing the energy consumption and the distance between UAVs.
In the problem, system utilities are closely related to the E2E system throughput, the energy consumption of UAVs, and the distance between UAVs, in each time slot. 
% However, since the correlation between system utility and the rate and the energy consumption is not linear, we adopt a sigmoid function to describe the correlation as $f(x) = \nicefrac{1}{\left( 1 + e^{-g(x)} \right)}$,
% where $g(x) = \frac{x - \mu}{\sigma}$ is the normalization function for $x$.
% Therefore, we adopt the normalization function for $x$, $g(x) = \frac{x - \mu}{\sigma}$.
% Note that in the system, $\mu$ is the achievable rate of the baseline (i.e., fixed ground-relaying in Sec.~\ref{Numerical Result}) and $\sigma$ is the normalization parameter which yields the outcome of $g(R_{\mathcal{S}, \mathcal{D}}[n])$, $g(\sum_{j\in \mathcal{J}} P_{j}[n]\delta_{t})$, and $g(d_{j, j^{'}}[n])$ between $-1$ and $1$ by considering $c_{\mathrm{R}}$ and $c_{\mathrm{E}}$.

To reflect the expressions in reward function, we adopt the normalization function for $x$, $g(x) = \frac{x - \mu}{\sigma}$. 
% rather than using linear function.
The mean value for the E2E system throughput $\mu_{\mathrm{R}}$ is the E2E system throughput of the baseline (i.e., SAT-Ground in Sec.~\ref{Numerical Result}), the mean value for the energy consumption $\mu_{\mathrm{E}}$ is the minimum energy for hovering \cite{SCA_YZ}, and the mean value for distance between UAVs $\mu_{\mathrm{D}}$ is $d_{\mathrm{max}}$.
The standard deviations for $g(R_{\mathcal{S}, \mathcal{D}}[n])$, $g(\sum_{j\in \mathcal{J}} P_{j}[n]\delta_{t})$, and $g(d_{j, j^{'}}[n])$ are denoted as $\sigma_{\mathrm{R}}$, $\sigma_{\mathrm{E}}$, and $\sigma_{\mathrm{D}}$, respectively, in which the output value of each function yields between $-1$ and $1$.
% Note that the normalization parameter for E2E system throughput ($\sigma_{\mathrm{R}}$) and for energy consumption ($\sigma_{\mathrm{E}}$) accord with $c_{\mathrm{R}}$ and $c_{\mathrm{E}}$.
Note that $\sigma_{\mathrm{R}}$ and $\sigma_{\mathrm{E}}$ accord with $c_{\mathrm{R}}$ and $c_{\mathrm{E}}$, respectively.
In addition, we further use a \textit{Relu} function of $h(x)=\max\{ 0, g(x) \}$, for the distance constraint in \eqref{P2_C_Delay}, such that the agent gets penalty when exceeding a threshold.

As such, the system reward in the $n$-th time slot induced by the current state $s[n]$ and action $a[n]$ is defined as
\begin{equation}
\begin{split}
r[n] = \ & g({\textstyle\sum}_{j \in \mathcal{J}} R_{\mathcal{S}, \mathcal{D}}[n]) - g({\textstyle\sum}_{j\in \mathcal{J}} E_{j}[n]) \\
& - h({\textstyle\sum}_{j\in \mathcal{J}, j^{'}\in \mathcal{J}/j} d_{j, j^{'}}[n]). \label{Reward}
\end{split}
\end{equation}
We note that the reward function remains positive if our network model outperforms a baseline; it will be a penalty, a negative reward, if not.

%%% ING
%%%% “이 SAR은 기존 방식의 네트워크 구조에 적용해 풀기 어렵다.” 
%%% ING

%%%%%%%%%%%%%%%%%%%%%%%%%%%%%%%%%%%%%%%%%%%%%%
\subsection{Centralized-Critic MARL} \label{Body_3}

% \tred{[JH: Carefully double check the $j$ index throughout the entire section, e.g., $\pi_j$, wihch is used in Fig. 3 and the centralized-Critic MARL descriptions; Why do we omit $j$ only in $\theta$ and $\phi$ but in $a$ and $s$? I suggest not to omit this, and for the single we can simply add `Hereafter, the index $j$ identifies different actors for multi-agent scenarios, which can be omitted for a single agent case' at the first place we use the index.]}
Following the standard RL settings, we consider an environment that is interacted with RL agents for a given number of discrete time steps. At each time step $n$, the agent $j$ receives a state $s_{j}[n]$ and selects an action $a_{j}[n]$ from some set of possible actions $A$ according to its policy $\pi_{\theta}$, where $\pi_{\theta}$ is a mapping from states $s_{j}[n]$ to actions $a_{j}[n]$. 
% For notational simplicity, we omit the dependence of $\theta$ on $j$.
% \footnote{For notational simplicity, we omit the dependence of $\theta$ on $j$. \tred{[JH: Let's minimize the use of footnotes. You can simply put this in line]}}
In return, the agent receives the next state $s_{j}[n+1]$ and receives a scalar reward $r[n]$. 
The process continues until the agent reaches a terminal state after which the process restarts. The return $R[n] = \sum\nolimits_{k=0}^{\infty} \gamma^{k} r[n+k]$ is the total accumulated return from time step $n$ with discount factor $\gamma \in (0, 1]$. 
Multiple agents act in environments with the goal of maximizing their shared utility in a cooperative manner.
% The goal of the agent $j$ is to maximize the expected return from each state $s_{j}[n]$.

In our prior work \cite{DRL_UAV}, we considered a single UAV agent operating a deep-Q network (DQN) framework wherein each action of the agent is evaluated as the Q-value that is approximated using the output of a neural network (NN). While effective, DQN may not guarantee the convergence due to its off-policy learning nature, particularly under complicated tasks with large state and action dimensions \cite{ActionState}. By contrast, there exist policy gradient methods in which an NN's output approximates each policy \cite{Sutton}, yet the training convergence is often too slow due to its ignoring the effort to find better actions associated with higher values. Alternatively, in this work we consider the Actor-Critic RL framework \cite{AC}, which combines the benefits of both policy gradient and value-based methods. In what follows, we first explain the basic Actor-Critic RL operations for a single agent, and then describe how to extend this to MARL settings, followed by presenting an action dimensionality reduction technique.
% \tred{that can reduce the computing complexity. [JH: Revise this if it's not about reducing the complexity]}

\subsubsection{Single-Agent Actor-Critic RL}

Actor-Critic framework comprises a pair of two NNs: an Actor NN seeking to take better actions to obtain higher rewards based on the policy gradient method; and its paired Critic NN aiming at approximating the value functions of the actions more accurately via the value-based method. We follow the training operations of the synchronous advantage Actor-Critic (A2C) framework \cite{A3C}, in which the Critic NN updates its model parameters $\phi$ according to the policy $\pi_{\theta}$ given by the Actor NN. Meanwhile, the Actor NN updates its model parameters $\theta$ according to the value functions $V^{\pi_{\theta}}(s[n]; \phi)$ approximated by the Critic NN. Specifically, the Critic NN aims to minimize the loss function
\begin{align}
    L_{\mathrm{Critic}}(\phi) = \kappa[n]^{2},
\end{align}
where $\kappa[n] = r[n+1] + \gamma V^{\pi_{\theta}}(s[n+1]; \phi) - V^{\pi_{\theta}}(s[n]; \phi)$ is referred to as the advantage of an action \cite{TDerror}. The Critic NN model parameters are then updated as
% $\phi \leftarrow \phi + \beta_{C} \kappa[n] \nabla_{\phi} V^{\psi_{\theta}}(s[n]; \phi)$, 
\begin{equation}
    \phi \leftarrow \phi + \beta_{C} \kappa[n] \nabla_{\phi} V^{\pi_{\theta}}(s[n]; \phi),
\end{equation}
where $\beta_{C}$ is the learning rate of the Critic NN. Meanwhile, the Actor NN aims to minimize the loss function
\begin{equation}
    L^{j}_{\mathrm{Actor}}(\theta_{j}) = -\kappa[n] \log\pi(a_{j}[n] | s_{j}[n];\theta_{j}).
\end{equation}
Hereafter, the index $j$ identifies different actors for multi-agent scenarios, which can be omitted for a single agent case. Consequently, the Actor NN model parameters are updated~as 
% $\theta \leftarrow \theta + \beta_{A}\kappa[n]\nabla_{\theta}\log\pi(a[n] | s[n];\theta)$, 
\begin{equation}
    \theta_{j} \leftarrow \theta_{j} + \beta_{A}\kappa[n]\nabla_{\theta_{j}}\log\pi(a_{j}[n] | s_{j}[n];\theta_{j}), 
\end{equation}
where $\beta_{A}$ is the learning rate of the Actor NN.

\subsubsection{Centralized-Critic MARL}
 
\begin{figure}
    % \vspace{-1em}
    \centering
    \includegraphics[width=.8\linewidth]{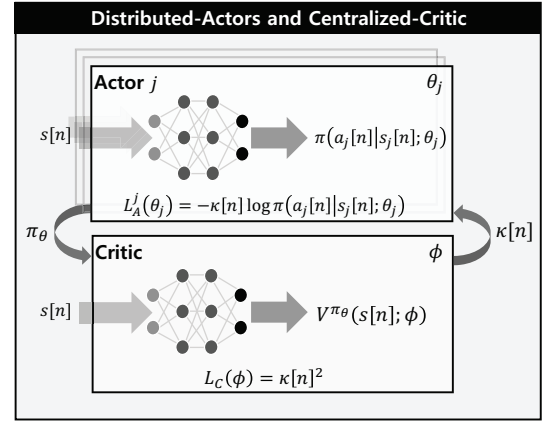}
    \caption{A schematic illustration of a centralized-Critic MARL wherein distributed Actor NNs exchange their learned policies with a single Critic~NN.}
    \label{Overview_MADRL}
    % \vspace{-1.em}
\end{figure}

  MARL frameworks are broadly categorized into cooperative and non-cooperative methods \cite{Tan93:MARL}. In non-cooperative methods such as independent Q-learning \cite{Independent_DQN}, each agent is independently and simultaneously trained without sharing its learned knowledge. Following this principle, the aforementioned single-agent Actor-Critic framework can be extended to its non-cooperative MARL version wherein each agent runs a pair of Actor and Critic NNs without exchanging any information. However, from our experimental observations, such a non-cooperative MARL extension does not guarantee the convergence, probably due to the dynamic nature of orbiting SATs and moving UAV agents. Alternatively, based on a cooperative MARL principle in \cite{MA_Foerster_1, Lowe, MA_Foerster_2}, we consider the centralized-Critic MARL framework wherein a single Critic NN guides the training of distributed Actor NNs that determine individual agents' policies, as visualized in Fig.~\ref{Overview_MADRL}. In this approach, while interacting with the environment, each agent $j$ storing an Actor NN uploads its learned policy $\pi(a_{j}[n] | s_{j}[n];\theta_{j})$, i.e., a pair of its action $a_j[n]$ and state $s_j[n]$, to the Critic NN located at a randomly chosen agent. The centralized-Critic NN takes the entire $J$ agents' actions $a[n] = \{ a_{1}[n],\ldots,a_{J}[n] \}$ and states $s[n] = \{ s_{1}[n], \ldots, s_{J}[n] \}$ as the input, and produces the advantage $\kappa[n]$ as the output that is a function of the values $V^{\pi_{\theta}}(s[n], a[n])$ obtained via the temporal difference method~\cite{Sutton}. Each agent downloads $\kappa[n]$, and thereby updates its Actor~NN.

% In the approach, one of agents operates as an aggregator, which collects other agents observation that is enabled by communication between agents. 
% Note that the agent as an aggregator is composed of Actor network and Critic network, while other agents are composed of Actor network.

% \begin{figure}
%     % \vspace{-1em}
%     \centering
%     \includegraphics[width=\linewidth]{Figure3_J}
%     \caption{Proposed multi-agent Actor-Critic structure for vertical multi-hop communication in LEO SAT constellation network.}
%     \label{Structure}
%     % \vspace{-1.em}
% \end{figure}

%%%% Action Dimension Reduction technique
\subsubsection{Action Dimensionality Reduction}
% To cope with a large number of possible actions due to time-varying network topology, a novel action dimension reduction technique is applied, which focuses only on a couple of SATs proximal to Src.
As identified in \cite{ActionState}, a task with large numbers of discrete actions is intractable or even often impossible to train.
% \tred{serious issues [JH: specify what the seirious issues are (within our scope)]}
Unfortunately, in our MARL settings, the space of association actions $a_{\mathcal{W}}$ is large due to the time-varying network topology with many orbiting SATs. To clarify how large the action space is, in what follows we describe the orbiting dynamics of SATs in an orbital lane. 
% Since SATs orbits along a predetermined orbital lane, SATs in the orbital lane periodically circulate on the surface of Earth.
% In other words, these SATs will come back to the same position after a certain amount of time (such as orbital period).
Since each SAT periodically completes an orbit, the position $\mathbf{q}^{i_{k}}_{\mathrm{L}}[n]\in\mathbb{R}^{I \times 3}$ of an SAT in the $k$-th orbital lane is updated~as
\begin{equation}
\mathbf{q}^{i_{k}}_{\mathrm{L}}[n] = \mathbf{q}^{i_{k}}_{\mathrm{L}}[n] \ \mathrm{mod} \ c_{\mathrm{E}},  \label{C_LEO_q&v_1}
\end{equation}
where $\mathrm{mod}$ represents the modulo operation, $c_{\mathrm{E}}=2\pi r_{\mathrm{E}}$ denotes the circumstance of orbital lane (such as circumstance of Earth), and $r_{\mathrm{E}}$ denotes the radius of orbit. 
% Note that each SAT $i$ , while the position of each SAT basically follows in a coordinate as in \eqref{C_LEO_q&v_i}. 
Note that by the operation in \eqref{C_LEO_q&v_1}, SAT~$i$ in orbit~$k$ returns to the original position after the orbital period.
% \tred{[JH: Revise this part that is not very clear. You simply wrote that the maximum possible number of SAT positions is fixed. Then is the number of SATs only matters? (I didn't thought so); The last sentence is confusing as (41) is also about the position]}

To ameliorate the problem incurred by the large action space in our MARL environment, we devise the following action dimension reduction method.
% It can be seen intuitively that one SAT in each two orbit which located between Src and Dst will be associated, rather than another SAT which is distant from Src or Dst.
The key idea is to utilize the fact that for maximizing the E2E sum throughput, only the SATs sufficiently close to Src or Dst are the candidates of optimal associations $a_{j}^{\mathcal{W}}[n]$.
Thereby we can filter out distant SATs from Src or Dst, and focus on only a few SATs in the action space.
%  narrowing the view to a specific coordinate space near Src and Dst (e.g., $4000\times6000\times550$ [km$^3$]), a certain pattern is observed that SATs in the coordinate space also periodically circulate.
% \tred{[JH: revise the following descriptions that are currently unclear and partly redundant]} 
% Considering that SATs in the coordinate space orbit at an equal velocity at an equal intervals, we focus on the position of the SAT located proximal to Src or Dst in the orbital circumference in coordinate space $c_{\mathrm{C}}$, instead of the position of all SATs in the entire orbital circumference.
Particularly, we focus only on a certain orbital line segment $c_{\mathrm{C}}$ which is located proximal to Src or Dst, instead of the entire orbital lane circumference $c_{\mathrm{E}}$.
The position of the selected SAT with $c_{\mathrm{C}}$ is thereby rewritten as follows 
\begin{equation}
\mathbf{q}^{i^{*}_{k}}_{\mathrm{L}}[n] = \mathbf{q}^{i^{*}_{k}}_{\mathrm{L}}[n] \ \mathrm{mod} \ c_{\mathrm{C}}. \label{C_LEO_q&v_2}
\end{equation}
Here, $c_{\mathrm{C}}$ depends on the size of considered coordinate space. 
Since $c_E$ is based on an Earth scale while $c_C$ is based on a distance of Src-Dst, it is clear that $c_C \ll c_E$.
As a result, the dimension of action space as well as the state space related to the number of SATs reduces from $I$ to $I'$, since we now only consider the position of selected SAT.
Note that it holds $I' < I$, where $I'$ denotes the maximum number of SATs that can be included in a given coordinate space which depends on $c_{\mathrm{C}}$.
For instance, the action space of $a^{\mathcal{W}}[n]$ reduces to $\mathbb{R}^{I' \times 2}$ and the state space of $\mathbf{q}^{i_{k}}_{\mathrm{L}}[n] \in I$ shrinks to $\mathbf{q}^{i^{*}_{k}}_{\mathrm{L}}[n], \in I'$.

% In the following Sec.~\ref{Numerical Result}, we present the numerical results of proposed algorithm for the EE maximization of SAT-UAV integrated NTN.

% \tred{[JH: we should first clarify that $c_C \ll c_E$, and in the following example, we may need to exemplify specific values of $c_C$ and $c_E$ (currently I cannot understand the following example)]} 

%%%%%%%%%%%%%%%%%%%%%%%%%%%%%%%%%%%%%%%%%%%%%%%%%%%%%%%%%%%%%%%%%%%%%%%%%%%%%%%%%%%%%%%%%%%%%%%%%%%%%%%%%
%%%%%%%%%%%%%%%%%%%%%%%%%%%%%%%%%%%%%%%%%%%%%%%%%%%%%%%%%%%%%%%%%%%%%%%%%%%%%%%%%%%%%%%%%%%%%%%%%%%%%%%%%%%%%%%%%%%%%%%%%%%%%%%%%%%%%%%%%%%%%%%%%%%%%%%%%%%%%%%%%%%%%%%%%%%%%%%%%%%

%%%%%%%%%%%%%%%%%%%%%%%%%%%%%%%%%%%%%%%%%%%%%%%%%%%%%%%%%%%%%%%%%%%%%%%%%%%%%%%%%%%%%%%%%%%%%%%%%%%%%%%%%%%%%%%%%%%%%%%%%%%%%%%%%%%%%%%%%%%%%%%%%%%%%%%%%%%%%%%%%%%%%%%%%%%%%%%%%
% Numerical Result %
%%%%%%%%%%%%%%%%%%%%%%%%%%%%%%%%%%%%%%%%%%%%%%%%%%%%%%%%%%%%%%%%%%%%%%%%%%%%%%%%%%%%%%%%%%%%%%%%%%%%%%%%%%%%%%%%%%%%%%%%%%%%%%%%%%%%%%%%%%%%%%%%%%%%%%%%%%%%%%%%%%%%%%%%%%%%%%%%%
\section{Numerical Evaluations} \label{Numerical Result}
%%%%%%%%%%%%%%%%%%%%%%%% Parameter Table

% \tred{[JH: throughout the paper (particularly in all preceding sections), be consistent with the following expressions: orbital lane (NOT orbital lane, constellation, ...), orbital line segment, orbital lane circumference (NOT circumstance)]; E2E sum throughput (either this or system throughut consistently); path (for an E2E route, given by up to the number of agents); link (for a single hop when specifying FSO/RF)}

In this section, we present the simulation results to demonstrate the performance of the proposed MARL method in terms of EE, E2E throughout, SAT associations, and UAV trajectories. To this end, we consider an area of $4000\times6000\times 550$ [km$^3$] wherein the ground Src and Dst located at $[0,0,0]$ and $[4000,4000,0]$ are connected through two LEO orbital lanes of SATs between which UAVs are flying at an altitude of $50$ [km] \cite{Intro_4}. Each orbital lane consists of $22$ SATs with the orbital lane circumference of $43486$ [km], resulting in an inter-SAT distance of $1977$ [km]. Since the orbital lane circumference is much larger than the area of interest, each orbital lane is approximated as a line segment at an altitude of $550$ [km] \cite{Starlink}, in which an infinite number of moving SATs are separated with the inter-SAT distance $1977$ [km] and re-numbered at an interval of $22$. In our MDP model, the objective function of (P1$^{\star}$) corresponds to the un-discounted accumulated rewards over an episode up to time slot $N$. The length of each episode is set as one orbital period, i.e., $N=572$, during which an SAT with the orbital speed $7.59$ [km/s] completes one orbit.
Here, the orbital period and speed are calculated using the relations $4\pi^{2}(r_{\mathrm{E}})^{3} = T^{2}GM$ and $V^{2}r_{\mathrm{E}} = GM$, respectively, where $r_{\mathrm{E}}$, radius of orbit in metres; $T$, orbital period in seconds; $V$, orbital speed in [m/s]; $G$, gravitational constant, approximately $6.673 \times 10^{-11}$ [m$^{3}$/kg$^{1}$/s$^{2}$]; $M$, mass of Earth, approximately $5.98 \times 1024$ [kg].

\begin{table}
  \centering
  \resizebox{1\columnwidth}{!}{\begin{minipage}[t]{0.5\textwidth}
  \caption{Simulation parameters.}
  % \vspace{1mm}
  \centering
  \begin{tabular} {l l}
    % \centering
	%\begin{tabular} {P{1.2cm} P{0.001cm} P{0.001cm} P{0.001cm} P{0.001cm}P{0.001cm} P{0.001cm}}
	\toprule[1pt]
	\textbf{Parameter} & \textbf{Value} \\
	\midrule
	Time slot size & $\delta_t=10$  \\
	Number of time slots per episode & $N=572$ \\
	Location of Src & $\mathbf{q}_{\mathcal{S}}=[0,0,0]^T$ \\  
	Location of Dst & $\mathbf{q}_{\mathcal{D}}=[4000,4000,0]^T$ \\
	\midrule
	Altitude of SAT & $H_{\mathrm{L}}=550$ {[}km{]} (from \cite{Starlink}) \\ 
	Velocity (Speed) of SAT1 ($k=1$) & $\mathbf{v}^{1}_{\mathrm{L}, \mathrm{I}}=[0, 7.59, 0]^T$ ($7.59$ {[}km/s{]}) \\
	Velocity of SAT2 ($k=2$) & $\mathbf{v}^{2}_{\mathrm{L}, \mathrm{I}}=[0, -7.59, 0]^T$\\
	Radius of an orbit & $r_{\mathrm{E}} = 6921$ {[}km{]} \\ 
	Circumference of an orbital lane & $c_{\mathrm{E}} = 43486$ {[}km{]} \\ 
	Number of SATs per orbital lane & $I = 22$ \\ 
	Inter-SAT distance & $1977$ {[}km{]}  \\
	Length an orbital line segment & $c_{\mathrm{C}} = 6000$ {[}km{]} \\ 
	Number of SATs per orbital line segment & $I' = 3$ \\ 
	\midrule
	Altitude of UAV & $H_{\mathrm{U}}=50$ {[}km{]} (from \cite{Intro_4}) \\
	Maximum acceleration of UAV & $A_{\mathrm{max}} = 5$ [m/s$^2$] \\
	Discretization level for $a_{j}^{\mathcal{A}}[n]$ & $D = 5$ \\
	\midrule
	Bandwidth for RF and FSO links & $B_{\mathrm{RF}} = B_{\mathrm{FSO}} = 10^{9}$ {[}Hz{]} \\   
	Reference SNR ($d=1$ {[}m{]}) & $\gamma_{0} = 10^{9}$ \\ 
	Visibility & $V = 15$ {[}km{]} \\
	ASNR &  $\gamma_{\mathrm{FSO}} = 25$ {[}dB{]} ($\alpha = 0.1$) \\
% 	\midrule
% 	Learning rate, Discount factor & 0.0001, 0.995 \\ 
% 	Batch size, \# of iterations/update & 1072, 1072 \\ 
% 	\# of iterations/episode & 536 \\ 
% 	Training iterations of episode & 50000 \\
	\bottomrule[1pt]
\end{tabular}

  \label{table_Paramter}
  \end{minipage}}
\end{table}

Given the aforementioned MARL environment, we consider up to $3$ UAVs that are agents. Following the centralized-Critic MARL framework \cite{MA_Foerster_1, Lowe, MA_Foerster_2}, each agent has an Actor~NN, while only one randomly selected agent stores a Critic~NN. For all Actor and Critic NNs, we identically consider a $4$-layer fully-connected multi-layer perceptron (MLP) NN architecture. Each MLP has $2$ hidden layers, each of which has the dimension $128\times 128$ with the rectified linear unit (ReLU) activation functions. Both Actor and Critic NNs are trained using RMSprop with the learning rate $0.0001$, batch size $1072$, training episodes $50000$, and $1072$ iterations per update. The simulations are implemented using TensorFlow Version 2.1. More details on simulation settings are provided in Table~\ref{table_Paramter}. Throughout this section, our proposed scheme is compared with three benchmark schemes as follows.
\begin{enumerate}
\item \textbf{(Proposed) SAT-UAV} is a scenario with the proposed framework, in which Src and Dst are connected through SATs and UAVs. Unless otherwise specified, by default we consider $2$ SAT orbital line segments having the opposite orbiting directions, between which $2$ UAVs are flying, i.e., $K=J=2$. The initial UAV positions are chosen as $[2000, 2667, 50]$ and $[2000, 1333, 50]$.
%  Src-SAT1-UAV-SAT2-Dst.

\begin{figure}
  \centering
  \includegraphics[width=\linewidth, trim = 0 1.5cm 0 1.5cm, clip]{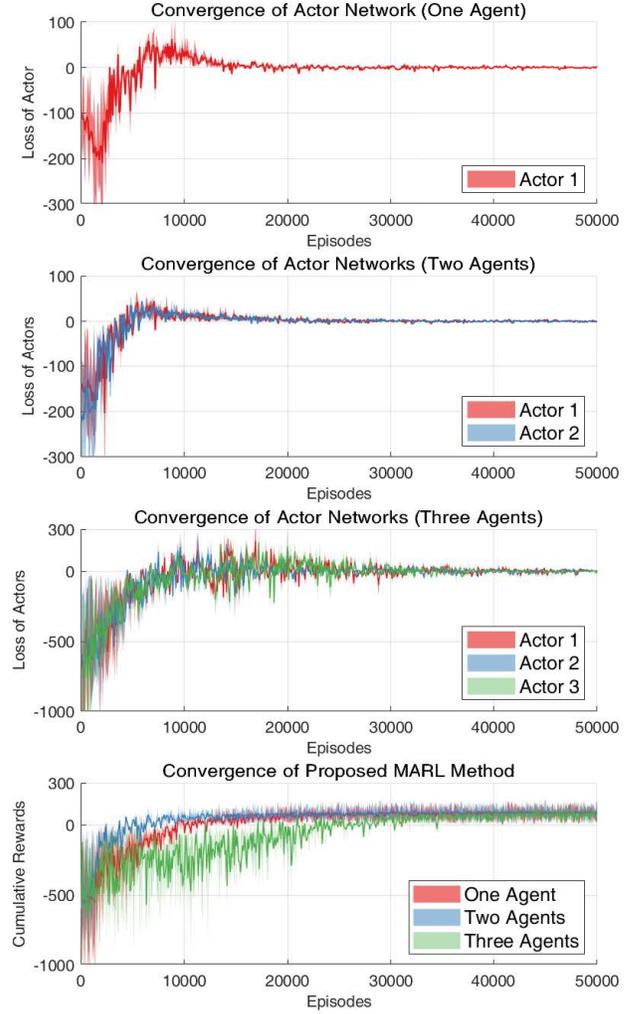}
  \caption{Learning curves of the proposed centralized-Critic MARL algorithm ($J=1$-$3$, $K=2$).}
  \label{fig_Convergence}
\vskip -5pt
\end{figure}

\item \textbf{SAT-Ground} implies a scenario wherein Src and Dst are connected through SATs and ground relay (GRs). The number of GRs and their initial positions are set as the same values of the UAV configurations, only except for their zero altitudes. GRs do not have any constraints and energy consumption incurred by UAV movements. Accordingly, in this case (P1$^*$) is reduced to 
\begin{eqnarray}
(\mathrm{P2)} \!&\!\displaystyle \max_{ \scriptsize \begin{array}{c} \scriptsize \mathcal{W} \end{array} } \!&\!
	\sum_{j \in \mathcal{J}}\sum_{n=1}^{N} \mathit{R}^{j}_{\mathcal{S}, \mathcal{D}}[n] \nonumber \label{P3} \\ 
\!&\!\textrm{s.t.}\!&\! \eqref{C_LEO_q&v_i}, \eqref{R_Si1}-\eqref{R_i2D}, \eqref{P1_C_i1j}-\eqref{P1_C_sum_ji2}. \nonumber
\end{eqnarray}
In (P2) for SAT-Ground, $\mathcal{J}=\{j=1,2,\ldots,J\}$ represents a set of ground terminals, which relay between the SAT constellation network.
% where $\mathcal{J}^{\star} \in \{j^{\star} = 1,2,\ldots,J^{\star}\}$ denotes a set of GRs.
% Note that multiple links support the E2E transmission between Src and Dst, if $J^{\star} \neq 1$.

\item \textbf{SAT-Only} is the case where Src and Dst are connected only through SATs without any UAVs or GRs. 
% We consider two orbital lanes, $K=2$, unless otherwise stated.
% In the case of $K=2$, two SATs serve as relay terminals between Src and Dst, e.g., Src-SAT1-SAT2-Dst.
% It is assumed that each orbital lane is spaced equally apart.

\item \textbf{Direct Transmission} refers to a scenario where Src directly communicates with Dst without any relays. For simplicity we assume that the LoS condition is always guaranteed, yielding the throughout upper bound performance.

% Note that we assume the LoS condition for this baseline, such that it is the upper bound of ground-to-ground communication.
% As identified in the following results, air-to-ground communication schemes exceed this upper bound of the ground-to-ground communication, in terms of E2E system throughput.
\end{enumerate}

% To investigate the effectiveness of an additional relay terminal in SAT-based NTN, we consider an additional UAV (or ground) relay terminal located in-between two SATs. 
%%% Episode, Done(terminal state) 관련

% Unless stated otherwise, the communication between Src and Dst is supported by two links through two relay terminals (See Fig. \ref{Illustration}), and all parameters are listed in Table \ref{table_Paramter}.

\subsection{MARL Training Convergence}

% \textbf{MARL Settings}.\quad

% \textbf{Training Convergence}.\quad
In the proposed centralized-Critic MARL framework, the training convergence of the Actor and Critic NNs is illustrated in Fig. \ref{fig_Convergence}, where the solid curves denote average values and the shared areas correspond to the maximum deviations during $3$ simulation runs. The first three figures show the loss values of the Actor NNs when the number $J$ of agents (i.e., the number of Actor NNs) increases from $1$ to $3$, respectively, validating the convergence of the Actor NNs. The last figure illustrates the cumulative rewards at the Critic NN when $J=3$, which implies the convergence of the overall MARL given the centralized-Critic architecture under study. 

The results show that by up to $2$ agents, each Actor NN and the overall MARL converge within about $10000$ episodes, whereas for $3$ agents, the convergence requires around $50000$ episodes. For large-scale systems, it is thus necessary to accelerate the training convergence. In this regard, periodically averaging the Actor NN parameters (i.e., federated learning \cite{Google:FL19}) or outputs (i.e., federated distillation \cite{MLPCD}) across agents could be an interesting topic for future study. Note that the cumulative reward does not directly represent the system utilities, i.e., E2E system throughput, energy, and EE, which are evaluated in the following subsections.

% such as E2E system throughput, consumption energy of UAVs.
% We present the detailed system utility of the proposed algorithm in the following subsection.

% The fluctuation of cumulative rewards decreases to less than $1$ \% after $50000$ episodes.
% It can be seen from Fig. \ref{fig_Convergence} that the proposed MARL method converges within certain episodes, while more iterations are required in which more agents are needed to be learned.

%%%%%%%%%%%%%%%%%%%%%%%% Table for E2E System Throughput and EE
% \begin{table*}
%     \centering
%     \resizebox{.95\columnwidth}{!}{\begin{minipage}[t]{.97\columnwidth}
%     \caption{End-to-end system throughput over NTN configuration types.}
%     \label{Table_1}
%     % \vspace{1mm}
%     % \centering\textbf{Network Configuration} \\[1pt]
%     \input{Table/Table1.tex}
%     \end{minipage}}
%   \centering    
%   \hspace{10pt}
%   \resizebox{1.\columnwidth}{!}{\begin{minipage}[t]{.98\columnwidth}
%   \centering
%     \caption{End-to-end system throughput of NTN over network configuration types and association decision types.}
% 	\label{Table_2,3}
% 	\vspace{4mm}
% 	\input{Table/Table_2,3}
%   \end{minipage}}
% \end{table*}
% \begin{table*}
% %   \hspace{5pt}
%   \centering    
%   \resizebox{1.25\columnwidth}{!}{\begin{minipage}[t]{1.21\columnwidth}
%   \centering
%     \caption{Comparison of proposed SAT-UAV over different objective.}
% 	\label{Table_Proposed}
% 	\input{Table/Table_Proposed}
%   \end{minipage}}
% \end{table*}

\subsection{Throughput and Energy Efficiency}

%%%%%%%%%%%%%%%%%%%%%%%% Figure and Table for E2E system throughput
\begin{table}[t]
    \centering
    \resizebox{1\columnwidth}{!}{\begin{minipage}[t]{.97\columnwidth}
    \caption{End-to-end throughput with or without relaying ($J=1$).}
    \label{Table_1}
    % \vspace{1mm}
    % \centering\textbf{Network Configuration} \\[1pt]
    \newcolumntype{R}{>{\raggedleft\arraybackslash}X}
\begin{tabularx}{1\linewidth}{l l}
    \toprule[1pt]
    {Network Configuration} & E2E Sum Txpt. [Mbps] \\     
    \cmidrule(lr){1-1} \cmidrule(lr){2-2}
    Direct Transmission & $0.4507$\hspace{3pt}  \tikz{
        % \fill[fill=lightgray,fill opacity=0.1] (0,0) rectangle (2.2,0.2);
        \fill[fill=color1] (0.0,0) rectangle (0.1,0.2);
        \fill[pattern=north west lines, pattern color=black!30!color1] (0.0,0) rectangle (0.1,0.2);
    } \\
    SAT-Only, $K=1$ & $0.6488$\hspace{3pt}  \tikz{
        \fill[fill=color2] (0.0,0) rectangle (0.3,0.2);
        \fill[pattern=north west lines, pattern color=black!30!color2] (0.0,0) rectangle (0.3,0.2);
    } \\
    SAT-Only, $K=2$ & $0.6839$\hspace{3pt}  \tikz{
        \fill[fill=color3] (0.0,0) rectangle (0.35,0.2);
        \fill[pattern=north west lines, pattern color=black!30!color3] (0.0,0) rectangle (0.35,0.2);
    } \\
    SAT-Only, $K=3$ & $1.1202$\hspace{3pt}  \tikz{
        \fill[fill=color4] (0.0,0) rectangle (1.1,0.2);
        \fill[pattern=north west lines, pattern color=black!30!color4] (0.0,0) rectangle (1.1,0.2);
    } \\
    \textbf{SAT-Ground, $K=2$} & {$\mathbf{1.9721}$}  \tikz{
        \fill[fill=color5] (0.0,0) rectangle (2.1,0.2);
        \fill[pattern=north west lines, pattern color=black!30!color5] (0.0,0) rectangle (2.1,0.2);
        % \fill[pattern=north east lines, pattern color=black!70!color3] (0.7910210041398044,0) rectangle (1.1728886640255558,0.2);
    } \\
    & \ \ \ \ \ \ \ \ \ \tikz{
        \draw[black] (1.1,0) -- (3.15,0);
        \draw[black] (1.1,-2pt) -- (1.1,2pt)node[anchor=north] {\tiny$0.5$};
        \draw[black] (1.9,-2pt) -- (1.9,2pt)node[anchor=north] {\tiny$1.0$};
        \draw[black] (2.5,-2pt) -- (2.5,2pt)node[anchor=north] {\tiny$1.5$};
        \draw[black] (3.15,-2pt) -- (3.15,2pt)node[anchor=north] {\tiny$2.0$};
        \draw[black] (3.6,-0.1pt) -- (3.6,0.1pt)node[anchor=north] {\tiny[Mbps]};
    } \vspace{-1mm} \\
    \bottomrule[1pt]
\end{tabularx}
    \end{minipage}}
\end{table}

\begin{table}[t]
  \centering    
  
  \resizebox{1\columnwidth}{!}{\begin{minipage}[t]{.98\columnwidth}
  \centering
  \caption{End-to-end throughput with GRs or UAVs ($J=K=2$).}
  \vskip -11pt  
	\label{Table_2,3}
	\vspace{4mm}
	\newcolumntype{R}{>{\raggedleft\arraybackslash}X}
\begin{tabularx}{1\linewidth}{l l}
    \toprule[1pt]
    Network Configuration & E2E Sum Txpt. [Mbps]\\     
    \cmidrule(lr){1-1} \cmidrule(lr){2-2}
    SAT-Ground, Non-cooperative & $17.9519$\hspace{4pt}  \tikz{
    \fill[fill=color1] (0.0,0) rectangle (0.3,0.2);
    \fill[pattern=north east lines, pattern color=black!50!color1] (0.0,0) rectangle (0.3,0.2);
    } \\
    SAT-Ground, Cooperative & $29.3159$\hspace{4pt}  \tikz{
        \fill[fill=color2] (0.0,0) rectangle (0.9,0.2);
        \fill[pattern=north east lines, pattern color=black!50!color2] (0.0,0) rectangle (0.9,0.2);
    } \\
    \textbf{SAT-UAV, Cooperative} & {$\mathbf{58.2921}$}  \tikz{
        \fill[fill=color3] (0.0,0) rectangle (2.0,0.2);
        \fill[pattern=north east lines, pattern color=black!50!color3] (0.0,0) rectangle (2.0,0.2);
    } \\
    & \ \ \ \ \ \ \ \ \ \tikz{
        \draw[black] (0,0) -- (2.2,0);
        \draw[black] (0.0,-2pt) -- (0.0,2pt)node[anchor=north] {\tiny$10$};
        \draw[black] (1.1,-2pt) -- (1.1,2pt)node[anchor=north] {\tiny$35$};
        \draw[black] (2.2,-2pt) -- (2.2,2pt)node[anchor=north] {\tiny$60$};
        \draw[black] (2.7,-0.01pt) -- (2.7,0.01pt)node[anchor=north] {\tiny[Mbps]};
    } \vspace{-1mm} \\
    \bottomrule[1pt]
\end{tabularx}
  \end{minipage}}
\end{table}

In our proposed NTN framework, UAVs and SATs play key roles for improving the system throughput and EE. In Tables~\ref{Table_1}-\ref{Table_Proposed}, we provide ablation studies to clarify their contributions in relation to long-distance, inter-orbit, and mobile relaying under different objectives, as elaborated next.

To identify the effectiveness of relaying in E2E throughput, Table~\ref{Table_1} compares SAT-Only relaying, SAT-Ground relaying, and Direct Transmission without relaying. With $K=2$, the result shows that SAT-Only achieves $1.51$x higher throughput compared to Direct Transmission. Next, it is remarkable that SAT-Ground achieves $4.38$x higher throughput than Direction Transmission. This throughput is $2.9$x higher than SAT-Only with the same $K=2$, which is even $1.88$x higher compared to SAT-Only with $K=3$. The result highlights the effectiveness of additional relaying in long-distance communication throughput. 

% Src and Dst associate with closest SAT, and time-varying network topology (such as SATs, GR) follows the association decision of Src and Dst.
% As shown in Table \ref{Table_1}, E2E throughput of the ISLs composed of multi SAT surpass E2E throughput of the direct transmission from Src to Dst.
% Besides, additional GR, which is located between two SATs, i.e., SAT-Ground (Two SATs), achieves $188.36$ \% throughput-gain compared to the result of ISLs composed with two SAT, i.e., SAT-Only (Two SATs). 
% These results confirm that the more SATs are, the more E2E system throughput can be achieved, and additional relay in-between SAT constellations can improve the throughput of SAT constellation.

\begin{table}[t]
  %   \hspace{5pt}
    \centering    
    \resizebox{1\columnwidth}{!}{\begin{minipage}[t]{1.13\columnwidth}
    \centering
      \caption{Comparison of proposed SAT-UAV over different objectives ($J=K=2$).}
    \label{Table_Proposed}
    \begin{tabularx}{1\linewidth}{l c c c}
			\toprule[1pt]
			{Objective} & E2E Sum Txpt. [Mbps] & Avg. Power [W] & EE [Kbits/J]\\
			\cmidrule(lr){1-1} \cmidrule(lr){2-2} \cmidrule(lr){3-3} \cmidrule(lr){4-4} 
% 			Baseline & $29.316$ \tikz{
% \draw[gray,line width=.3pt] (0,0) -- (1.2,0);
% \draw[white, line width=0.01pt] (0,-2pt) -- (0,2pt);
% \draw[black,line width=3pt] (0.25200451660156203,0) -- (0.2751595458984377,0);} & - & - \\
Rate-max & {$\mathbf{73.231}$} \tikz{
\draw[gray,line width=.3pt] (0,0) -- (1.1,0);
\draw[white, line width=0.01pt] (0,-2pt) -- (0,2pt);
\draw[black,line width=1pt] (0.69,0) -- (0.77,0);
\draw[black,line width=1pt] (0.69,-2pt) -- (0.69,2pt);
\draw[black,line width=1pt] (0.77,-2pt) -- (0.77,2pt);} & $1614.2$\; \tikz{
\draw[gray,line width=.3pt] (0,0) -- (1.1,0);
\draw[white, line width=0.01pt] (0,-2pt) -- (0,2pt);
\draw[black,line width=1pt] (0.73,0) -- (0.87,0);
\draw[black,line width=1pt] (0.73,-2pt) -- (0.73,2pt);
\draw[black,line width=1pt] (0.87,-2pt) -- (0.87,2pt);} & $45.367$\; \tikz{
\draw[gray,line width=.3pt] (0,0) -- (1.1,0);
\draw[white, line width=0.01pt] (0,-2pt) -- (0,2pt);
\draw[black,line width=1pt] (0.42,0) -- (0.48,0);
\draw[black,line width=1pt] (0.42,-2pt) -- (0.42,2pt);
\draw[black,line width=1pt] (0.48,-2pt) -- (0.48,2pt);} \\

Energy-min & $18.223$\; \tikz{
\draw[gray,line width=.3pt] (0,0) -- (1.1,0);
\draw[white, line width=0.01pt] (0,-2pt) -- (0,2pt);
\draw[black,line width=1pt] (0.11,0) -- (0.25,0);
\draw[black,line width=1pt] (0.11,-2pt) -- (0.11,2pt);
\draw[black,line width=1pt] (0.25,-2pt) -- (0.25,2pt);} 
& {$\mathbf{314.24}$} \tikz{
\draw[gray,line width=.3pt] (0,0) -- (1.1,0);
\draw[white, line width=0.01pt] (0,-2pt) -- (0,2pt);
\draw[black,line width=1pt] (0.11,0) -- (0.19,0);
\draw[black,line width=1pt] (0.11,-2pt) -- (0.11,2pt);
\draw[black,line width=1pt] (0.19,-2pt) -- (0.19,2pt);} 
& $57.991$\; \tikz{
\draw[gray,line width=.3pt] (0,0) -- (1.1,0);
\draw[white, line width=0.01pt] (0,-2pt) -- (0,2pt);
\draw[black,line width=1pt] (0.5,0) -- (0.64,0);
\draw[black,line width=1pt] (0.5,-2pt) -- (0.5,2pt);
\draw[black,line width=1pt] (0.64,-2pt) -- (0.64,2pt);} \\ 

\textbf{EE-max}  & $58.292$ \tikz{
\draw[gray,line width=.3pt] (0,0) -- (1.1,0);
\draw[white, line width=0.01pt] (0,-2pt) -- (0,2pt);
\draw[black,line width=1pt] (0.52,0) -- (0.64,0);
\draw[black,line width=1pt] (0.52,-2pt) -- (0.52,2pt);
\draw[black,line width=1pt] (0.64,-2pt) -- (0.64,2pt);}
& $570.62$\; \tikz{
\draw[gray,line width=.3pt] (0,0) -- (1.1,0);
\draw[white, line width=0.01pt] (0,-2pt) -- (0,2pt);
\draw[black,line width=1pt] (0.23,0) -- (0.35,0);
\draw[black,line width=1pt] (0.23,-2pt) -- (0.23,2pt);
\draw[black,line width=1pt] (0.35,-2pt) -- (0.35,2pt);}
& $\mathbf{102.16}$ \tikz{
\draw[gray,line width=.3pt] (0,0) -- (1.1,0);
\draw[white, line width=0.01pt] (0,-2pt) -- (0,2pt);
\draw[black,line width=1pt] (0.97,0) -- (1.07,0);
\draw[black,line width=1pt] (0.97,-2pt) -- (0.97,2pt);
\draw[black,line width=1pt] (1.07,-2pt) -- (1.07,2pt);} \\
			\bottomrule[1pt]
	\end{tabularx}
    \end{minipage}}
  \end{table}

  %%%%%%%%%%%%%%%%%%%%%%%% Figure for Optimal Trajectory and Association
\begin{table*}
  %   \vspace{-1.em}
    \centering
    \begin{minipage}{\textwidth}
      \caption{Best-Effort: SAT associations and UAV trajectories during time slots $n =$ 1 - 572 under (P1$^*$) ($J=K=2$).}
      \label{table_best-effort}
      % \vspace{1mm}
      % Generated with rank_based_methods.py
\newcolumntype{R}{>{\raggedleft\arraybackslash}X}
% \newcolumntype{C}{ >{\centering\arraybackslash} m{4cm} }
% \newcolumntype{D}{ >{\centering\arraybackslash} m{1cm} }

% \tablefontsize
\begin{tabularx}{1\linewidth}{Xcccc}
    \toprule[1pt] 
    & $n=200$ ($t = 2000$ [sec]) & $n=300$ & $n=400$ & $n=500$ \\
    \cmidrule(lr){2-2} \cmidrule(lr){3-3} \cmidrule(lr){4-4} \cmidrule(lr){5-5}
    Association (Best-Effort)
        & \raisebox{-.5\height}{\includegraphics[width=.2\linewidth]{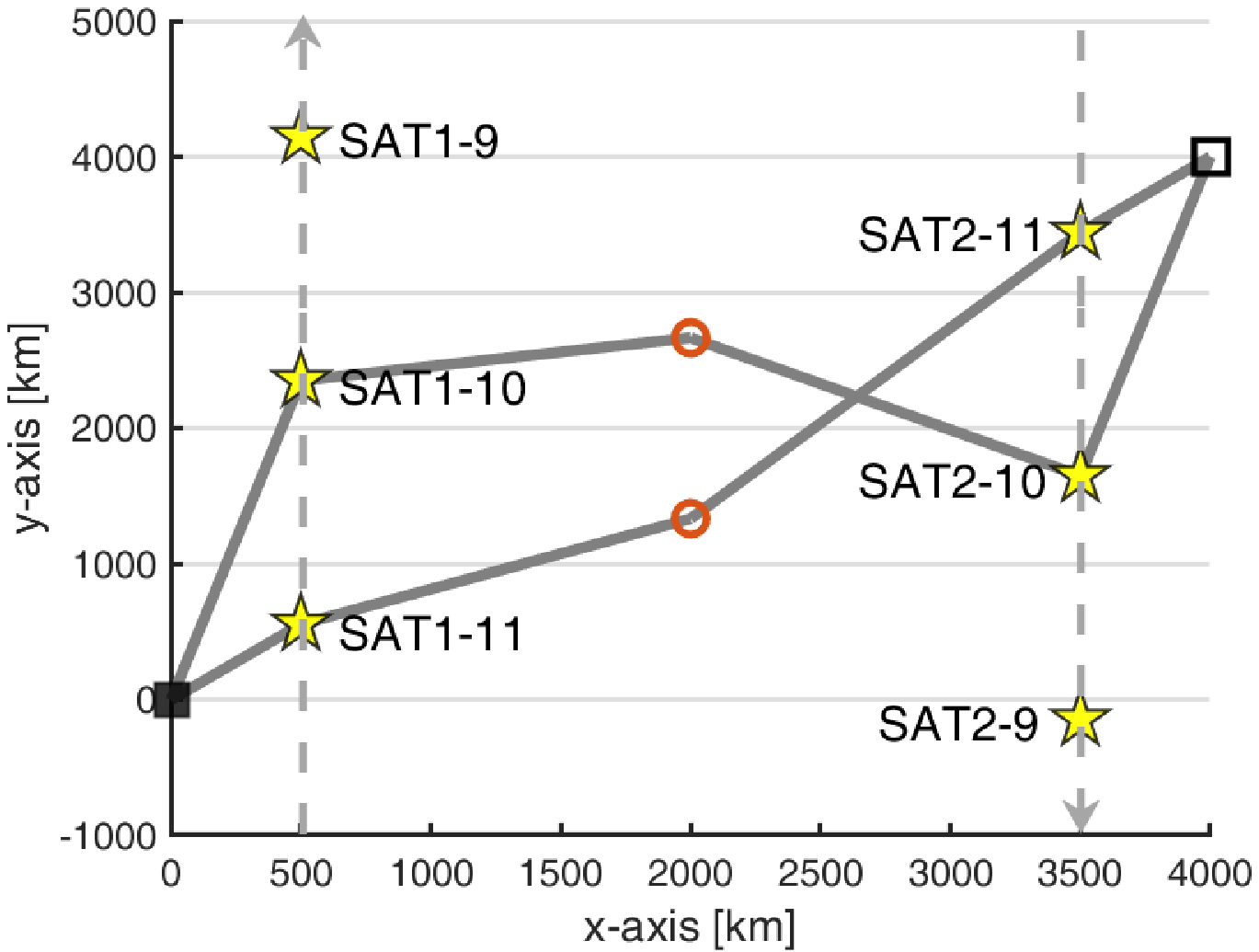}}
        & \raisebox{-.5\height}{\includegraphics[width=.2\linewidth]{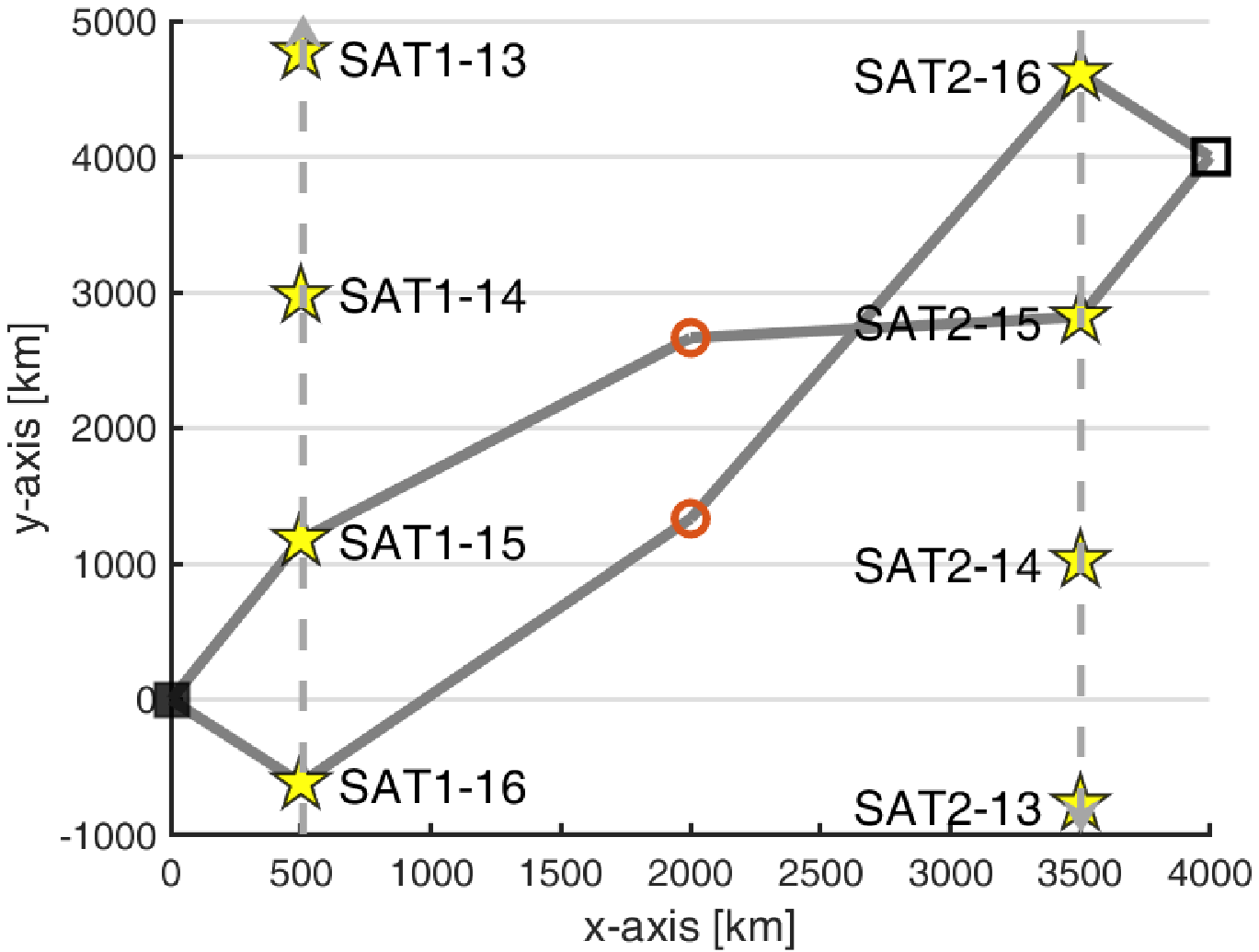}}
        & \raisebox{-.5\height}{\includegraphics[width=.2\linewidth]{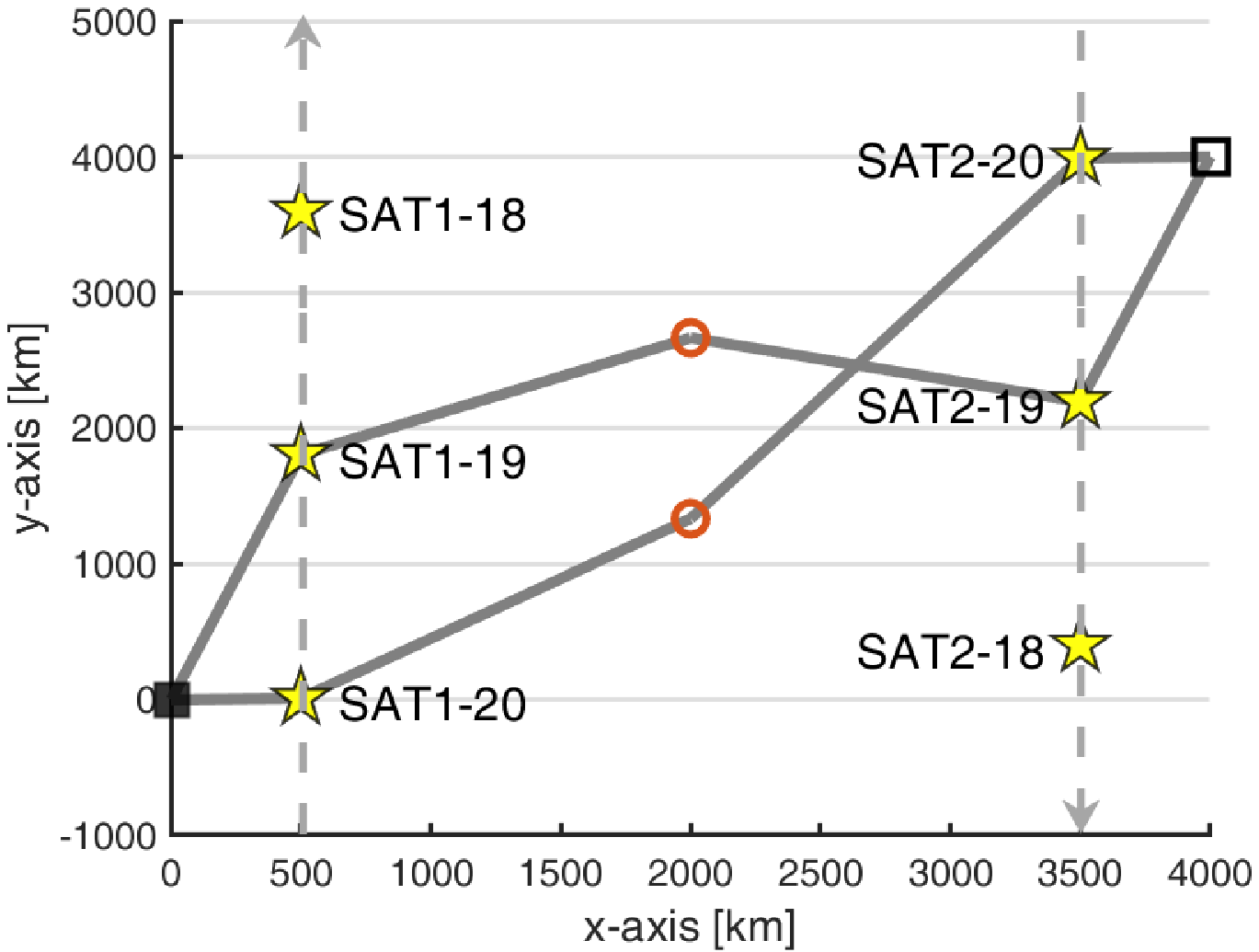}}
        & \raisebox{-.5\height}{\includegraphics[width=.2\linewidth]{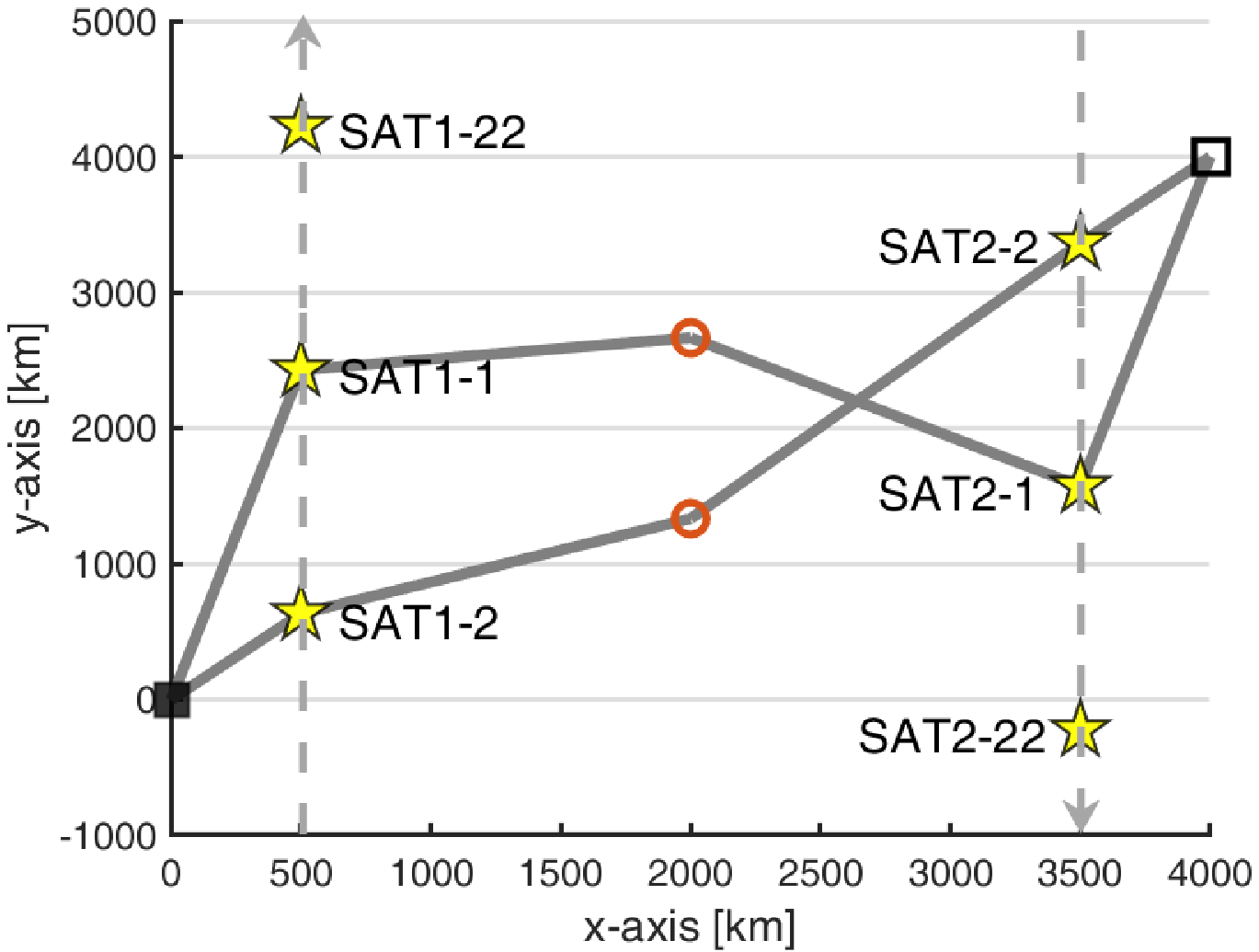}}
        \\
    \cmidrule(lr){2-2} \cmidrule(lr){3-3} \cmidrule(lr){4-4} \cmidrule(lr){5-5}
    Trajectory (Best-Effort)
        & \raisebox{-.5\height}{\includegraphics[width=.2\linewidth]{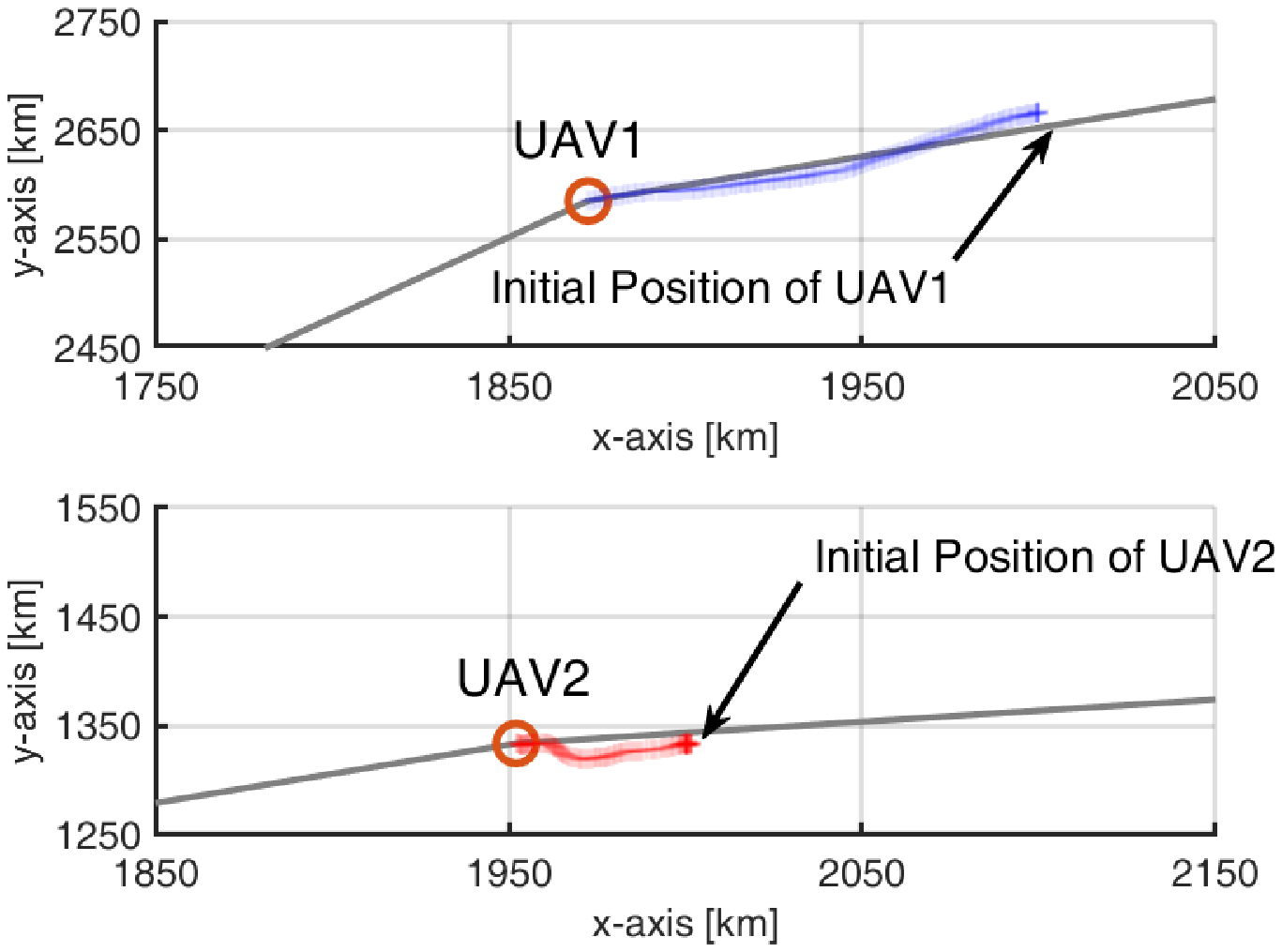}}
        & \raisebox{-.5\height}{\includegraphics[width=.2\linewidth]{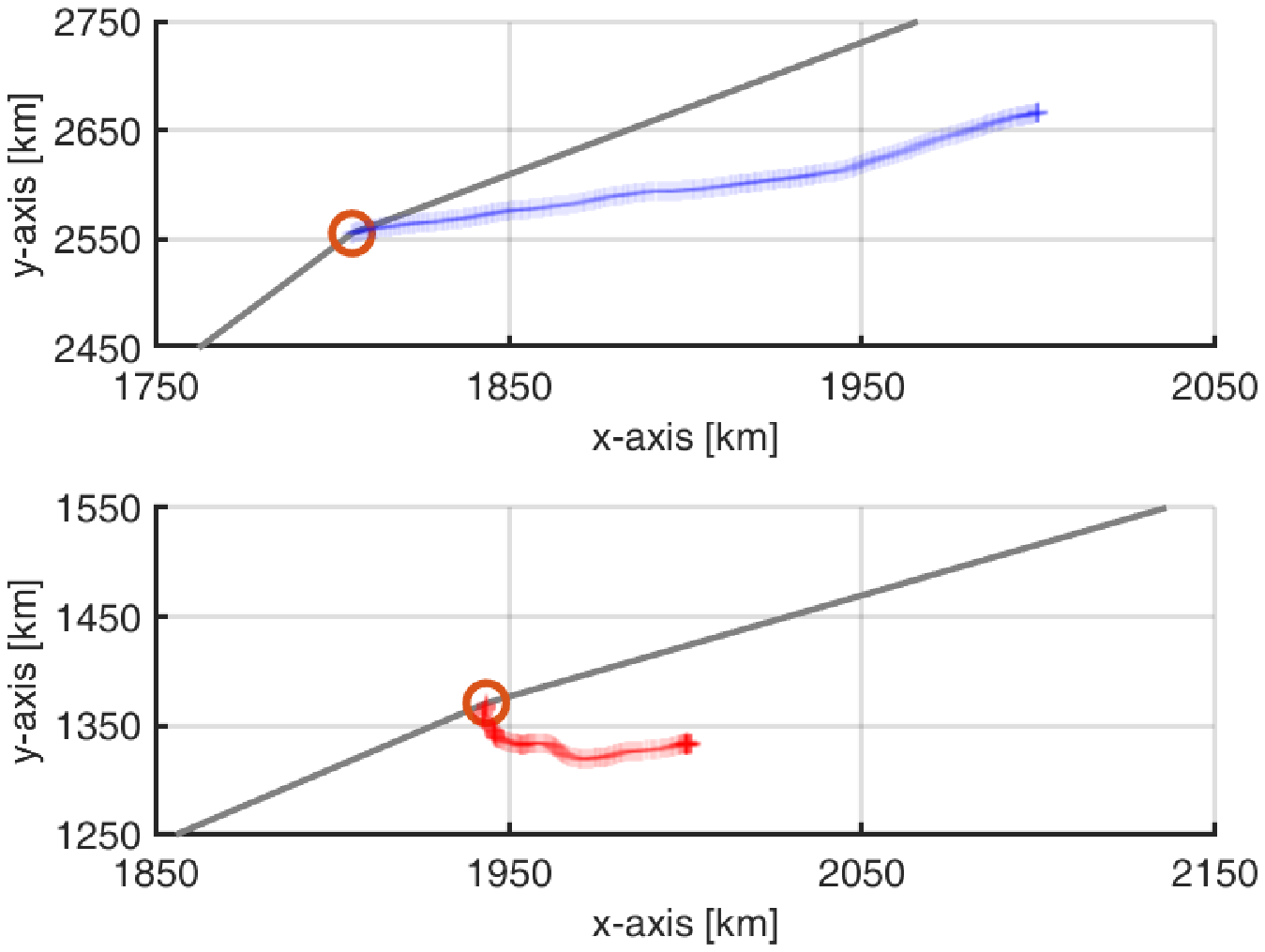}}
        & \raisebox{-.5\height}{\includegraphics[width=.2\linewidth]{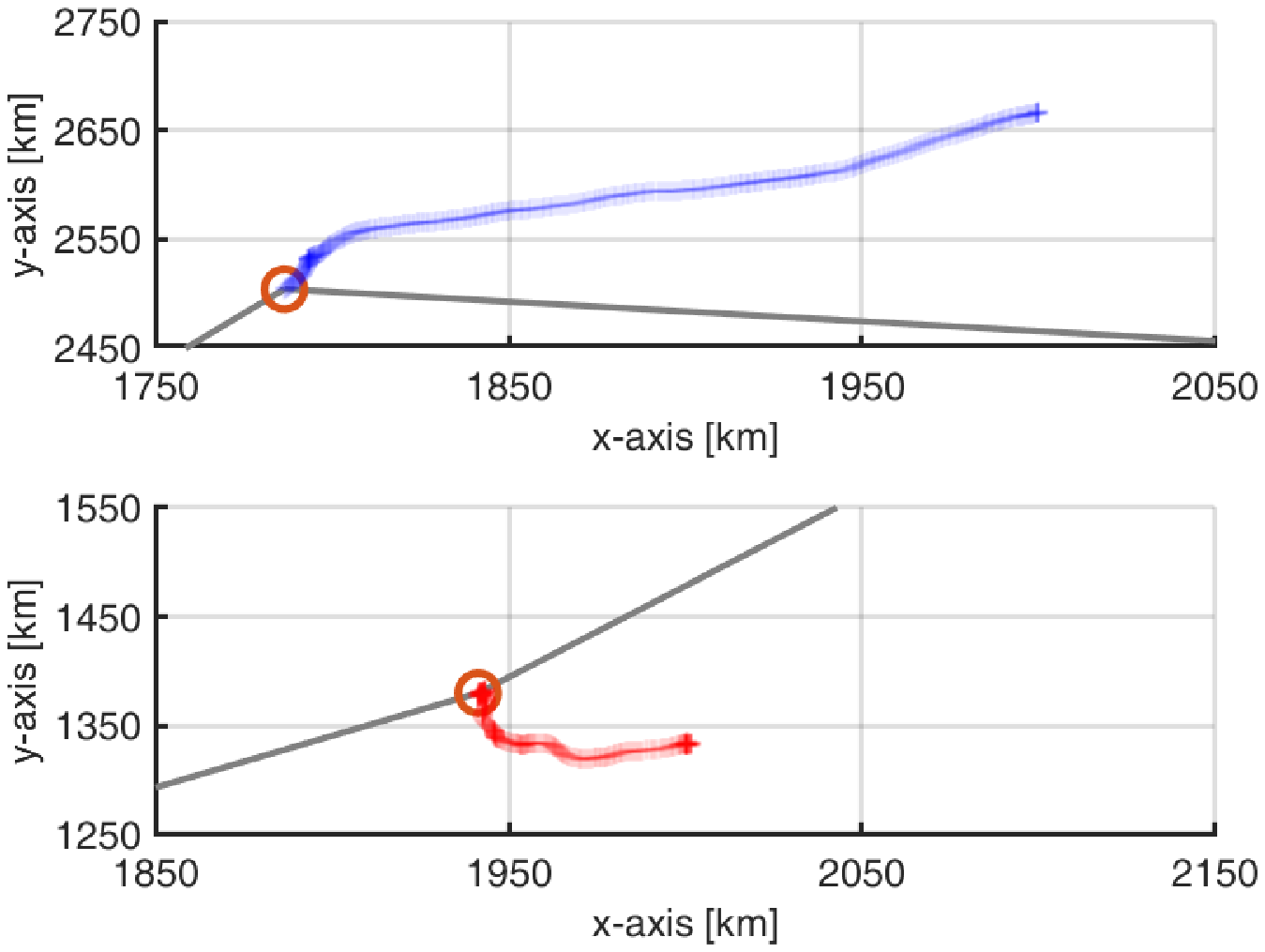}}
        & \raisebox{-.5\height}{\includegraphics[width=.2\linewidth]{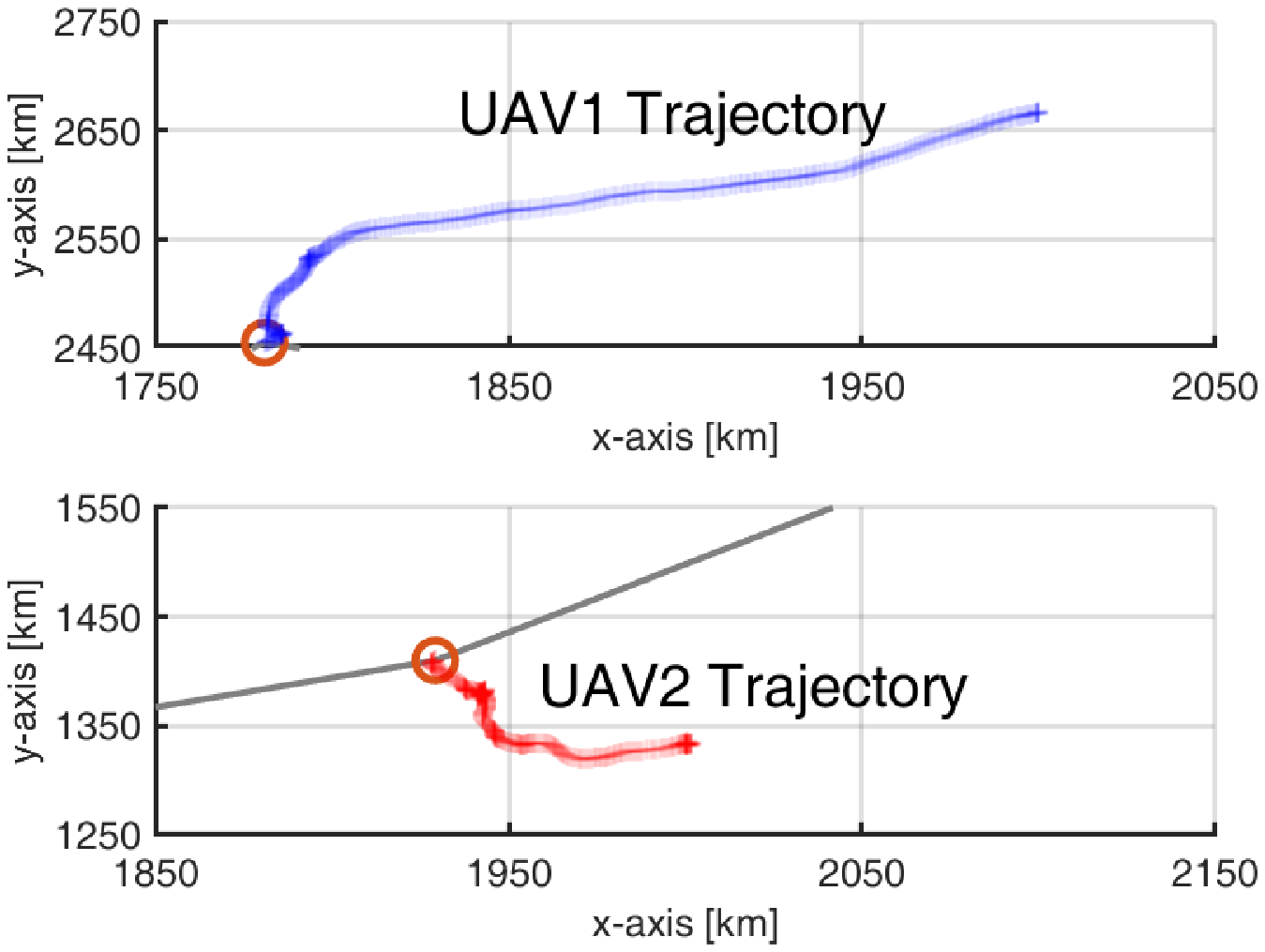}}
        \\
    \bottomrule[1pt] 
\end{tabularx}
    \end{minipage}
  %   \vspace{-1.5em}
  \end{table*}

  \begin{table*}
    % \vspace{-1.em}
    \centering
    \begin{minipage}{\textwidth}
      \caption{Fairness-Oriented: SAT associations and UAV trajectories during time slots $n =$ 1 - 572 under the objective function in \eqref{Reward_reliability} ($J=K=2$).}
      \label{table_fairness}
      % \vspace{1mm}
      % Generated with rank_based_methods.py
\newcolumntype{R}{>{\raggedleft\arraybackslash}X}
% \newcolumntype{C}{ >{\centering\arraybackslash} m{4cm} }
% \newcolumntype{D}{ >{\centering\arraybackslash} m{1cm} }

% \tablefontsize
\begin{tabularx}{1\linewidth}{Xcccc}
    \toprule[1pt] 
    & $n=200$ ($t = 2000$ [sec]) & $n=300$ & $n=400$ & $n=500$ \\
    \cmidrule(lr){2-2} \cmidrule(lr){3-3} \cmidrule(lr){4-4} \cmidrule(lr){5-5}
    Association (Fairness)
        & \raisebox{-.5\height}{\includegraphics[width=.2\linewidth]{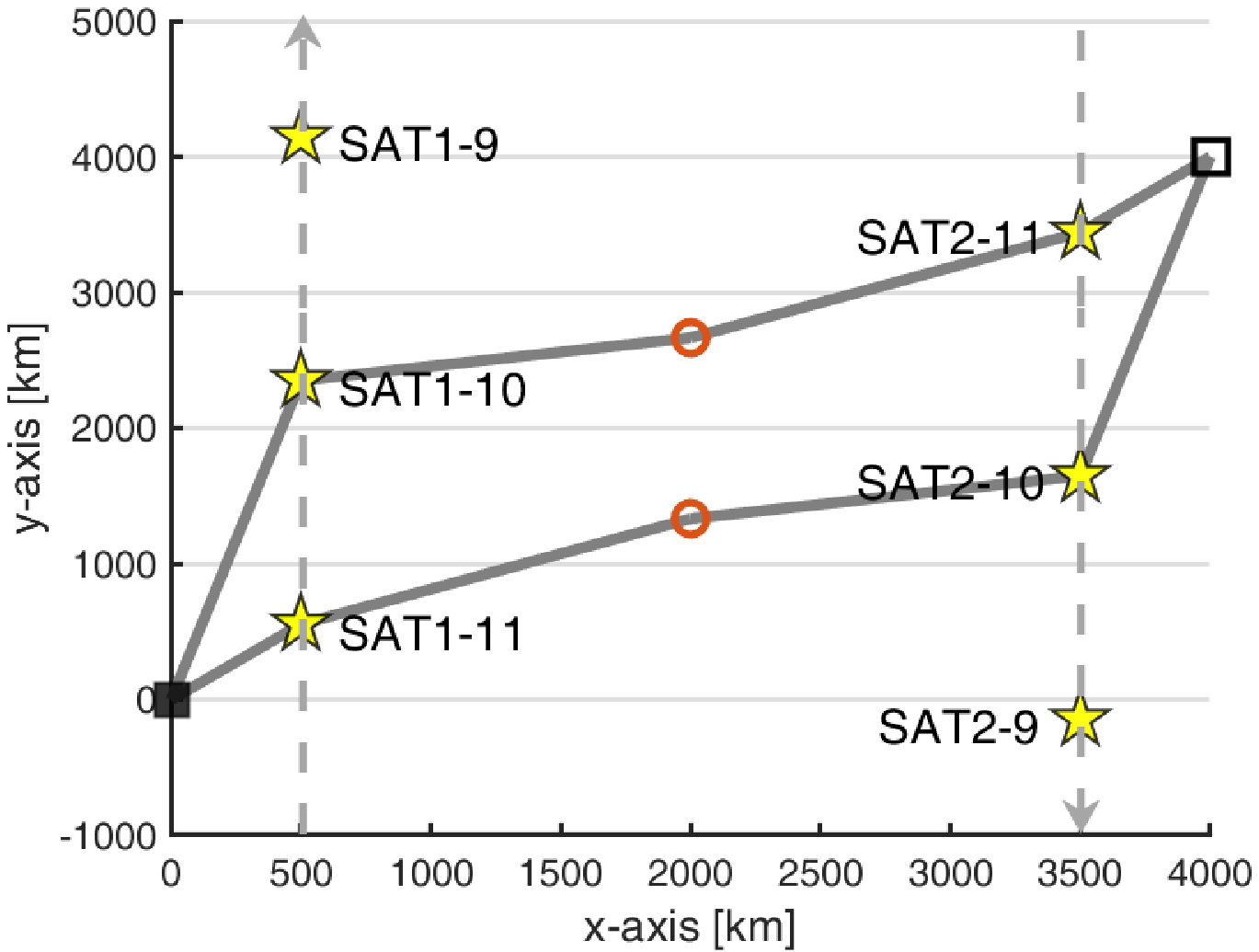}}
        & \raisebox{-.5\height}{\includegraphics[width=.2\linewidth]{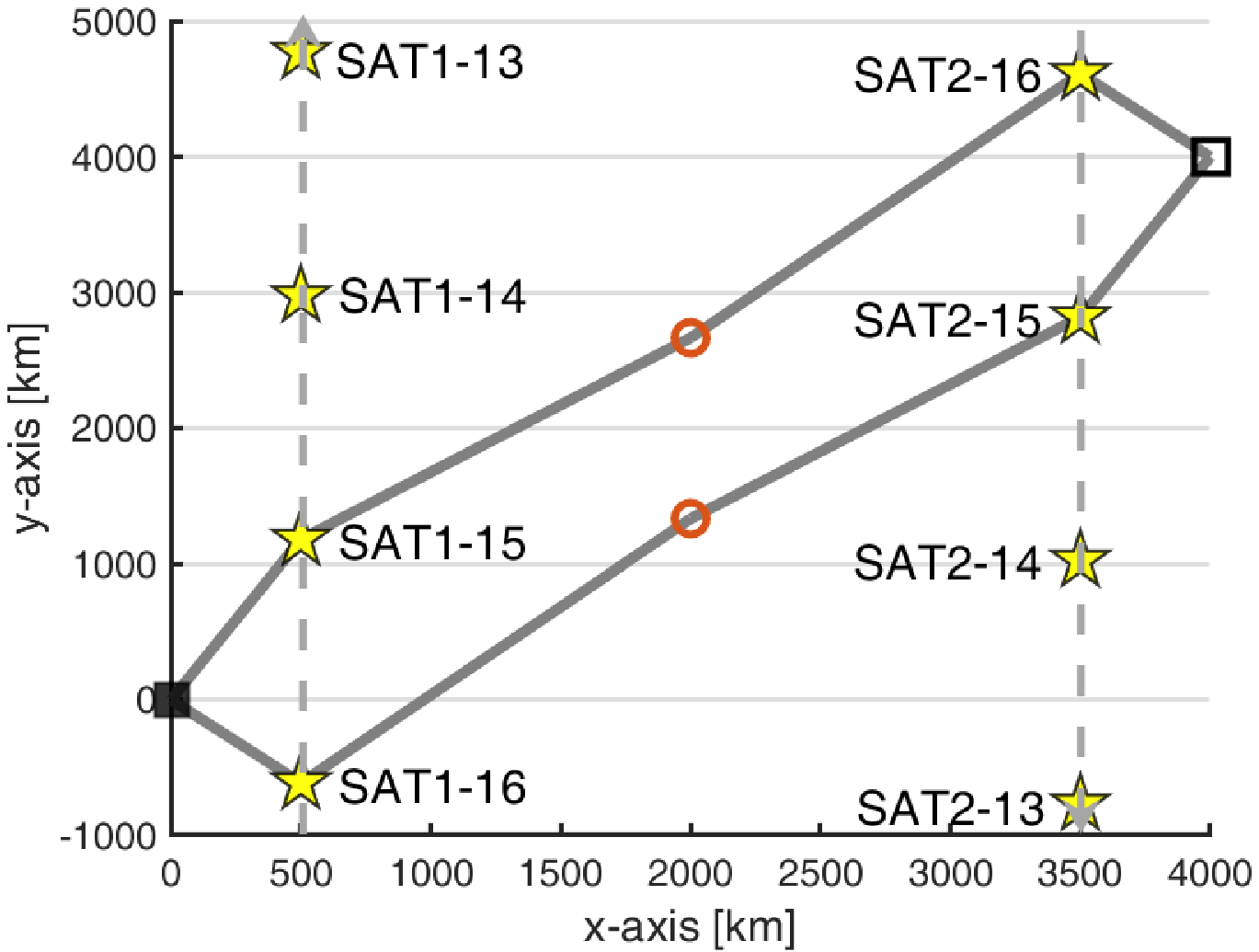}}
        & \raisebox{-.5\height}{\includegraphics[width=.2\linewidth]{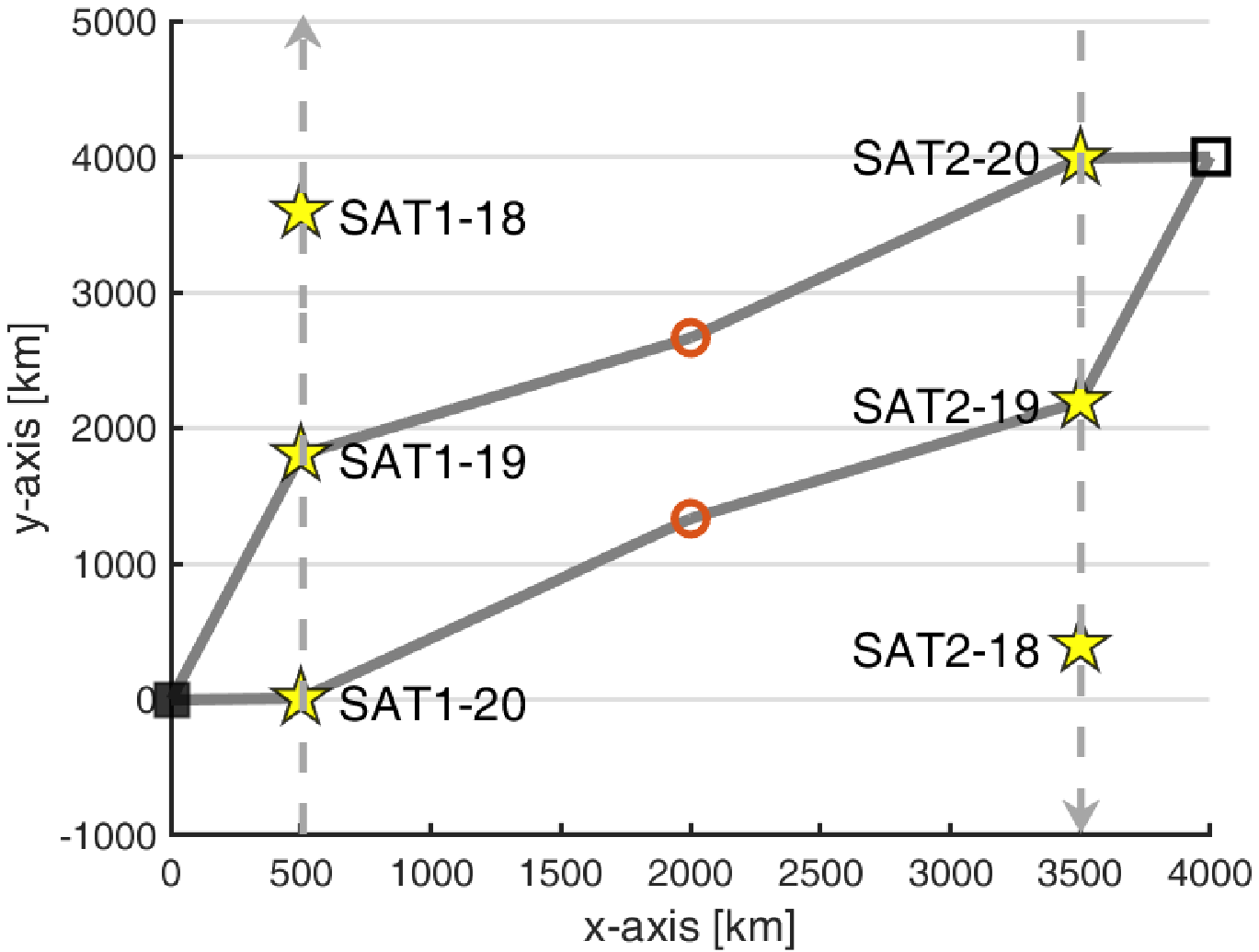}}
        & \raisebox{-.5\height}{\includegraphics[width=.2\linewidth]{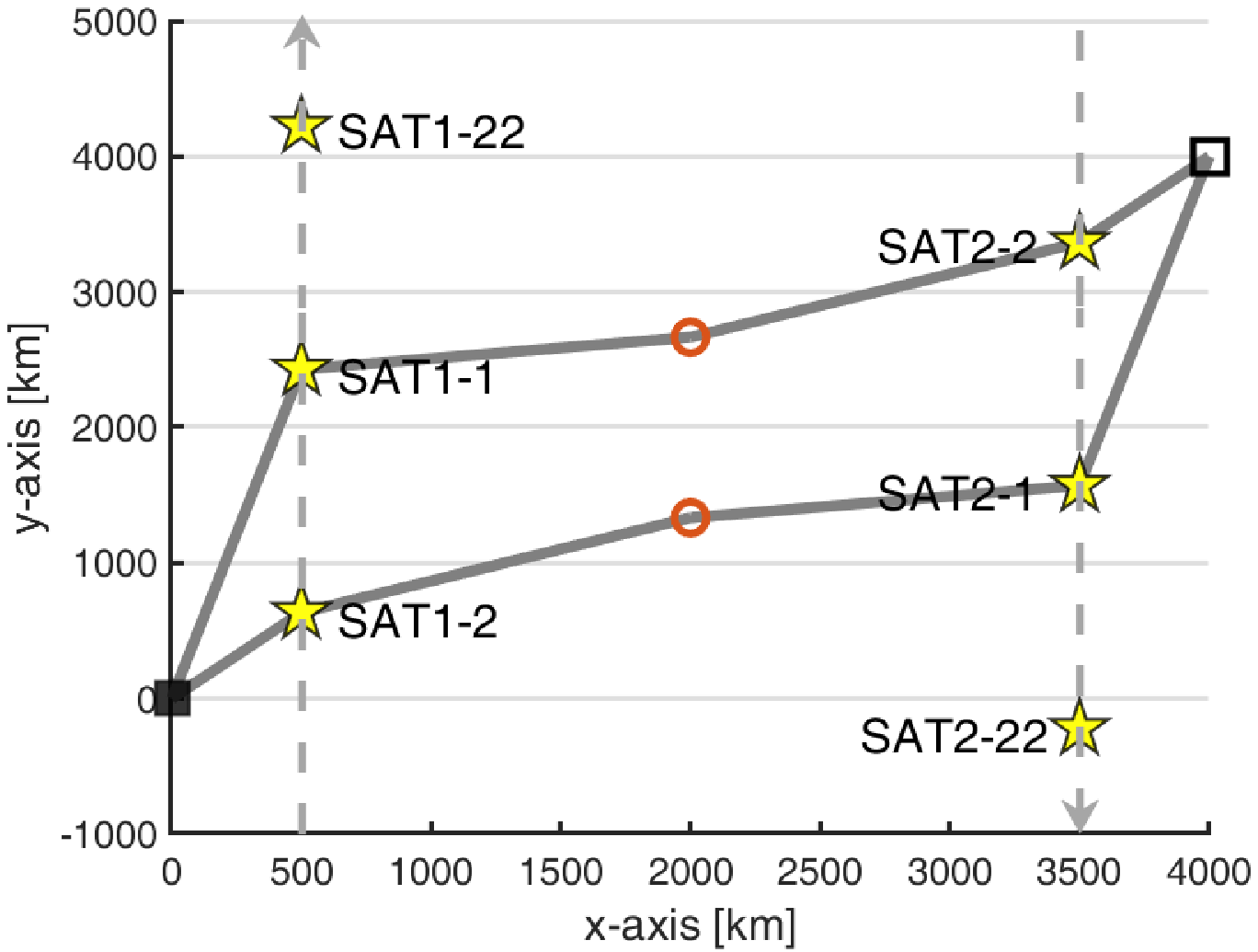}}
        \\
    \cmidrule(lr){2-2} \cmidrule(lr){3-3} \cmidrule(lr){4-4} \cmidrule(lr){5-5}
    Trajectory (Fairness)
        & \raisebox{-.5\height}{\includegraphics[width=.2\linewidth]{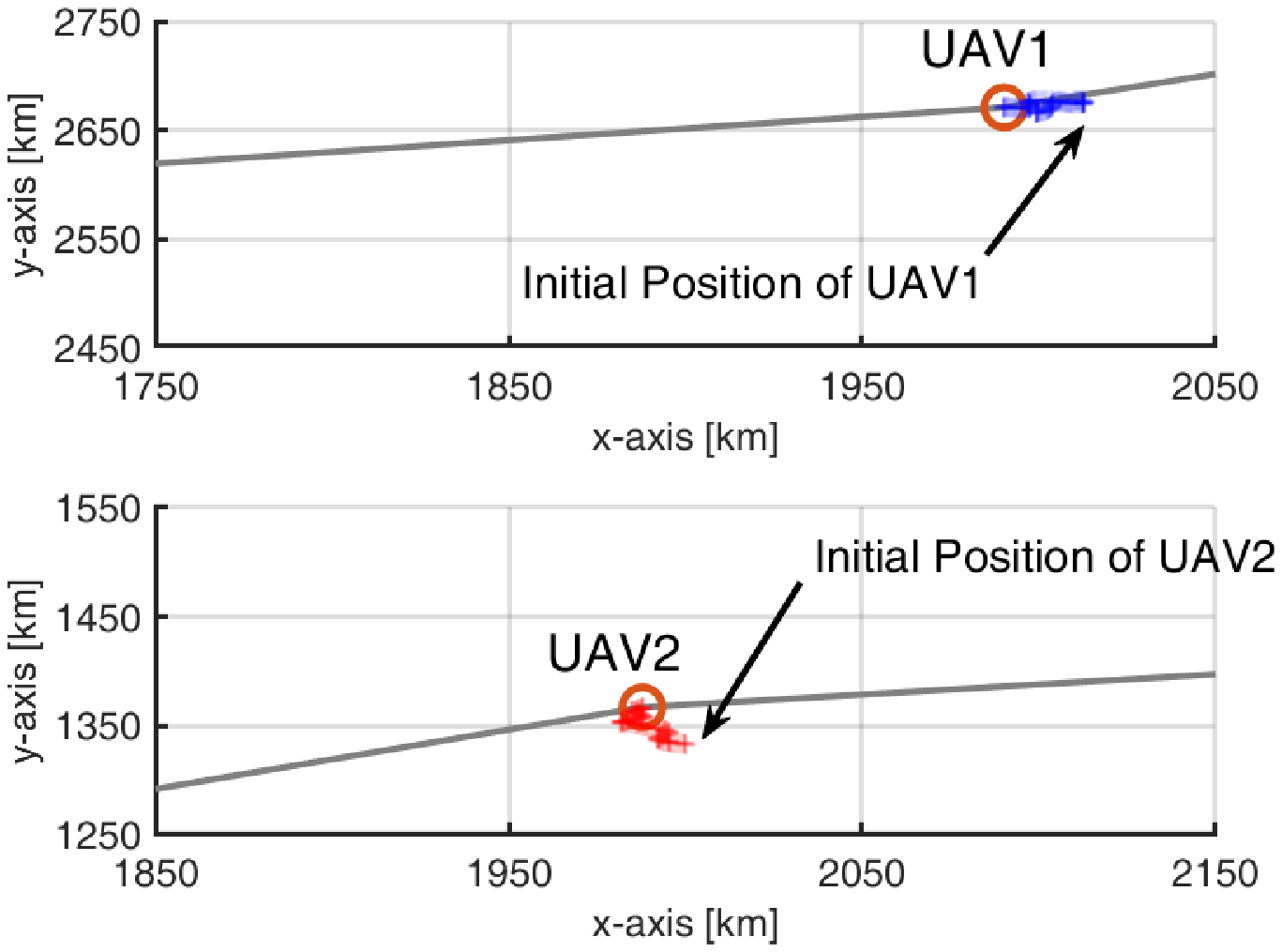}}
        & \raisebox{-.5\height}{\includegraphics[width=.2\linewidth]{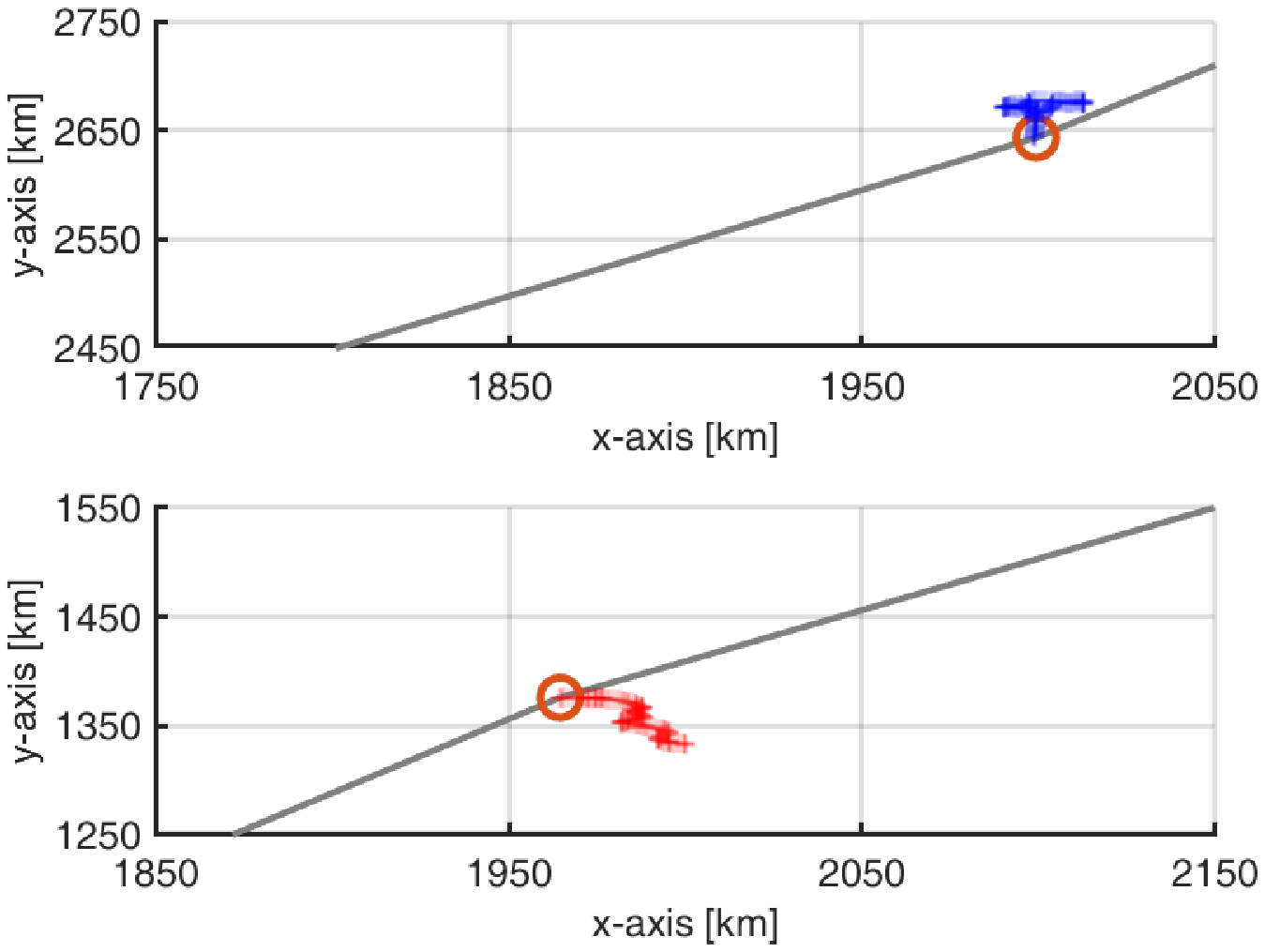}}
        & \raisebox{-.5\height}{\includegraphics[width=.2\linewidth]{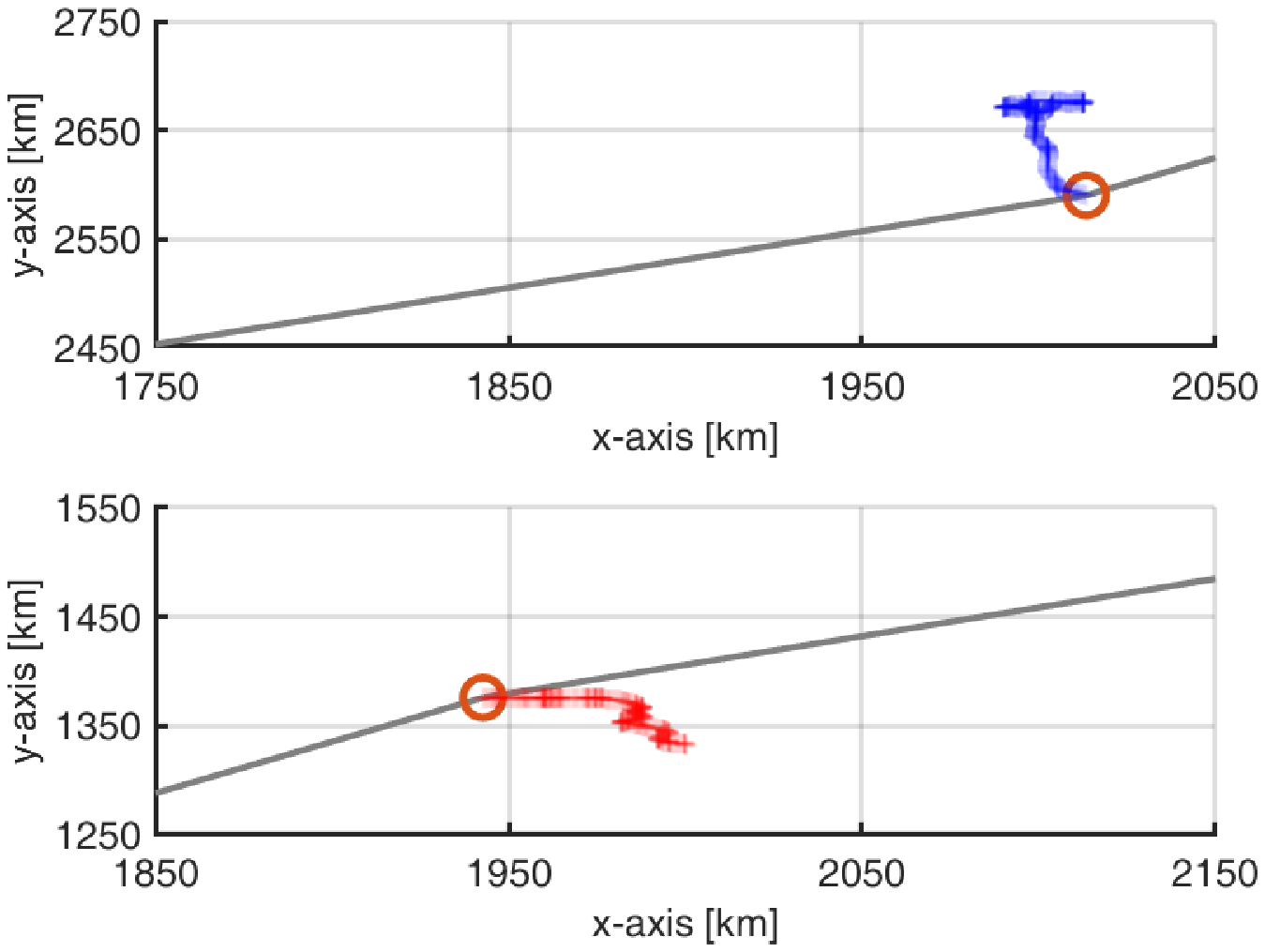}}
        & \raisebox{-.5\height}{\includegraphics[width=.2\linewidth]{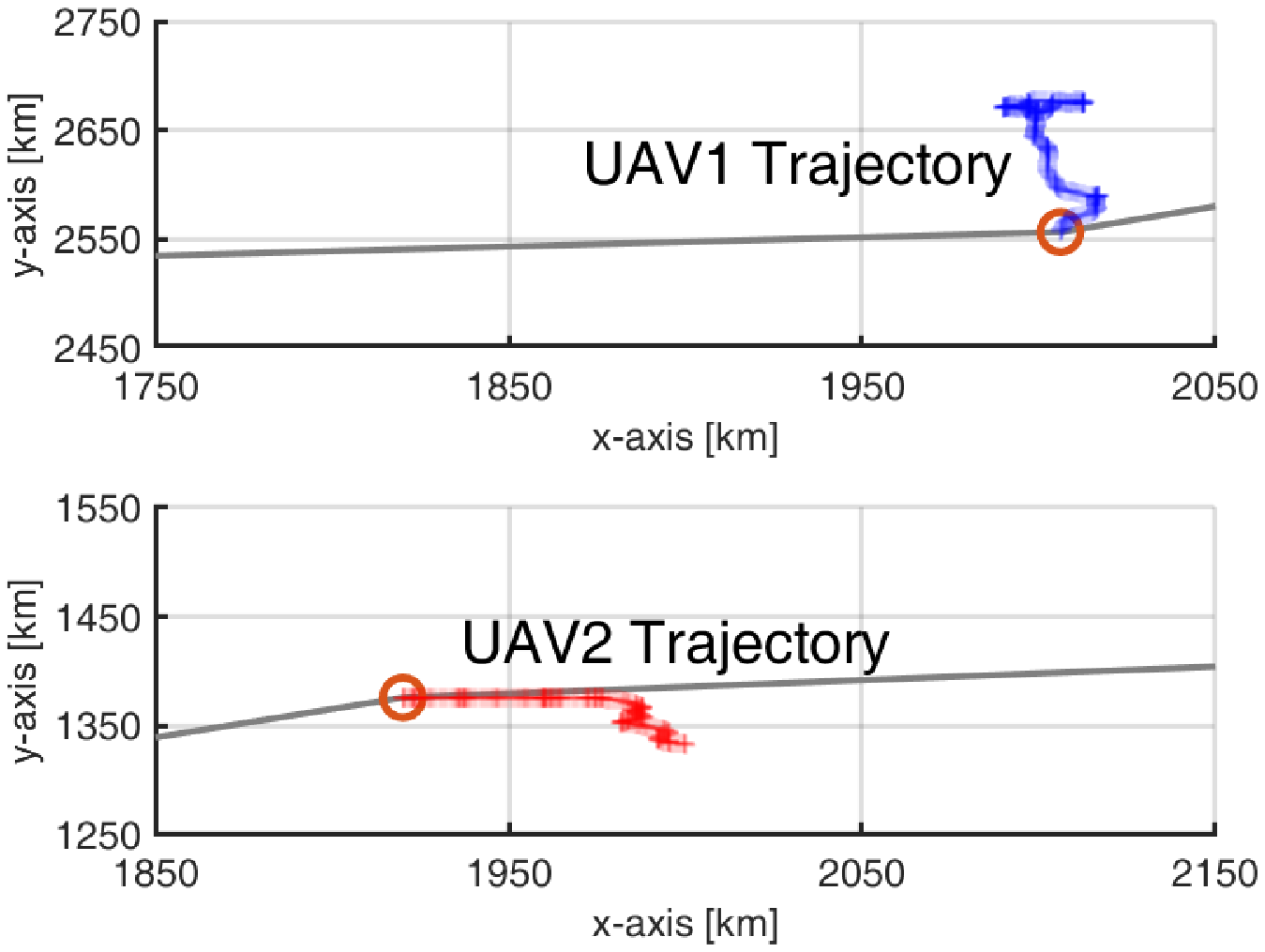}}
        \\ 
        % \vspace{-1mm}
    \bottomrule[1pt] 
\end{tabularx}
    \end{minipage}
  %   \vspace{-1.5em}
  \end{table*}

%%% Table 3
The previous result in Table ~\ref{Table_1} demonstrates that even a single fixed GR can provide a significant throughput gain. The gain can be further improved by multiple-UAV mobile relaying. In this regard, Table \ref{Table_2,3} illustrates the effectiveness of multi-agent relaying with or without cooperation across agents (i.e., UAVs or GRs). With multiple relaying agents, there exist multiple E2E communication paths by up to the number of agents. Then, cooperative scenarios correspond to (P1$^*$) with $\sigma_{\mathrm{R}} = 1 \times 10^{9}$ and $\sigma_{\mathrm{E}} = 3 \times 10^{4}$ and (P2) for SAT-UAV and SAT-Ground, respectively, in which multiple agents aim to maximize their sum-path E2E throughput. By contrast, non-cooperative scenarios are the cases where each agent aims to maximize its own-path E2E throughput. The result shows that cooperative SAT-Ground yields $1.63$x higher E2E sum throughput than its non-cooperative counterpart. This is mainly thanks to avoiding the overlapping associations with the same SAT that orthogonally splits the total bandwidth, significantly degrading the E2E sum throughput. Next, cooperative SAT-UAV achieves $1.99$x higher E2E sum throughput than cooperative SAT-Ground, corroborating the effectiveness of the mobile UAV relaying compared to the fixed GRs.

Achieving the throughput gain in Table~\ref{Table_2,3} comes at the cost of UAV movement energy consumption. To carefully identify its impact, Table \ref{Table_Proposed} focuses on SAT-UAV with $J=2$ for different objective functions, and evaluates the E2E sum throughput, the power consumption averaged across UAVs, and EE, where the numbers are average values and the error bars indicate the maximum deviations during $3$ simulation runs.
% by adjusting the weights $c_R$ and $c_E$ in the original objective function \eqref{P2_C_Delay} of (P1$^*$). 
Here, EE-max corresponds to considering the original objective function of (P1$^*$) in \eqref{P2_C_Delay}, while Rate-max and Energy-min consider only the first term or the second term of the objective function in \eqref{P2_C_Delay}, respectively. The result shows that EE-max is well-designed, in that EE-max yields the highest EE while achieving about $80$\% of Rate-max's E2E sum throughput and consuming less than $2$x average power compared to Energy-min. As observed by these results, the proposed framework well excludes undesirable extreme outcomes of MARL, e.g., motionless UAVs to maximize the EE (i.e., sum throughput divided by total energy consumption) by making the denominator zero, which also advocates the usefulness of the objective reformulation from (P1) to (P1$^*$) avoiding such extreme cases.

\begin{figure*}
  % \vspace{-1.2em}
  \centering
  \includegraphics[width=\linewidth]{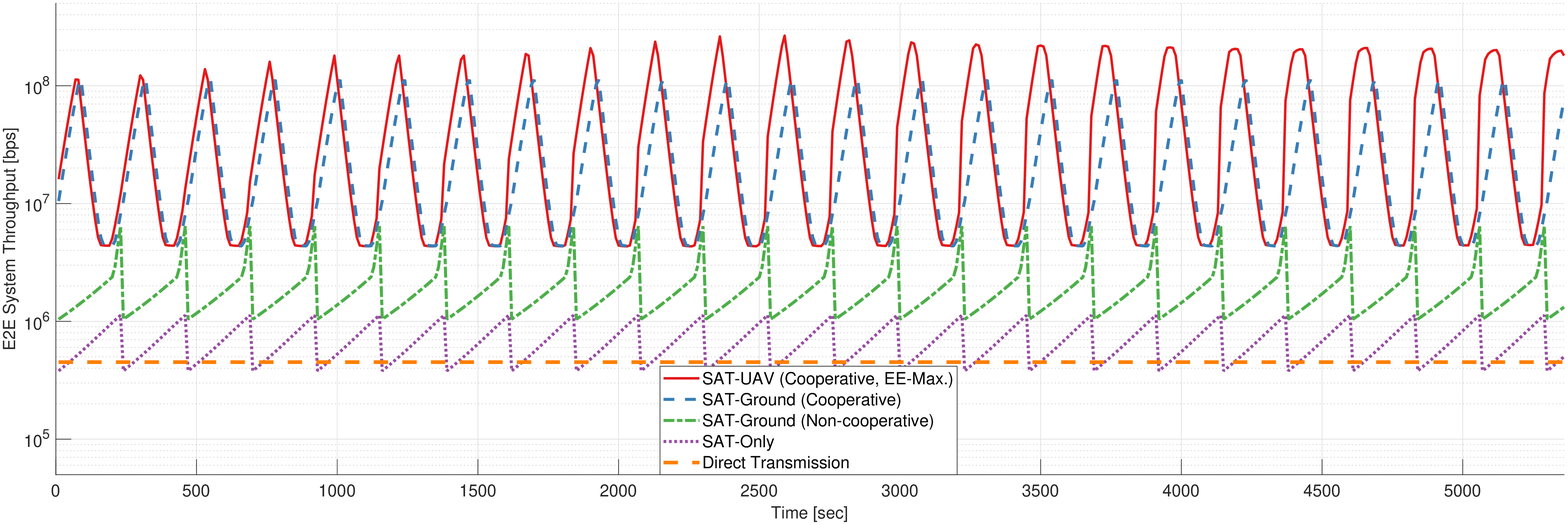}
  \caption{End-to-end system throughput of a time-varying network topology for different NTN configurations ($J=K=2$, Best-Effort).}
  \label{fig_E2E}
% 	\vspace{-1.em}
\end{figure*}

\subsection{Best-Effort Associations and Trajectories}
% \textbf{Optimal Action for SAT-UAV assisted Non-Terrestrial Network}.\quad
% \tred{[JH: The left qutation mark should be `` in LaTeX; For readability, if not necessary let's not use quatation marks throughout the paper (for some abbreviated notations, you can simply describe them via capitalization (as defined) without quatation)]}
In Table \ref{table_best-effort}, we present the results of association and trajectory in SAT-UAV during one orbital period $n=1-572$.
In this table, the association for Src-SAT1-UAV1(or UAV2)-SAT2-Dst and the trajectory of UAV1 and UAV2 is shown over time slots, for Best-Effort scenario which refers to the proposed SAT-UAV (EE-max).

Over time slots, the association for the two UAV relays in-between SAT1 and SAT2 changes according to time-varying SATs position.
Accordingly, in the optimization result, the handover occurs at a certain period as SATs circulate on an orbital lane at a constant velocity.
Overall, UAV1 selects the association with SATs relatively close to itself, while UAV2 selects the association with SATs relatively close to Src and Dst.
Specifically, in the cooperative system of SAT-UAV, one UAV occupies the optimal link, and the other UAV occupies the remaining link, which is more advantageous in terms of the E2E system throughput.
This is because the optimal policy decides not to optimize specific links, but to improve the weakest link according to \eqref{Rate_E2E}.

The optimized trajectory for UAV1 and UAV2 is shown in the bottom of this table.
Note that the blue dots represent the traces of the UAV1 trajectory, and the red dots represent the traces of the UAV2 trajectory.
The optimal trajectories draw like a straight line.
It seems likely that each UAV goes to better location for connecting with moving SATs.
We note that the E2E system throughput of the SAT-based NTN is efficiently improved with mobile UAV relaying, as identified in Table~\ref{Table_2,3}.

The best-effort scenario focuses on the E2E system throughput, rather than fairness.
If one mainly considers the fairness or link loss case, the policy should take account of proportional fairness.
In the following subsection, we further investigate the fairness-oriented scenario.

\subsection{Fairness-Oriented Associtions and Trajectories}

LEO SAT communication, which propagates through quite long link distance, may suffer severe link deterioration.
Hence, consideration for link loss can be another significant issue.
Accordingly, we further present the simulation results of the fairness-oriented scenario.
In this scenario, the objective concerns fairness, such that the policy aims to the throughput of each agent, not the entire system.

Particularly, the fairness-oriented optimization is implemented with only minor modification to the proposed reward function. 
% \tred{[JH: Throughout the paper, first introduce the clear reasoning (for which purpose, in more general), followed by its applications and details; currently sigmoid function is first applied, followed by some discussions on sigmoid]}
We re-design the reward function to improve the performance of each link fairly rather than maximizing one superior link to improve the E2E system throughput, with a \textit{Sigmoid} function.
Adopting the non-linear \textit{Sigmoid} function of $f(x) = \nicefrac{1}{\left( 1 + e^{-g(x)} \right)}$, the reward function in \eqref{Reward} slightly changes as follows.
\begin{align}
r'[n] = & f({\textstyle\sum}_{j=1}^{2} R_{\mathcal{S}, \mathcal{D}}[n]) - g({\textstyle\sum}_{j=1}^{2} P_{j}[n]\delta_{t}) \\
& - h(d_{j=1, j^{'}=2}[n]). \label{Reward_reliability}
\end{align}
% \tred{[JH: for the next time, you can use only `align,' without any `equation' and `eqnarray'; see the details: https://tex.stackexchange.com/questions/196/eqnarray-vs-align]}
% \tred{[JH: Elaborate more clearly; non-linearity is too broad]}
% We note that the modified reward function improves the performance of each link fairly rather than maximizing one superior link to improve E2E system throughput, thanks to the non-linearity of \textit{Sigmoid} function.
Note that in the optimization for a multi-agent system, the \textit{Sigmoid} function, which saturates at large values, allows each agent to achieve the reward evenly, rather than a particular agent obtaining a large reward alone.
% \tred{performance [JH: Unless necessary, let's try to avoid such an ambiguous expression]}.

Table \ref{table_fairness} shows the association and trajectory as a result of the fairness-oriented optimization.
Two agents of UAV1 and UAV2 are connected to the SATs close to them, considering their own E2E throughput.
Both links with UAV1 and UAV2 achieves almost the same E2E throughput.
In the worst case of losing one link, the fairness scenario achieves a 9.4372x higher E2E throughput than the best-effort scenario.
% Worst case for Best-effort case: 1.9865[Mbps]
% We note that the spectral efficiency of $18.747$ in [bits/Hz] is achieved with the reliability-optimization, which still achieves higher rate than fixed-ground relaying scheme.

\subsection{Impact of Time-Varying Network Topology}
% .\quad
% \tred{[JH: Highlighting best-effort may be confusing as the fairness counterpart is apparently missing]}
Fig. \ref{fig_E2E} compares the E2E system throughput over time slots with the baselines.  
As shown in this figure, the E2E system throughput result of relaying schemes fluctuates over time slots due to handover between SATs, e.g., $w_{i_{1}, j}[n] \neq w_{i_{1}, j}[n+1]$.
Direct Transmission achieves stable throughput but relatively very low throughput, since no relay terminal supports its link.
SAT-Only, which leverage LEO SAT constellation between Src and Dst, outperforms the Direct Transmission. 
SAT-Ground and SAT-UAV achieve the significantly more E2E system throughput than SAT-Only.
For all baselines except Direct Transmission, the gap between high and low peak rates is huge overall, suggesting that it may require frequent and fast handover. 

It is interesting to note that the low peak rate of SAT-Ground and SAT-UAV are similar, while SAT-UAV is superior at the high peak rate.
In other words, the E2E system throughput of mobile UAV relaying is higher but more fluctuating than that of the fixed ground relaying.
We investigate this fact in detail in the following subsection.
% For SAT-Ground (Cooperative) and SAT-UAV (Cooperative) in Fig. \ref{fig_E2E}, the behavior of both low and high peak rates is remarkable, which in detail is covered in the following subsection.

%%%%%%%%%%%%%%%%%%%%%%%% Table for Comparison study for possible link options

\subsection{Impact of Hybrid RF/FSO Links}

% \tred{[JH: throughout the paper, satellite, aerial, SAT, UAV appear many times. Please make them consistent; I prefer to only present SAT and UAV; aerial links are not consistent with satellite links, but space links; accepting such unavoidable discrepancy, I think there's no problem when we replace all `aerial's with `UAV's]}
% \textbf{Optimal Link Types for Inter SAT-UAV Link (RF vs FSO)}.\quad
In order to investigate the candidates of the inter SAT-UAV link and the ground-to-SAT link, i.e., RF and FSO, we present simulation results for the hybrid FSO/RF.

\begin{figure}
%   \vspace{-.5em}
    \centering
    \includegraphics[width=\linewidth]{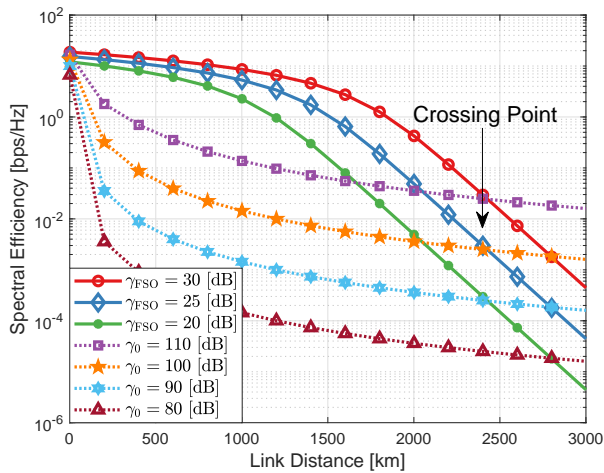}
    \caption{Spectral efficiency over distance for a given RF/FSO link.}
    \label{fig_Rate_Distance_RFFSO}
% 	\vspace{-1.em}
\end{figure}

In Fig. \ref{fig_Rate_Distance_RFFSO}, to compare RF and FSO links in hybrid FSO/RF, we firstly show the spectral efficiency over link distance for RF and FSO links.
As shown, the spectral efficiency of FSO and RF show different trends over link distance.
The spectral efficiency of FSO follows the equation \eqref{Rate_FSO}, and the spectral efficiency of RF follows \eqref{Rate_RF}.
Comparing the FSO link of $\gamma_{\mathrm{FSO}}=25$ [dB] and the RF link of $\gamma_{0}=100$ [dB], the FSO link achieves a higher rate up to around $2400$ [km] while the RF link achieves a higher rate after around $2400$ [km].
Note that the crossing point of $2400$ [km] is highlighted in Fig. \ref{fig_Rate_Distance_RFFSO}. 

\begin{figure*}
  \vspace{-.5em}
  \centering
  \includegraphics[width=\linewidth]{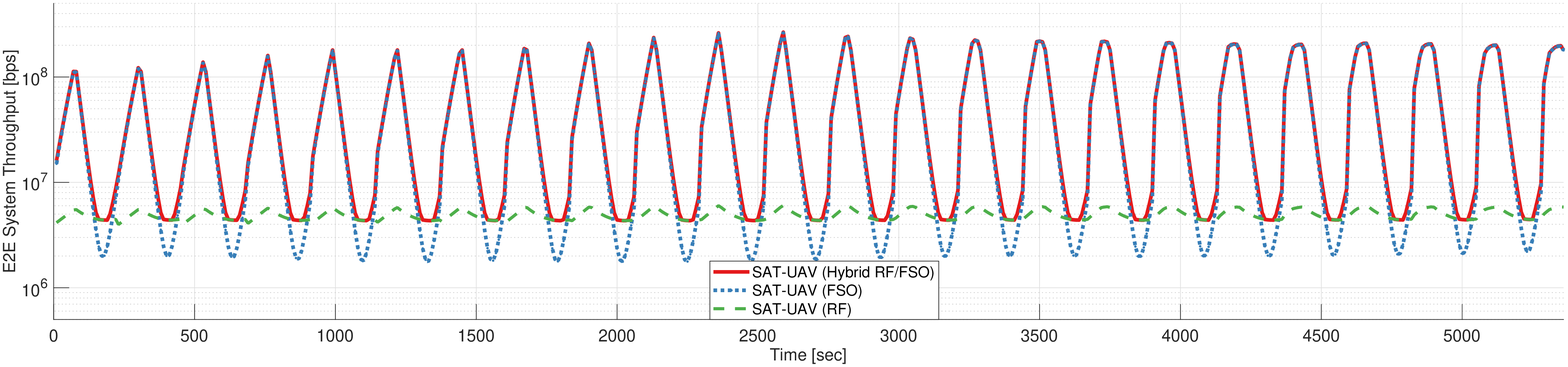}
  \caption{End-to-end system throughput of proposed SAT-UAV for different FSO/RF link types ($J=K=2$, Best-Effort).}
  \label{fig_E2E_HybridRFFSO}
% 	\vspace{-1.em}
\end{figure*}

% \tred{[JH: Specify the numerics showing that how much hybrid RF/FSO increases the avg (or peak) and worst-case throughput comared to FSO only and RF only baselines]} 
Narrowing the scope to our network scenario, we compare the system throughput of RF and FSO links in more detail.
% As identified in Fig. \ref{fig_E2E}, there is a gap between high and low peak rates, particularly in SAT-Ground and SAT-UAV.
% We present a detailed analysis of this result in Fig. \ref{fig_E2E_HybridRFFSO}.
% For a detailed analysis of this result, 
In Fig. \ref{fig_E2E_HybridRFFSO}, we show the E2E system throughput of RF, FSO, and hybrid FSO/RF in our proposed scenario of SAT-UAV.
Note that the result of SAT-UAV (Hybrid FSO/RF) in Fig. \ref{fig_E2E_HybridRFFSO} is equivalent result of SAT-UAV (Cooperative, EE-max) in Fig. \ref{fig_E2E}.
In the proposed scenario, the hybrid FSO/RF link achieves up to $62.56$x higher peak throughput and $21.09$x higher worst-case throughput, compared to RF and FSO links, respectively.
As shown in Fig. \ref{fig_E2E_HybridRFFSO}, the FSO link shows a high rate performance overall, but a very low peak rate at a specific point.
On the other hand, the RF link shows a low rate performance overall, but the difference between the high peak rate and the low peak rate is relatively small. 
% that is, it shows stable rate performance over time slots.
These facts explain that of Fig. \ref{fig_E2E}, an RF link is dominant on the low peak rate of SAT-Ground and SAT-UAV and an FSO link is dominant on the high peak rate.
% As identified in Fig. \ref{fig_E2E}, there is a gap between high and low peak rates, particularly in SAT-Ground and SAT-UAV.
% these facts explain the behavior of SAT-Ground and SAT-UAV in Fig. \ref{fig_E2E}.
% It is worth noting that these trade-off for RF and FSO are roughly ascertained, even though this results depend on network scenarios.
% Clearly, the hybrid FSO/RF link, which can use both RF and FSO links, shows the best throughput performance. 

In Table \ref{Table_RFFSO}, we show which link type is selected in each link.
Note that the result of this table is based on three replications of the trained algorithm for SAT-UAV, and more favorable link (RF or FSO) is selected in each links.
In our network scenario which involves mobile topology (such as SAT and UAV), no certain link type has an absolute dominance since the link distance for each links is time-varying.

\begin{table}
  \centering
 \resizebox{1\columnwidth}{!}{\begin{minipage}[t]{0.6\textwidth}
  \centering
    \caption{Selected link types ($J=K=2$, Best-Effort).}
	\label{Table_RFFSO}
	\begin{tabular} {l c c c c}
		\toprule[1pt]
		\textbf{Path with UAV1} & \textbf{Src-LEO1} & \textbf{LEO1-UAV1} & \textbf{UAV1-LEO2} & \textbf{LEO2-Dst} \\
		\cmidrule(lr){1-1} \cmidrule(lr){2-2} \cmidrule(lr){3-3} \cmidrule(lr){4-4}  \cmidrule(lr){5-5} 
		FSO:RF Ratio [\%] & $69.9:30.1$ & $100:0$ & $100:0$ & $69.9:30.1$  \\
		Avg. Link Distance [km] & $2099.4$  & $1643.6$  & $1842.5$  & $2099.4$  \\
		Link Distance Range & \tikz{
\draw[gray,line width=.3pt] (0,0) -- (1.0,0);
\draw[white, line width=0.01pt] (0,-2pt) -- (0,2pt);
\draw[black,line width=1pt] (0.2217,0) -- (0.4859,0);
\draw[black,line width=1pt] (0.2217,-2pt) -- (0.2217,2pt);
\draw[black,line width=1pt] (0.4859,-2pt) -- (0.4859,2pt);
\draw[red,densely dotted,line width=1pt] (0.4,-4pt) -- (0.4,4.5pt);
}  
& \tikz{
\draw[gray,line width=.3pt] (0,0) -- (1.0,0);
\draw[white, line width=0.01pt] (0,-2pt) -- (0,2pt);
\draw[black,line width=1pt] (0.2259,0) -- (0.3701,0);
\draw[black,line width=1pt] (0.2259,-2pt) -- (0.2259,2pt);
\draw[black,line width=1pt] (0.3701,-2pt) -- (0.3701,2pt);
\draw[red,densely dotted,line width=1pt] (0.4,-4pt) -- (0.4,4.5pt);}  
& \tikz{
\draw[gray,line width=.3pt] (0,0) -- (1.0,0);
\draw[white, line width=0.01pt] (0,-2pt) -- (0,2pt);
\draw[black,line width=1pt] (0.2635,0) -- (0.3691,0);
\draw[black,line width=1pt] (0.2635,-2pt) -- (0.2635,2pt);
\draw[black,line width=1pt] (0.3691,-2pt) -- (0.3691,2pt);
\draw[red,densely dotted,line width=1pt] (0.4,-4pt) -- (0.4,4.5pt);}
& \tikz{
\draw[gray,line width=.3pt] (0,0) -- (1.0,0);
\draw[white, line width=0.01pt] (0,-2pt) -- (0,2pt);
\draw[black,line width=1pt] (0.2217,0) -- (0.4859,0);
\draw[black,line width=1pt] (0.2217,-2pt) -- (0.2217,2pt);
\draw[black,line width=1pt] (0.4859,-2pt) -- (0.4859,2pt);
\draw[red,densely dotted,line width=1pt] (0.4,-4pt) -- (0.4,4.5pt);}  \\
% & \tikz{\draw[black] (0.6,-0.1pt) -- (0.6,0.1pt)node[anchor=north] {\tiny{2400 [km]}}} &  &  &  \\ 
\midrule
\textbf{Path with UAV2} & \textbf{Src-LEO1} & \textbf{LEO1-UAV2} & \textbf{UAV2-LEO2} & \textbf{LEO2-Dst} \\
\cmidrule(lr){1-1} \cmidrule(lr){2-2} \cmidrule(lr){3-3} \cmidrule(lr){4-4}  \cmidrule(lr){5-5} 
FSO:RF Link Ratio [\%] & $100:0$ & $88.1:11.9$ & $13.1:86.9$ & $100:0$  \\
Avg. Link Distance [km] & $907.6$  & $2027.1$  & $2949.3$  & $907.6$  \\
Link Distance Range & 
\tikz{
\draw[gray,line width=.3pt] (0,0) -- (1.0,0);
\draw[white, line width=0.01pt] (0,-2pt) -- (0,2pt);
\draw[black,line width=1pt] (0.1239,0) -- (0.2110,0);
\draw[black,line width=1pt] (0.1239,-2pt) -- (0.1239,2pt);
\draw[black,line width=1pt] (0.2110,-2pt) -- (0.2110,2pt);
\draw[red,densely dotted,line width=1pt] (0.4,-4pt) -- (0.4,4.5pt);}    
& \tikz{
\draw[gray,line width=.3pt] (0,0) -- (1.0,0);
\draw[white, line width=0.01pt] (0,-2pt) -- (0,2pt);
\draw[black,line width=1pt] (0.2607,0) -- (0.4426,0);
\draw[black,line width=1pt] (0.2607,-2pt) -- (0.2607,2pt);
\draw[black,line width=1pt] (0.4426,-2pt) -- (0.4426,2pt);
\draw[red,densely dotted,line width=1pt] (0.4,-4pt) -- (0.4,4.5pt);}  
& \tikz{
\draw[gray,line width=.3pt] (0,0) -- (1.0,0);
\draw[white, line width=0.01pt] (0,-2pt) -- (0,2pt);
\draw[black,line width=1pt] (0.3669,0) -- (0.6229,0);
\draw[black,line width=1pt] (0.3669,-2pt) -- (0.3669,2pt);
\draw[black,line width=1pt] (0.6229,-2pt) -- (0.6229,2pt);
\draw[red,densely dotted,line width=1pt] (0.4,-4pt) -- (0.4,4.5pt);}  
& \tikz{
\draw[gray,line width=.3pt] (0,0) -- (1.0,0);
\draw[white, line width=0.01pt] (0,-2pt) -- (0,2pt);
\draw[black,line width=1pt] (0.1239,0) -- (0.2110,0);
\draw[black,line width=1pt] (0.1239,-2pt) -- (0.1239,2pt);
\draw[black,line width=1pt] (0.2110,-2pt) -- (0.2110,2pt);
\draw[red,densely dotted,line width=1pt] (0.4,-4pt) -- (0.4,4.5pt);}  \\
\bottomrule[1pt]
\end{tabular}
  \end{minipage}}
\end{table}

%%%%%%%%%%%%%%%%%%%%%%%%%%%%%%%%%%%%%%%%%%%%%%%%%%%%%%%%%%%%%%%%%%%%%
%%%%%%%%%%%%%%%%%%%%%%%%%%%%%%%%%%%%%%%%%%%%%%%%%%%%%%%%%%%%%%%%%%%%%
%%%%%% Conclusion %%%%%%
%%%%%%%%%%%%%%%%%%%%%%%%%%%%%%%%%%%%%%%%%%%%%%%%%%%%%%%%%%%%%%%%%%%%%
%%%%%%%%%%%%%%%%%%%%%%%%%%%%%%%%%%%%%%%%%%%%%%%%%%%%%%%%%%%%%%%%%%%%%

\section{Conclusion} \label{conclusion}

In this article, we investigated a satellite-UAV integrated hybrid FSO/RF NTN, and proposed a novel centralized-Critic MARL solution to maximize the E2E sum throughput while minimizing the UAV energy consumption, i.e., EE maximization. Numerical results corroborated the effectiveness of jointly controlling the UAV movements and associations with SATs. Not only the EE performance but also the UAV trajectories depend highly on the objective function choices. Optimizing the objective function for a specific task could thus be an interesting topic for future research. Besides, the MARL simulations revealed that the training convergence time increases with the number of agents. To scale up our proposed framework particularly under dynamic environments requiring frequent retraining, it is promising to apply federated learning methods that accelerate the training convergence. 
% Last but not least, our simulations showed that FSO links are effective in improving the maximum throughput of SAT-LEO integrated NTNs for long-distance communications, while RF links are useful for improving the worst-case throughput for relatively short-distance communications. 
Last but not least, our simulations showed that FSO links are effective in improving the maximum throughput of SAT-LEO integrated NTNs, while RF links are useful for improving the worst-case throughput.
Depending on application-specific and spatial channel characteristics, optimizing FSO/RF link types is therefore mandated for enabling fast and reliable NTNs in beyond 5G.

\bibliographystyle{IEEEtran} 
\bibliography{Reference_Paper4_Journal}
%%%%%%%%%%%%%%%%%%%%%%%%%%%%%%%%%%%%%%%%%%%%%%%%%%%%%%%%%%%%%%%%%%%%%%%%%%%%%%%%%%%%%%%%%%%%%%%%%%%%%%%%%%%%%%%%%%%%%%%%%%%%%%%%%%%%%%%%%%%%%%%%%%%%%%%%%%%%%%%%%%%%%%%%%%%%%%%%%

%%%%%%%%%%%%%%%%%%%%%%%%%%%%%%%%%%%%%%%%%%%%%%%%%%%%%%%%%%%%%%%%%%%%%%%%%%%%%%%%%%%%%%%%%%%%%%%%%%%%%%
%%%%%%%%%%%%%%%%%%%%%%%%%%%%%%%%%%%%%%%%%%%%%%%%%%%%%%%%%%%%%%%%%%%%%%%%%%%%%%%%%%%%%%%%%%%%%%%%%%%%%%

\end{document}